\documentclass{jfm}
\usepackage{graphicx,color}
\usepackage{float}
\usepackage{epstopdf, epsfig}
\usepackage{caption,subcaption}
\usepackage[intlimits]{amsmath}
\usepackage{lscape}

\usepackage[utf8]{inputenc} 
\usepackage[T1]{fontenc}    
\usepackage{hyperref}       
\usepackage{url}            
\usepackage{booktabs}       
\usepackage{amsfonts}       
\usepackage{nicefrac}       
\usepackage{microtype}      
\usepackage{lipsum}		
\usepackage{graphicx}
\usepackage{natbib}
\usepackage{amsmath}
\usepackage{mathtools}
\usepackage{empheq}
\usepackage[most]{tcolorbox}

\newcommand{\hats}{\hat{\mathbf{s}}}
\newcommand{\hatn}{\hat{\mathbf{n}}}
\newcommand{\hatt}{\hat{\mathbf{t}}}
\newcommand{\hatb}{\hat{\mathbf{b}}}
\newcommand{\om}{\boldsymbol{\omega}}
\newcommand{\uvec}{\mathbf{u}}

\newcommand\Web{\mbox{\textit{We}}}
\newcommand\Mar{\mbox{\textit{Ma}}}

\shorttitle{Turbulent surfactant-laden jets}
\shortauthor{C. R. Constante-Amores}

\title{
Direct numerical simulations of  
turbulent jets:
vortex-interface-surfactant interactions}

\shortauthor{C. R. Constante-Amores}
\author[Constante-Amores {\it et al.}]
{C. R. Constante-Amores$^{1,2}$, 
\ns T. Abadie$^1$, 
\ns L. Kahouadji$^1$,\ns S. Shin$^3$,\ns J. Chergui$^4$, \ns D. Juric$^{4,5}$,
\ns A. A. Castrejon-Pita$^2$
and O. K. Matar$^1$
\corresp{\email{o.matar@imperial.ac.uk}}}%

\affiliation{
$^1$Department of Chemical Engineering, Imperial College London,  London SW7 2AZ, UK \\[\affilskip]
$^2$Department of Engineering Science, University of Oxford, Oxford OX1 3PJ, UK\\[\affilskip]
$^3$Department of Mechanical and System Design Engineering, Hongik University, KR
\\[\affilskip]
$^4$ Universit\'e Paris Saclay, Centre National de la Recherche Scientifique (CNRS), Laboratoire Interdisciplinaire des Sciences du Num\'erique (LISN), 91400 Orsay, France
\\[\affilskip]
$^5$Department of Applied Mathematics and Theoretical Physics, University of Cambridge,  Cambridge CB3 0WA, UK

}

\begin{document}

\maketitle

\begin{abstract}
We study the effect of insoluble surfactants on the spatio-temporal evolution  
of  turbulent jets. We use three-dimensional numerical simulations and employ an interface-tracking/level-set method that accounts for surfactant-induced Marangoni stresses. The present study builds on our previous work (Constante-Amores {\it et al.}, 2021, J. Fluid Mech., {\bf 922}, A6) in which we examined in detail the vortex-surface interaction in the absence of surfactants.  
Numerical solutions are obtained for a wide range of Weber  and elasticity numbers in which vorticity production is  generated by surface deformation and surfactant-induced Marangoni stresses. The present work demonstrates, for the first time, the crucial role of Marangoni stresses, brought about by surfactant concentration gradients, in the formation of coherent, hairpin-like vortex structures. 
These structures have a profound influence on the development of the three-dimensional interfacial dynamics. 
We also present theoretical expressions for the mechanisms that influence the rate of production of circulation in the presence of surfactants for a general, three-dimensional, two-phase flow and highlight the dominant
contribution surfactant-induced Marangoni stresses.

\end{abstract}

\section{Introduction}
The atomisation of a liquid jet has driven 
interest in the fluid mechanics community because of its occurrence in both natural and industrial applications (e.g., propellant combustion, pharmaceutical sprays, etc.).  
The process  results in a `cascade mechanism' for fluid fragmentation \citep{Plateau_1873,Eggers_rmp_1997,Marmottant_jfm_2004,Constante-Amores_prf_2020}: from the growth of linear modes through a Kelvin-Helmholtz instability to the development of nonlinearities leading to capillary breakup events via long filament pinch-off that can be modulated by a Rayleigh-Plateau instability or controlled by an `end-pinching' mechanism.
The understanding of the interfacial dynamics relies on the characterisation of the vortex-interface interactions. For instance, \citet{Jarrahbashi_jfm_2016}, \citet{Zandian_jfm_2018,zandian_sirignano_hussain_2019} and \citet{constante_jets}
reported that their interplay determines the interfacial dynamics for turbulent jets;  \citet{Hoepffner_jfm_2013}  showed that vorticity production results in a change in  the capillary retraction of a liquid thread.
Theoretically, \citet{longuet_1992}, \citet{wu_1995}, \citet{lundgren_koumoutsakos_1999} demonstrated that  vorticity production  depends 
on the velocity field and  the interfacial curvature for the condition of zero shear stress at a free surface. Additionally, 
\citet{brons_2014} and \citet{terrington_hourigan_thompson_2020,terrington_jfm_2021} extended the previous results to 
show that interfacial curvature effects, viscosity and  density difference across the interface are the only  mechanisms driving  vorticity production. Recently, \citet{Fuster_2021} also demonstrated
the role of interfacial curvature and density differences across the interface with identical dynamical viscosity via two-dimensional, non-axisymmetric numerical studies. 

We note that the studies mentioned in the foregoing involve a constant surface tension and therefore do not support the formation of Marangoni gradients. Liquid streams, however, are invariably 
contaminated with surface-active-agents (surfactants), deliberately-placed or naturally-occurring, which give rise to surface tension gradients, and subsequently Marangoni-induced flow  \citep{manikantan_squires_2020}. 
While the atomisation of uncontaminated liquid jets has received significant attention in the literature  
\citep{Herrmann_jcp_2010,Desjardins_as_2010,Jarrahbashi_pof_2014, Jarrahbashi_jfm_2016,Zandian_jfm_2018,zandian_sirignano_hussain_2019,constante_gfm,constante_jets}, the effect of surfactant on their dynamics remains far less studied. 
The multi-scale nature of the flow, and the  
complex coupling between the surfactant concentration fields and interfacial topology  complicate its  experimental scrutiny. 
This can be alleviated via
the use of high-fidelity simulations which can unravel 
the delicate interplay among the different physical mechanisms across the relevant scales. 

Through the use of state-of-the-art imaging techniques,  
\citet{Kooij_2018}, \citet{Sijs_2019}, and \citet{Sijs_2021}  showed that the presence of surfactants influences the interfacial fragmentation during atomisation and decreases the mean-droplet size 
in agreement with \citet{Butler} and  \citet{Ariyapadi}.  
All the previous studies, however, have not reported the role of Marangoni stresses 
which the present paper will address for the case of an insoluble surfactant. 
Although the presence of surfactants can also induce both shear and dilatational surface rheological effects \textcolor{black}{(discussed below)}, these effects will not be considered 
in this study. Nonetheless, we will use transient numerical simulations to demonstrate that the Marangoni stresses influence the production of vorticity near the interface, and 
modify the interface-vortex interactions and the three-dimensional destabilisation of the jet. 
In order to focus on the role of Marangoni stresses in the jet dynamics, we will study the case of a jet of one fluid issuing into another characterised by equal densities and viscosities. 

\textcolor{black}{There has been significant scientific interest in studying the role of surfactants in the destabilization and fragmentation of non-turbulent liquid jets of pure Newtonian fluids (see for example \citet{Eggers_prl_1993, Lister_Stone_pf_1998,craster_pof_2002,Liao_2004,Craster_jfm_2009}). Those authors have shown the existence of multiple intermediate or transient scaling regimes which are not altered by the presence of surfactants as they are convected away from the pinch-off region. However, \citet{Mcgough_prl_2006} and \cite{Kamat_prf_2018} showed the formation of micro threads, which connect drops during the surfactant-induced thinning. Additionally, the presence of surfactants not only give rise to gradients in surface tension and hence tangential interfacial stresses, but also induce both shear and dilatational surface rheological effects. Recently, work by \citet{wee_2021} and \citet{martinez_calvo} have analysed theoretically the influence of surface viscosities on the pinch-off dynamics of a jet of an incompressible Newtonian liquid that is surrounded by a passive gas.}

The rest of this paper is structured as follows: in Section \ref{Numerical}, the problem formulation, governing dimensionless parameters, and numerical method are introduced. 
Section \ref{sec:Results} provides a discussion of the results, and concluding remarks are given in Section \ref{sec:Con}.

\section{Problem formulation and numerical method\label{Numerical}}
Since the aim here is to shed light on the different mechanisms that influence the production of vorticity near the interface in the presence of surfactants, we present a general theoretical description of vorticity and circulation in a three-dimensional control volume enclosing an interface using Lighthill’s and Lyman’s flux definitions \citet{terrington_jfm_2021}.
We also provide a brief description of the numerical technique which is used to carry out the computations. Finally, we provide motivation for the choice of physical and physico-chemical parameters made in the present work. 
%

\begin{figure}
\begin{center} 
\begin{tabular}{cc}
\includegraphics[width=0.40\linewidth]{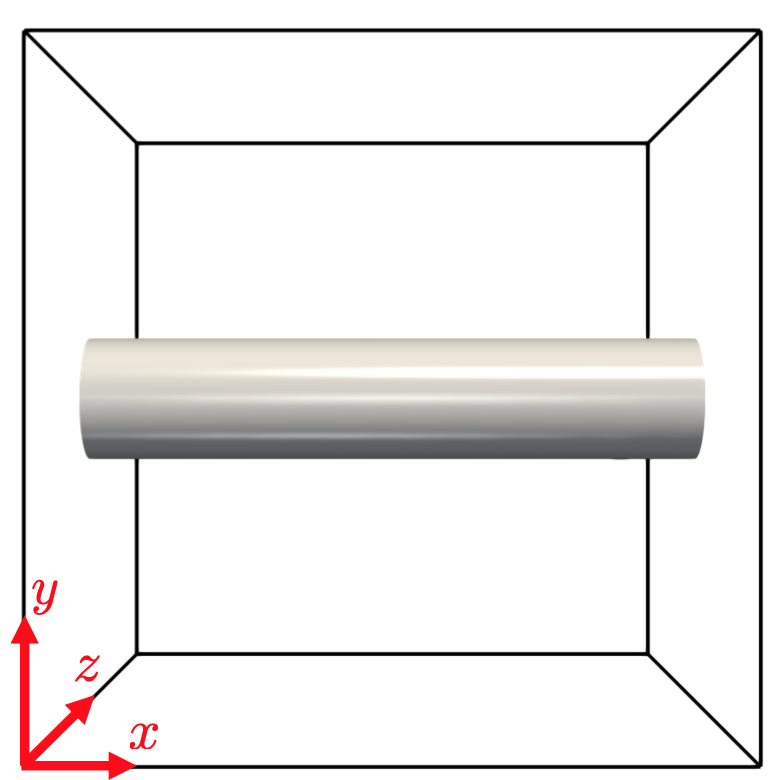}     & \includegraphics[width=0.45\linewidth]{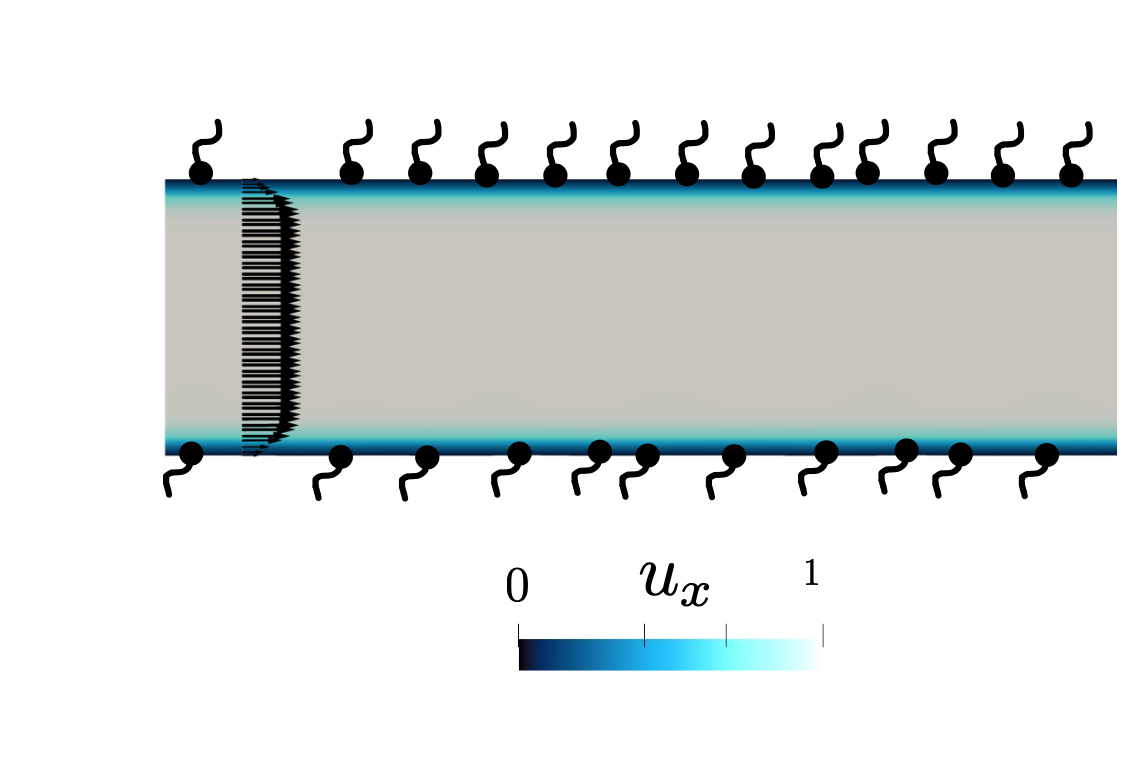}\\
(a) & (b)
\end{tabular}
\end{center} 
\caption{(a) Initial interfacial shape, highlighting the computational domain of size $(5D)^3$  in a three-dimensional Cartesian space $\mathbf{x} = (x, y, z)$; 
(b) schematic representation of the problem in the $x-y$ ($z=2.5D$) plane showing the initial ($t=0$)  streamwise velocity profile, $u_x$, and a representation of a monolayer of an insoluble surfactant.\label{configuration}} 
\end{figure}

\subsection{Problem formulation and numerical method}
Figure  \ref{configuration} shows a representation of the flow configuration  considered in this study
in a three-dimensional Cartesian domain $\mathbf{x} = \left(x, y, z \right)$: a liquid segment is initialised as a cylinder of diameter $D$, with a finite length, i.e. $5D$, in the positive $x-$(streamwise) direction. Such an approach has been used by \citet{Desjardins_as_2010}, \citet{Jarrahbashi_jfm_2016}, 
and \citet{Zandian_jfm_2018}  for planar and cylindrical jets.  
We will focus on the case of insoluble surfactants, which enables us to isolate the surfactant-induced Marangoni dynamics during the atomisation of the jet. We acknowledge, however, that experimental studies feature 
soluble surfactants which are dissolved in the liquid that issues from a nozzle to form the jet 
and that the sorption kinetics control the surfactant interfacial concentration  
adding extra layers of complexity. 

The dimensional governing equations, which can be found in the work of \cite{Shin_jcp_2018}, are rendered dimensionless  
using the following scalings:
\begin{equation}
\quad \tilde{\mathbf{x}}=\frac{\mathbf{x}}{D},
\quad \tilde{t}=\frac{t}{t_{r}}, 
\quad \tilde{\textbf{u}}=\frac{\textbf{u}} {U},
\quad \tilde{p}=\frac{p}{\rho U^2}, 
\quad \tilde{\sigma}=\frac{\sigma}{\sigma_s},
\quad \tilde{\Gamma}=\frac{\Gamma}{\Gamma_\infty},
\label{scalings}
\end{equation}
\noindent	
where, $t$, $\textbf{u}$, and $p$ stand for time, velocity, and pressure, respectively; here, the dimensionless variables are designated using tildes. The physical parameters correspond to the liquid density $\rho$, viscosity, $\mu$, surface tension, $\sigma$, surfactant-free surface tension, $\sigma_s$, initial jet diameter, $D$, and injection velocity,
$U$. Hence, the characteristic time scale based on the injection velocity is $t_r=D/U$. 
The interfacial surfactant concentration, $\Gamma$, is scaled with the saturation interfacial concentration, $\Gamma_{\infty}$. 

Using the relations in Eq. (\ref{scalings}), the dimensionless form of the continuity and momentum equations is respectively expressed as:
\begin{equation}\label{div}
 \nabla \cdot \tilde{\textbf{u}}=0,
\end{equation}
\begin{eqnarray}\label{NS}
 \tilde{\rho}\left(\frac{\partial \tilde{\textbf{u}}}{\partial \tilde{t}}+\tilde{\textbf{u}} \cdot\nabla \tilde{\textbf{u}}\right)  & = & - \nabla \tilde{p}  +  \frac{1}{\Rey}\nabla\cdot  \left [ \tilde{\mu} (\nabla \tilde{\textbf{u}} +\nabla \tilde{\textbf{u}}^T) \right ] \nonumber\\
 && +\frac{1}{\Web} \int_{\tilde{A}(\tilde{t})} \left(
\tilde{\sigma} \tilde{\kappa} \hats 
 +  
 \nabla_s  \tilde{\sigma}\right)\delta \left(\tilde{\textbf{x}} - \tilde{\textbf{x}}_{_f}  \right) d\tilde{A} , 
\end{eqnarray}
\noindent where $\tilde\kappa$ represents the interface curvature, $\nabla_s$ the surface gradient operator, and $\hats$ the outward-pointing unit normal to the interface. Here, $\tilde{\textbf{x}}_f$ is the parametrization of the time-dependent interface area $\tilde{A}(\tilde{t})$, where $\delta   (\tilde{\textbf{x}}-\tilde{\textbf{x}}_f)$ is the three-dimensional Dirac delta function. 
The density, $\tilde{\rho}$, and viscosity, $\tilde{\mu}$, are given by the following expressions
    \begin{equation}
    \label{prop}
    \tilde{\rho}\left( \tilde{\textbf{x}},\tilde{t}\right)=\frac{\rho_g}{\rho_l} + \left(1 -\frac{\rho_g}{\rho_l}\right) H\left( \tilde{\textbf{x}},\tilde{t}\right),~
    \tilde{\mu}\left( \tilde{\textbf{x}},\tilde{t}\right)=\frac{\mu_g}{\mu_l}+ \left(1 -\frac{\mu_g}{\mu_l}\right) H\left( \tilde{\textbf{x}},\tilde{t}\right),
    \end{equation}
where $H\left( \tilde{\textbf{x}},\tilde{t}\right)$ represents a smoothed Heaviside function; this is zero in the gas phase and unity in the liquid phase, while the subscripts $l$ and $g$ designate the individual liquid and gas phases, respectively. 

The dimensionless surfactant transport is given by:
 \begin{equation} \label{Gamma_Eq}
 \frac{\partial \tilde{\Gamma}}{\partial \tilde{t}}+\nabla_s \cdot (\tilde{\Gamma}\tilde{\textbf{u}}_{\text{t}})=\frac{1}{Pe_s} \nabla^2_s \tilde{\Gamma},
 \end{equation}
 \noindent where $\tilde{\textbf{u}}_{\text{t}}=(\tilde{\textbf{u}}_{\text{s}}\cdot\textbf{t})\textbf{t}$ is the tangential velocity vector in which $\tilde{\textbf{u}}_{\text{s}}$ is the surface velocity and ${\mathbf{t}}$ is the unit tangent to the interface. 

The scaling results in the following 
dimensionless groups: 
\begin{equation}
Re =\frac{\rho U D}{\mu}, ~~~
\quad We =\frac{\rho U^2 D}{\sigma_s},  ~~~
Pe_s=\frac{ U D}{\mathcal{D}_s},~~~
\beta_s= \frac{\Re \mathcal{T} \Gamma_\infty}{\sigma_s}, 
\end{equation}
where $Re$, $We$, and $Pe_s$ denote the Reynolds, Weber, and (interfacial) Peclet numbers, respectively, while 
$\beta_s$ is a surfactant elasticity number which represents a measure of the sensitivity of $\sigma$ to $\Gamma$; here,
$\Re$ is the ideal gas constant value $8.314$ J K$^{-1}$ mol$^{-1}$,  $T$ denotes temperature and $\mathcal{D}_s$ refers to the diffusion coefficient. 

To describe the relation between $\tilde{\sigma}$ and $\tilde{\Gamma}$, 
we use the non-linear Langmuir equation: 
\begin{equation}
    \tilde{\sigma}=1 + \beta_s \ln{ (1 -\tilde{\Gamma})}.
\end{equation}
Surface tension gradients 
are expressed as a function of $\tilde{\Gamma}$ as
\begin{equation}
    \nabla_s \tilde{\sigma} /\Web =-\Mar / (1-\tilde{\Gamma}) \nabla_s\tilde{\Gamma}, \label{eq:def_tau}
\end{equation}
where $Ma=\beta_s/We=\Re T \Gamma_\infty/\rho U^2 D$ is a Marangoni parameter.  

The three-dimensional numerical simulations were performed by solving the two-phase Navier-Stokes equations in the Cartesian domain $\mathbf{x} = \left(x, y, z \right)$. 
A hybrid front-tracking/level-set method was used to treat the interface where surfactant transport was resolved in the plane of the interface 
\citep{Shin_jcp_2018}. The simulations are initialised with a turbulent velocity profile in the liquid jet segment (i.e., 
$u(r) = 15/14 ~U (1- (r/ (D/2))^{28})$ 
\citep{constante_jets}. Solutions are sought subject to Neumann boundary conditions on all variables at the lateral boundaries, and periodic boundary conditions in the $x-$(streamwise) direction. The computational domain is a cube with dimensions $(5D)^3$ globally resolved by a uniform grid of $(786)^3$ cells; see Appendix of \cite{constante_jets} for details of  
mesh-refinement studies and validation of the numerical method. 
This method has also been widely tested for surfactant-laden flows 
\citep{Shin_jcp_2018,Constante-Amores_prf_2020,constanteamores2020bb,constante_2022,batchvarov2020threedimensional} and
the numerical simulations in this study conserve fluid volume and surfactant mass with a relative error of less than  $10^{-3}\%$.

Next, we motivate the values of material properties by looking into the sources for vorticity production at an interface in a three-dimensional framework. 
These sources are due to 
differences in density (i.e., baroclinic effect) and viscosity, surface tension forces (due to gradients of curvature along the interface), and Marangoni stresses. 
Thus, to unravel the importance of the surfactant-induced Marangoni stresses on the vortex-surface-surfactant interactions, we focus on situations in which surface tension forces and Marangoni stresses are the only physical mechanisms responsible for vorticity production at the interface,  i.e., the jump  in material properties across the interface is zero \citep{Fuster_2021}.  
This  is a realistic assumption for immiscible liquid-liquid systems exemplified by 
the silicone oil-water pairing used  by  \citet{Ibarra_2017} and \citet{ibarra_jfm_2020} in their two-phase, stratified pipe flow experiments.

The values of the dimensionless quantities are consistent with experimentally-realisable systems and are chosen to  ensure a full coupling between surfactant-induced Marangoni stresses and interfacial diffusion, and inertia. 
We set $Re=5000$ to ensure a rich dynamics 
\citep{constante_jets} and  focus on the range $50 <We< 1000$ to account for realistic values of $\sigma_s$, i.e. $\mathcal{O}(10^{-3}) <\sigma_s<\mathcal{O}(10^{-1}) ~{\rm N}~{\rm m}^{-1}$. 
The parameter $\beta_s$ is related to 
$\Gamma_\infty$ and therefore the critical micelle concentration (CMC), i.e. $\Gamma_\infty \sim \mathcal{O}(10^{-6})$ mol m$^{-2}$  for NBD-PC (1-palmitoyl-2-{12-[(7-nitro-2-1,3-benzoxadiazol-4-yl)amino]dodecanoyl}-sn-glycero-3 -phosphocholine) \citep{Strickland2015}; thus, we have explored the range of $0.1 <\beta_s<0.9$ which corresponds to CMC in the range $ \mathcal{O}(10^{-7})<$ CMC $<\mathcal{O}(10^{-6})$ mol m$^{-2}$, for typical values of $\sigma_s$.  
We have set $Pe_s=10^2$ following 
\citet{batchvarov2020effect} and \citet{Constante-Amores_prf_2020} who showed that the interfacial dynamics are weakly-dependent on $Pe_s$ beyond this value.

\subsection{Vorticity and circulation}

\begin{figure}
	\centering
	\includegraphics[scale=0.37]{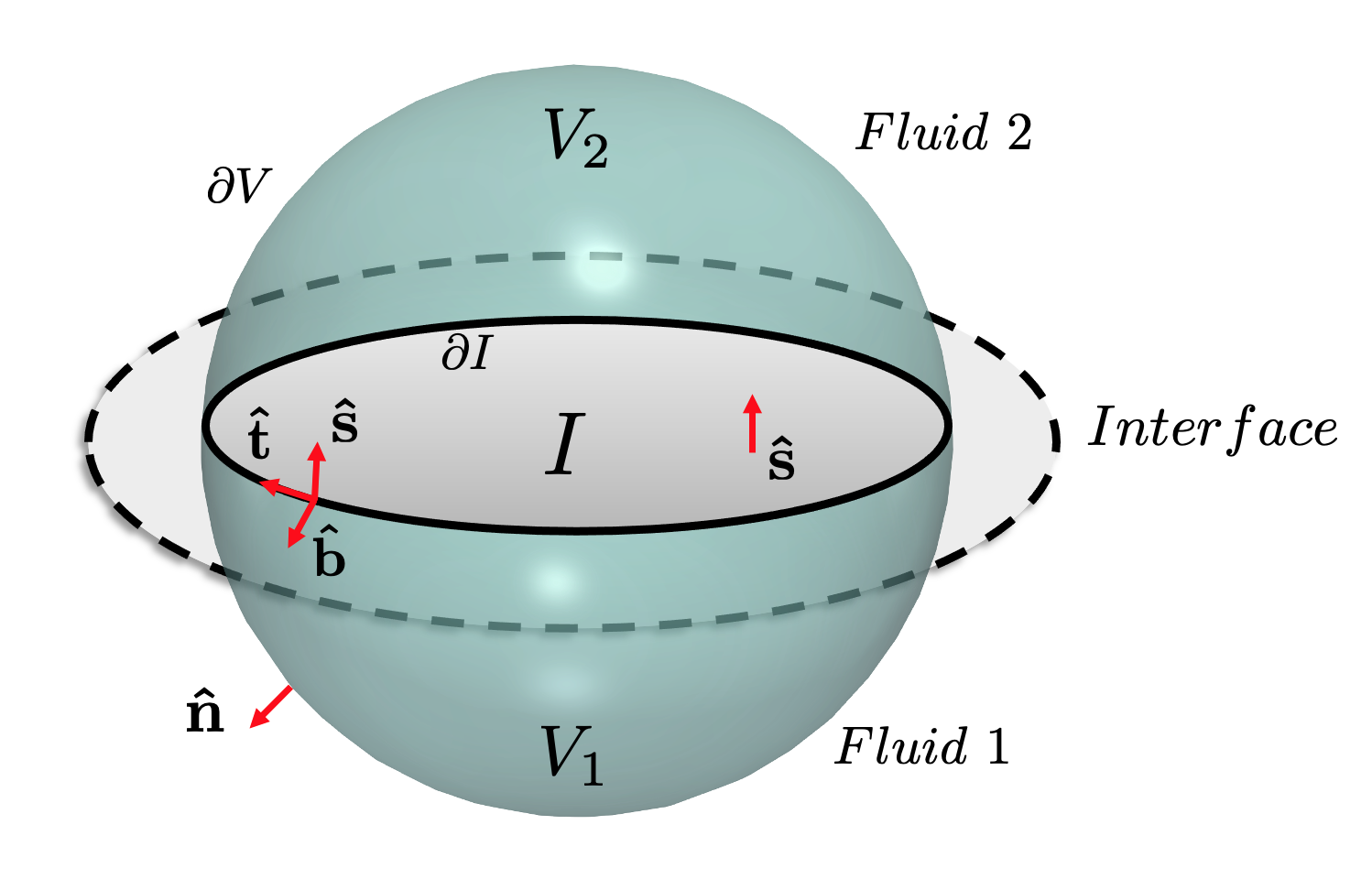}
	\caption{Schematic showing a volume $V$ with a surface  $\partial V$ which encloses two fluids separated by an interface surface $I$. \textcolor{black}{Here, the two smaller control volumes $V_1$ and $V_2$ refer to the control volume of each fluid. Local unit vectors to the interface are $\hatb$ , $\hats$ and $\hatt$; $\hatn$ corresponds to the unit normal vector to the control volume $\partial V$, $\hatb$ is a vector tangent to $I$, but orthogonal to $\partial I$, and $\hatt$ is the unit tangent vector to the boundary curve $\partial I$.    }} 
	\label{fig:fig3}
\end{figure}


\textcolor{black}{This section aims to present a general description of vorticity generation in a three-dimensional framework. We present a  theoretical formulation which builds upon the inviscid theory presented by \citet{Morton}  for near-interface vorticity generation  
in three dimensions. 
For inviscid fluids, the rate of generation of vorticity is a result of  the relative tangential acceleration of fluid on each side of the interface, which is caused by tangential pressure gradients or body forces. 
The present theoretical formulation is expressed as a conservation law for circulation in a control volume that  includes a general surface. The total circulation is expressed as the vorticity from the fluids from both sides of the interface as well as circulation contained in the interface.  }

\textcolor{black}{It is well known that curvature induces the generation of vorticity as
the normal viscous stress at an interface is balanced by the capillary  pressure.  However,  the presence of surfactant leads to a reduction in surface tension, which influences this mechanism. Furthermore, surfactant interfacial concentration variations induce 
surface tension gradients, and, as we will show,  lead to a new route for vorticity generation near the interface.
Once we have presented our theoretical expressions for a general three-dimensional surface, we will simplify them for the limiting case in which the jump in the tangential and normal components of the velocity across the interface vanish; this is the case for identical material properties such as density and viscosity. This assumption  will help to  shed some light on the crucial role of the Marangoni-induced vorticity generation mentioned above. Future studies should extend our work to situations featuring density and viscosity contrasts.}

In order to examine the effect of the surfactant on the vorticity near the interface, we consider 
a fixed, three-dimensional (3D) control volume $V$ bounded by a closed surface of area $\partial V$ with an outward-pointing unit normal $\hatn$ (see figure \ref{fig:fig3}). This volume encloses regions of the incompressible fluids 1 and 2, of volumes, $V_1$ and $V_2$, separated by an interfacial surface $I$ whose intersection with $V$ defines the curve $\partial I$. The vector $\hats$ is the outward-pointing unit normal to the surface $I$ while $\hatt$ and $\hatb$ are two orthogonal unit tangent vectors to the interface. We proceed below using dimensional variables and then apply the scalings in equation \ref{scalings} to render the final equations dimensionless.

For fluid `i', it is possible to write down expressions for $\omega_{b,i}$ and $\omega_{t,i}$, which represent the components of the vorticity $\boldsymbol{\omega}_i$ in the $\hatb$ and $\hatt$ directions, respectively:
%
\begin{eqnarray}
    \omega_{b,i}=(\hats\times\hatt)\cdot(\nabla\times\uvec_i),\\
    \omega_{t,i}=(\hats\times\hatb)\cdot(\nabla\times\uvec_i),
\end{eqnarray}
where $\mathbf{u}_i$ denotes the velocity fields. These expressions may be recast as follows
\footnote{Using 
    $(\mathbf{a}\times\mathbf{b})\cdot(\mathbf{c}\times\mathbf{d})=(\mathbf{a}\cdot\mathbf{c})(\mathbf{b}\cdot\mathbf{d})-(\mathbf{a}\cdot\mathbf{d})(\mathbf{b}\cdot\mathbf{c})$,
valid for any vector $\mathbf{a}$, $\mathbf{b}$, $\mathbf{c}$, and $\mathbf{d}$.} 
%
\begin{eqnarray}
    \omega_{b,i}= \hats\cdot\nabla \mathbf{u}_i\cdot \hatt -\hatt\cdot\nabla\mathbf{u}_i\cdot\hats,
    \label{eq:omega_b_2}\\
    \omega_{t,i}=\hats\cdot\nabla \mathbf{u}_i\cdot \hatb -\hatb\cdot\nabla\mathbf{u}_i\cdot\hats.
    \label{eq:omega_t_2} 
\end{eqnarray}

In the presence of interfacial stresses arising from gradients of surface tension $\sigma$ due to surfactant concentration gradients, the interfacial shear stress conditions are given by
\begin{eqnarray}
    \left[\left[\hatt\cdot\mathbf{T}\cdot\hats\right]\right]&=&-\hatt\cdot\nabla\sigma,\label{eq:shear_stress_t}\\
    \left[\left[\hatb\cdot\mathbf{T}\cdot\hats\right]\right]&=&-\hatb\cdot\nabla\sigma,\label{eq:shear_stress_b}
\end{eqnarray}
$[[q]]=q_2-q_1$ represents the jump across the interface of a quantity $q$, $\mathbf{T}_i=-p_i+\mu_i \mathbf{D}_i$ is the total stress in fluid `i' in which $p_i$ is the pressure,  $\mathbf{D}_i=(\nabla\mathbf{u}_i+\nabla\mathbf{u^{T}}_i)/2$ is the rate of deformation tensor, and $\mu_i$ denote the viscosities, whence 
\begin{eqnarray}
    \left[\left[\mu\left(\hatt\cdot \nabla\mathbf{u}\cdot\hats+\hats\cdot \nabla\mathbf{u}\cdot\hatt\right)\right]\right]=-2\hatt\cdot\nabla\sigma, \\
    \left[\left[\mu\left(\hatb\cdot \nabla\mathbf{u}\cdot\hats+\hats\cdot \nabla\mathbf{u}\cdot\hatb\right)\right]\right]=-2\hatb\cdot\nabla\sigma. 
\end{eqnarray}
Substitution of these results into Eqs. (\ref{eq:omega_b_2}) and (\ref{eq:omega_t_2}) yields
%
%
%
\begin{eqnarray}
    \left[\left[\mu\left(\omega_b+2\hatt\cdot\nabla\uvec\cdot\hats\right)\right]\right]=-2\hatt\cdot\nabla\sigma,\\
    \left[\left[\mu\left(\omega_t+2\hatb\cdot\nabla\uvec\cdot\hats\right)\right]\right]=-2\hatb\cdot\nabla\sigma.
\end{eqnarray}

For the case $[[\mu]]=0$, which is the focus of this paper, we obtain
\begin{eqnarray}
    \left[\left[w_b\right]\right]&=&-\frac{2}{\mu}\nabla\sigma\cdot\hatt - 2 \left[\left[\hatt\cdot \nabla \mathbf{u}\cdot\hats\right]\right],\\
    \left[\left[w_t\right]\right]&=&-\frac{2}{\mu}\nabla\sigma\cdot\hatb - 2 \left[\left[\hatb\cdot \nabla \mathbf{u}\cdot\hats\right]\right],
\end{eqnarray}
where $\mu_2=\mu_1=\mu$. 
Noting that $\hatt\cdot \nabla  = \partial/\partial s$ and $\hatb\cdot\nabla = \partial/\partial b$, it can be shown that 
\begin{equation}
    \left[\left[\omega_b\right]\right]=-\frac{2}{\mu}\frac{\partial\sigma}{\partial s}-2\left[\frac{\partial}{\partial s}\left[\left[\uvec\cdot\hats\right]\right]-\kappa_1\left[\left[\uvec\cdot\hatt\right]\right]\right],
    \label{eq:omega_b_surf}
\end{equation}
\begin{equation}
    \left[\left[\omega_t\right]\right]=-\frac{2}{\mu}\frac{\partial\sigma}{\partial b}-2\left[\frac{\partial}{\partial b}\left[\left[\uvec\cdot\hats\right]\right]-\kappa_2\left[\left[\uvec\cdot\hatb\right]\right]\right],
    \label{eq:omega_t_surf}
\end{equation}
where the curvatures $\kappa_1$ and $\kappa_2$ are defined as follows
\begin{equation}
    \kappa_1=\hatt\cdot\frac{\partial\hats}{\partial s}, ~~~~~
    \kappa_2=\hatb\cdot\frac{\partial\hats}{\partial b}.
    \label{eq:kappa1_kappa2}
\end{equation}
From continuity of the normal and tangential components of the velocity at the interface, i.e., $[[\mathbf{u}\cdot\hats]]=0$, and $[[\mathbf{u}\cdot\hatt]]=[[\mathbf{u}\cdot\hatb]]=0$, respectively, it is seen that the interfacial jumps in the vorticity components are directly related to the Marangoni stresses:
\begin{eqnarray}
    \left[\left[\omega_b\right]\right]&=&-\frac{2}{\mu}\frac{\partial\sigma}{\partial s}, \label{eq:omega_b_dim}\\
    \left[\left[\omega_t\right]\right]&=&-\frac{2}{\mu}\frac{\partial\sigma}{\partial b}.\label{eq:omega_t_dim}
\end{eqnarray}

We now consider the circulation vector $\boldsymbol{\Omega}$ for 3D flows given by
\begin{equation}
    \boldsymbol{\Omega}=\int_V \boldsymbol{\omega}dV,
\end{equation}
for the fixed 3D control volume $V$ shown in figure \ref{fig:fig3}. The 3D vorticity equation is given by
\begin{equation}
    \frac{\partial\boldsymbol{\omega}}{\partial t}+\nabla\cdot (\mathbf{u}\boldsymbol{\omega})=\nabla \cdot (\boldsymbol{\omega}\mathbf{u})+\nu \nabla^2\boldsymbol{\omega},
\end{equation}
and the total rate of change of $\boldsymbol{\Omega}$ is then expressed by
\begin{eqnarray}
    \frac{D\boldsymbol{\Omega}}{Dt}&=&\int_V \frac{D\boldsymbol{\omega}}{Dt}dV=\frac{D}{Dt}\int_V\boldsymbol{\omega}dV=\int_V \nabla \cdot \left( \boldsymbol{\omega}\mathbf{u}+\nu\nabla\boldsymbol{\omega}\right)dV\nonumber\\
    &=&\int_{\partial V} \hatn \cdot (\boldsymbol{\omega}\mathbf{u})dS+\int_{\partial V}\hatn\cdot (\nu\nabla\boldsymbol{\omega})dS.
    \label{eq:roc_circulation_3D}
\end{eqnarray}
The first term on the RHS of Eq. \textcolor{black}{ (\ref{eq:roc_circulation_3D})} corresponds to vortex stretching/tilting and is present only in 3D. 
%
We now write
\begin{eqnarray}
    \frac{D}{Dt}\int_{V_1 U V_2} \boldsymbol{\omega}dV=&&\oint_{\partial V_1}\hat{\mathbf{n}}\cdot \left(\boldsymbol{\omega}\mathbf{u}+\nu\nabla\boldsymbol{\omega}\right)dS+\oint_{\partial V_2}\hatn\cdot \left(\boldsymbol{\omega}\mathbf{u}+\nu\nabla\boldsymbol{\omega}\right)dS\nonumber\\
    &+&\oint_{\partial V_1'}\hatn\cdot \left(\boldsymbol{\omega}_1\mathbf{u}_1+\nu_1\nabla\boldsymbol{\omega}_1\right)dS+\oint_{\partial V_2'}\hatn\cdot \left(\boldsymbol{\omega}_2\mathbf{u}_2+\nu_2\nabla\boldsymbol{\omega}_2\right)dS,
\end{eqnarray}
and let $V_1 U V_2 \rightarrow V$, $\hatn\rightarrow\hats$ from fluid 1, $\hatn\rightarrow-\hats$ from fluid 2, and $(\partial V_1,\partial V_2) \rightarrow I$, then it follows that
\begin{equation}
    \frac{D}{Dt}\int_{V} \boldsymbol{\omega}dV=\oint_{\partial V}\hatn\cdot \left(\boldsymbol{\omega}\mathbf{u}+\nu\nabla\boldsymbol{\omega}\right)dS -\left[\oint_I [[\hats\cdot(\om\uvec)]]dS+\oint_I [[\nu \hats\cdot \nabla\om]]dS
    \right].
    \label{eq:circulation_jump_3D}
\end{equation}

It is important to establish a connection between $\oint_I[[\nu\hats\cdot \nabla\om]]dS$, which represents the jump across the plane of the interface of the vorticity flux, and the momentum conservation equation given by
\begin{equation}
    \frac{D\mathbf{u}}{Dt}=-\frac{\nabla p}{\rho}-\nu \nabla\times \boldsymbol{\omega}.
    \label{eq:momentum_2}
\end{equation}
In order to relate this term to the $\nu\nabla\times\om$ term in Eq. (\ref{eq:momentum_2}), we first write down the following general result 
\footnote{We have used the vector identity
$    \int_V\nabla\times\mathbf{A}dV=-\oint_{\partial V}\mathbf{A}\times d\mathbf{S}=-\oint_{\partial V}\mathbf{A}\times \mathbf{n}dS=\oint_{\partial V}\mathbf{n}\times\mathbf{A}dS,    
$ 
for any vector $\mathbf{A}$, and volume $V$ enclosed by a surface $\partial V$ with a unit normal $\mathbf{n}$.}
\begin{eqnarray}
    -\oint_{\partial V}\hats\cdot\nabla\om dS&=&-\int_V\nabla^2\om dV=-\int_V\left(\nabla(\nabla\cdot\om)-\nabla\times\nabla\times\om\right)dV=\int_V \nabla\times\nabla\times\om dV\nonumber\\
    &=&-\oint_{\partial V}\left(\nabla\times\om\right)\times\hats dS=\oint_{\partial V}\hats\times\nabla\times\om dS.
\end{eqnarray}
Note that this relation links  {\it Lighthill's} vorticity flux to  {\it Lyman's} flux, the latter being another form of the former (see \cite{terrington_jfm_2021} and references therein). 

Inspired by the form of Lyman's flux, the natural way to proceed is to take the cross product of $\hats=\hatt\times\hatb$ with the LHS of Eq. (\ref{eq:momentum_2}) 
and its pressure gradient term \footnote{We have exploited the fact that $\hatt\times\hatb\times\mathbf{c}=\hatb(\hatt\cdot\mathbf{c})-\mathbf{c}(\hatt\cdot\hatb)=\hatb(\hatt\cdot\mathbf{c})$ since $\hatt\cdot\hatb=0$.} and a cross product of $\hats$ with its  $\nu\nabla\times\om$ term to arrive at
\begin{empheq}{align*}
    -\nu\hats\times\nabla\times\om&=\hatb\hatt\cdot \frac{D\uvec}{Dt}-\hatb\hatt\cdot\nabla\left(\frac{p}{\rho}\right)\nonumber\\
    &=\hatb\left[\left(\frac{D}{Dt}(\uvec\cdot\hatt)-\uvec\cdot\frac{D\hatt}{Dt}\right)+\hatt\cdot\nabla\left(\frac{p}{\rho}\right)\right]\nonumber\\
    &=\nu\hats\cdot\nabla\om;
\end{empheq}
here, we note that the sources of vorticity are due to acceleration in the plane of the interface, which we can think of as a vortex sheet, and interfacial pressure gradients. 
Making use of this relation in Eq. (\ref{eq:circulation_jump_3D}), we arrive at 
\begin{eqnarray}
    \frac{D}{Dt}\left[\int_V\om dV+\hatb\oint_I [[\uvec\cdot\hatt]]dS\right]&=&\oint_{\partial V}\hatn\cdot\left(\om\uvec+\nu\nabla\om\right)dS\nonumber\\
    &&-\oint_I[[\hats\cdot(\om\uvec)]]dS
    +\oint_I\hatb[[\uvec\cdot\frac{D\hatt}{Dt}]]dS-\oint_I\hatb\frac{\partial}{\partial s}[[\frac{p}{\rho}]]dS,\nonumber\\
    \label{eq:circulation_jump_3D_2}
\end{eqnarray}
where we have set $\hatt\cdot\nabla(p/\rho)=\partial (p/\rho)/\partial s$.
An expression for $\uvec\cdot (D\hatt/Dt)$ 
can be developed given by (the details are in \textcolor{black}{Appendix} \ref{sec:app_kin}) 
\begin{equation}
    \uvec\cdot\frac{D\hatt}{Dt}=\frac{1}{2}\frac{\partial}{\partial s}\left[(\uvec\cdot\hats)^2+(\uvec\cdot\hatb)^2\right] + \frac{1}{2}\frac{\partial}{\partial b}\left[(\uvec\cdot\hats)^2+(\uvec\cdot\hatb)^2\right]  -\kappa_1(\uvec\cdot\hatt)(\uvec\cdot\hats).
    \label{eq:uDtDt_appendix}
\end{equation}
Furthermore, for $[[\rho]]=0$, the remaining term required to close equation \ref{eq:circulation_jump_3D_2} is one for $[[p]]$ (the details are in \textcolor{black}{Appendix} \ref{sec:app_pressure}):
\begin{equation}
    [[p]]=-\sigma(\kappa_1+\kappa_2)-2[[\mu\left(\frac{\partial}{\partial s}\left[(\uvec\cdot\hatt)+(\uvec\cdot\hatb)\right]+(\kappa_1+\kappa_2)(\uvec\cdot\hats)\right)]].
    \label{eq:pjump_3D_appendix}
\end{equation}

To collapse these equations to their two-dimensional (2D) equivalents, we first note that $\hats\cdot\om=\hatn\cdot\om=\uvec\cdot\hatb=0$ in 2D, and set $\partial/\partial b =0$; the latter leads to $\kappa_2=0$. We then take a dot product of Eq. (\ref{eq:circulation_jump_3D_2}) with $\hatb$ (and convert the volume and area integrals to area and line integrals, respectively) to arrive at a 2D analogue involving 
the vorticity scalar $\omega$. 
Moreover, in the case studied here, characterised by $[[\mu]]=0$, $[[\mathbf{u}\cdot\hats]]=0$,  $[[\mathbf{u}\cdot\hatt]]=0$, and $[[\mathbf{u}\cdot\hatb]]=0$,
equation (\ref{eq:circulation_jump_3D_2}) reduces to
\begin{equation}
    \frac{D}{Dt}\left[\int_V\om dV\right]=\oint_{\partial V}\hatn\cdot\left(\om\uvec+\nu\nabla\om\right)dS-\oint_I[[\hats\cdot(\om\uvec)]]dS+\frac{1}{\rho}\oint_I\hatb\frac{\partial}{\partial s}\left(\sigma\left[\kappa_1+\kappa_2\right]\right)dS.
    \label{eq:circulation_jump_3D_simplified}
\end{equation}

We note that the term involving $[[\hats\cdot(\om\uvec)]]$ on the right-hand-side of this equation is zero. To see this, we first note that $[[\hats\cdot\om \uvec]]$ can be re-expressed as
\begin{eqnarray}
    [[\hats\cdot\om \uvec]]&=&(\hats\cdot\om_2)\uvec_2-(\hats\cdot\om_1)\uvec_1\nonumber\\
    &=&(\hats\cdot\om_2-\hats\cdot\om_1)\uvec_1\nonumber\\
    &=&(\hats\cdot\om_2-\hats\cdot\om_1)\uvec_2\nonumber\\
    &=&[[\hats\cdot\om]]\uvec_1=[[\hats\cdot\om]]\uvec_2,
\end{eqnarray}
since $[[\uvec]]=0$.
We also note that $\hats \cdot \om=(\hatb \times \hatt)\cdot  (\nabla\times\uvec)$, 
which can be re-written as 
\begin{eqnarray}
    \hats\cdot\om&=&\hatb\cdot\nabla\uvec\cdot\hatt - \hatt \cdot \nabla\uvec\cdot\hatb\nonumber\\
    &=&\hatb\cdot\frac{\partial\uvec}{\partial s}-\hatt\cdot\frac{\partial \uvec}{\partial b}\nonumber\\
    &=&\frac{\partial}{\partial s}(\hatb\cdot\uvec)-\frac{\partial}{\partial b}(\hatt\cdot\uvec),
\end{eqnarray} 
since $\hatb\neq\hatb(s)$ and $\hatt\neq\hatt(b)$.
Thus, we can write
\begin{eqnarray}
    [[\hats\cdot\om]]&=&[[\frac{\partial}{\partial s}(\hatb\cdot\uvec)]]-[[\frac{\partial}{\partial b}(\hatt\cdot\uvec)]]\nonumber\\
    &=&\frac{\partial}{\partial s}[[\hatb\cdot \uvec]]-\frac{\partial}{\partial b}[[\hatt\cdot\uvec]]=0,
\end{eqnarray}
since $[[\hatb \cdot\uvec]]=0$ and $[[\hatt\cdot\uvec]]=0$, whence 
$[[\hats\cdot\om\uvec]]=0$.
Inspection of the terms remaining in equation \ref{eq:circulation_jump_3D_simplified} suggests that circulation is influenced by vorticity diffusion, vortex tilting/stretching, and gradients of curvature and interfacial tension. 

The dimensionless versions of equations (\ref{eq:omega_t_dim}) and (\ref{eq:omega_b_dim}) are then expressed by
\begin{eqnarray}
    \left[\left[\tilde{\omega}_t\right]\right]&=&-2ReMa\frac{1}{(1-\tilde{\Gamma})}\frac{\partial\tilde{\Gamma}}{\partial b},\label{eq:omega_t_dless}\\
    \left[\left[\tilde{\omega}_b\right]\right]&=&-2ReMa\frac{1}{(1-\tilde{\Gamma})}\frac{\partial\tilde{\Gamma}}{\partial s}, \label{eq:omega_b_dless}
\end{eqnarray}
and the dimensionless equation (\ref{eq:circulation_jump_3D_simplified}) reads
\begin{equation}
    \frac{D}{D\tilde{t}}\left[\int_{\tilde{V}}\tilde{\om} d\tilde{V}\right]=\oint_{\partial \tilde{V}}\hatn\cdot\left(\tilde{\om}\tilde{\uvec}+\frac{1}{Re}\nabla\tilde{\om}\right)d\tilde{S}
    +\frac{1}{We}\oint_I\hatb\frac{\partial}{\partial \tilde{s}}\left(\tilde{\sigma}\left[\tilde{\kappa}_1+\tilde{\kappa}_2\right]\right)d\tilde{S},
    \label{eq:circulation_jump_3D_simplified_dless}
\end{equation}
and the tildes are dropped henceforth. 

We note that in the limit of small variations of surfactant concentration around its initial value (i.e., so-called diluted systems), \textcolor{black}{ $\tilde{\Gamma}= \Gamma_0 + \delta\tilde{\Gamma}$}, with $\delta \textcolor{black}{ \tilde{\Gamma}} \ll \Gamma_0$, leads to $\textcolor{black}{ \tilde{\Gamma} } =1 + \delta \textcolor{black}{ \tilde{\Gamma}}$, and the equation of state can be linearized to result in $\tilde{\sigma}=1 - \beta_s \textcolor{black}{ \tilde{\Gamma}}$. Note that, in the case of non-isothermal systems, 
$\tilde{\sigma}$ has a linear  dependency  on the local temperature ( $\mathcal{T}$), and a  linear equation of state describes $\tilde{\sigma}(\tilde{ \mathcal{T}})$ (see for example \citet{williams_2021}). Therefore, surface tension gradients  in equation (\ref{eq:circulation_jump_3D_simplified_dless}) can also arise due to thermal gradients.

\section{Results\label{sec:Results}}

\begin{figure}
\begin{center} 
\includegraphics[trim =  1 10 1 40, clip,width=0.9\linewidth]{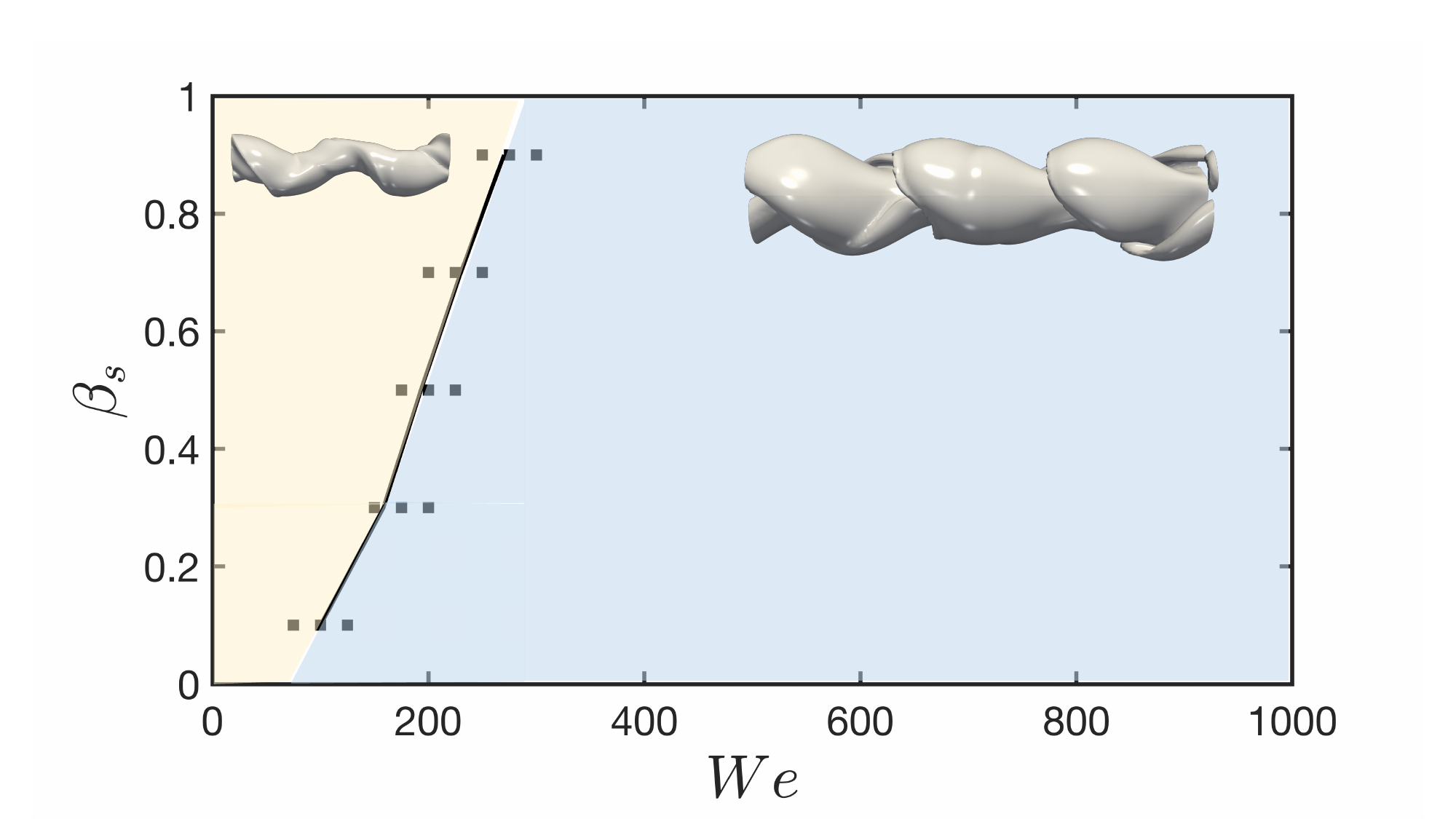} 
\end{center} 
 \caption{Regime map of the 
 interfacial morphology in the $\beta_s-We$ space for $Re=5000$, $Pe_s=100$, and $\Gamma_o=\Gamma_\infty/2$. The capillary-dominated and
 inertia-dominated  regimes, and their boundaries are clearly demarcated; the symbols represent simulations carried out at the transition lines separating these regimes. Three-dimensional  representations of the interface for both regimes are also shown.
 \label{regime}} 
\end{figure}

\begin{figure}
\begin{center} 
 \includegraphics[trim =  1 1 1 1, clip, width=1\linewidth]{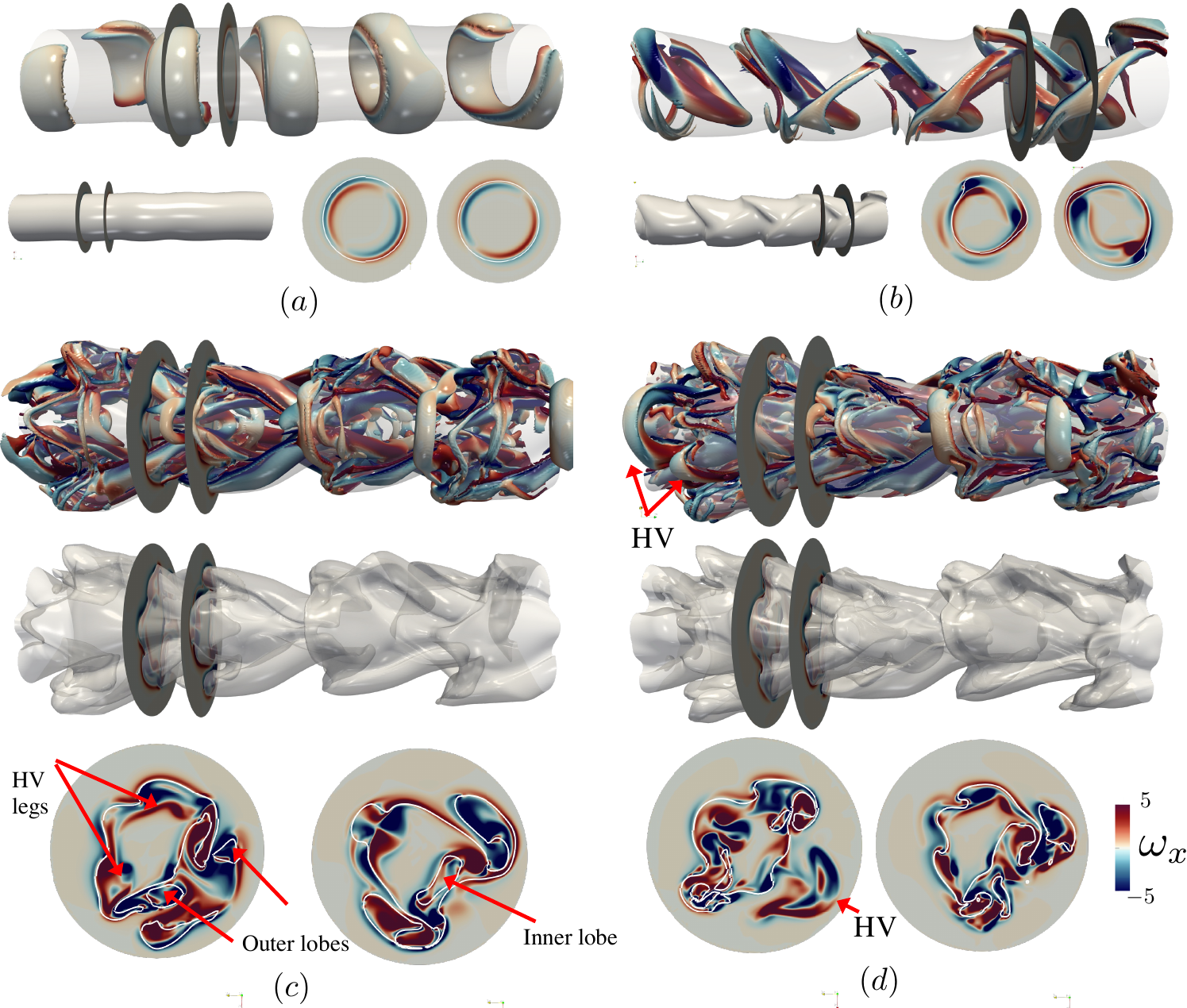}
\end{center} 
\caption{
Spatio-temporal representation of the interfacial dynamics and the coherent vortical structures for $Re=5000$ and $We=500$  at $t=(23.28,~ 28.12,~ 31.25, ~31.50)$ shown in (a)-(d), respectively. 
\textcolor{black}{For (c) and (d), the first row represents the  3D coherent structures with the location of the interface, the middle row represents only the interface location, and the bottom row shows two transversal cuts of the interface coloured by the magnitude of $\omega_x$ (the white lines represent the interface location).}
The three-dimensional coherent structures are visualised by the $Q$-criterion with values of $Q =(3,~220,~320,~320)$, where the colour represents the streamwise vorticity field, \textcolor{black}{$\omega_x$}.
In each panel, we also show \textcolor{black}{$\omega_x$} in the $y$–$z$ plane for each sampling location.
\label{clean_temporal}} 
\end{figure}


Figure \ref{regime} shows a flow regime map 
for $Re=5000$ that depicts the interfacial morphology associated with various regions of the $\beta_s-We$ parameter space generated by over 100 transient simulations performed 
in the ranges $100<We<1000$ and $0.1<\beta_s<0.9$.
We have divided the map  into two distinct regions depending on the morphology: 
for small $We$,  capillary forces control the interfacial dynamics  preventing the development of lobes which could result in the formation of large droplets; for large $We$, inertial forces dominate the dynamics  triggering the formation of interfacial lobes  whose thinning eventually results in the generation of holes 
and eventually droplets.
\textcolor{black}{The resulting non-uniform surfactant distribution generates gradients in surface tension affecting the local dynamics. Surfactant accumulation takes places in high-curvature regions giving rise to  Marangoni stresses that drive surfactant redistribution from high- to low-concentration regions. 
Marangoni stresses, therefore, oppose the  shear stresses produced by the flow field, the former exerting a restoring effect and the latter a perturbing effect in the local surfactant concentration field. The dimensionelss Marangoni velocities induced by surface tension differences $\Delta \sigma$ are of $O(Re We ^{-1} (\Delta \sigma/\sigma_s)$. Similarly,   
the dimensionless Marangoni stresses, $\tau$, are of $O(We^{-1}\tilde{\nabla}\tilde{\sigma})$, or, equivalently, $O(\beta_s We^{-1}\tilde{\nabla}\tilde{\Gamma})$, viz. equation (\ref{eq:def_tau}), while 
capillary forces and shear stresses are of $O(We^{-1})$ and 
$O(Re^{-1})$, respectively. Furthermore, from equations (\ref{eq:omega_t_dless}) and (\ref{eq:omega_b_dless}), it is clear that the Marangoni-induced vorticity jumps across the interface are of $O(Re~\beta_s We^{-1})$.  
Inspection of figure \ref{regime}, which was generated for a fixed $Re$ value, reveals that the presence of Marangoni stresses counteracts the transition from the low- to high-We regimes as the critical $We$ increases with $\beta_s$ with a quasi-linear dependence. The latter is consistent with the scaling highlighted above, $\tau \sim \beta_s We^{-1}$, which demonstrates that increasing $\beta_s$ and decreasing $We$ serve to enhance the restoring influence of the Marangoni stresses.}

To assess the effect of Marangoni-induced flow, 
we have analysed the flow physics of the surfactant-free and surfactant-laden flows characterised by 
$Re=5000$ and $We=500$. 
We start with the surfactant-free case depicted in Figure \ref{clean_temporal} which shows the spatio-temporal interfacial dynamics for the surfactant-free case through the $Q$-criterion (e.g.,  a measure of the dominance  of vorticity $\boldsymbol{\omega}$ over strain $\bf{s}$, i.e., $Q = (||\boldsymbol{\omega}||^2 -||{\bf s}||^2)/\textcolor{black}{2}$ \citep{Hunt_CTR_1988}). 
At early times, we observe the formation of a periodic array of 
quasi-symmetric Kelvin-Helhomltz (KH)-driven vortex rings as a result of the difference in velocity in the shear layer located under the interface (see figure \ref{clean_temporal}a). 
With increasing time, the three-dimensional instability starts with the deformation of the vortex-rings leading to a 
mutual-induction  between two consecutive vortex rings resulting in their  `knitting' (see figure \ref{clean_temporal}b); similar vortex-pairing has been reported by \citet{broze_hussain_1996}
and \citet{da_Silva}. With increasing time, we observe the formation of inner and outer  hairpin vortices whose pairing brings about a region where both overlap. 
The  cascade mechanism resulting in the formation of 
hairpin-vortices from  KH-rings  is triggered by   the magnitude of the streamwise vorticity, $\omega_x$,  which becomes comparable  to its azimuthal counterpart, $\omega_y$, in agreement with \citet{Jarrahbashi_jfm_2016} and \citet{constante_jets}, as shown in figure \ref{clean_temporal}b. 

To provide more conclusive evidence of the existence of inner/outer hairpin vortices in the jet dynamics, a careful study of the distribution of vortex signs shows the assembling into counter-rotating  vortex pairs (see $\omega_x$
 in the $y$-$z$ plane for each sampled location of the  panels in figure \ref{clean_temporal}). 
By analysing the distribution of streamwise  vorticity between the ring and braid regions of the jet core (see figure \ref{clean_temporal}a), we observe that their distribution is  $\pi$-out-of-phase.
The arrangement of the vorticity comes from vortex induction arguments, similar to those explained by \citet{Jarrahbashi_jfm_2016}, \citet{Zandian_jfm_2018} and \citet{constante_jets}, i.e., the upstream hairpin vortex from the ring overtakes the upstream hairpin vortex from the braid as the mutual induction takes place. 
Finally, the vortex-surface interaction triggers the formation of the interfacial structure as the interface adopts the shape of the vortex which is in its vicinity (see figure \ref{clean_temporal}b-d, \textcolor{black}{`HV' stands for hairpin vortices}).
The mutual induction between outer and inner hairpin vortices eventually leads to the
thinning of the lobes to ultimately form
inertia-induced holes whose capillary-driven expansion gives rise to the formation of droplets \citep{Jarrahbashi_jfm_2016,Zandian_jfm_2018,constante_jets}.

\begin{figure}
\begin{center} 
\begin{tabular}{c}
\includegraphics[ width=0.46\linewidth]{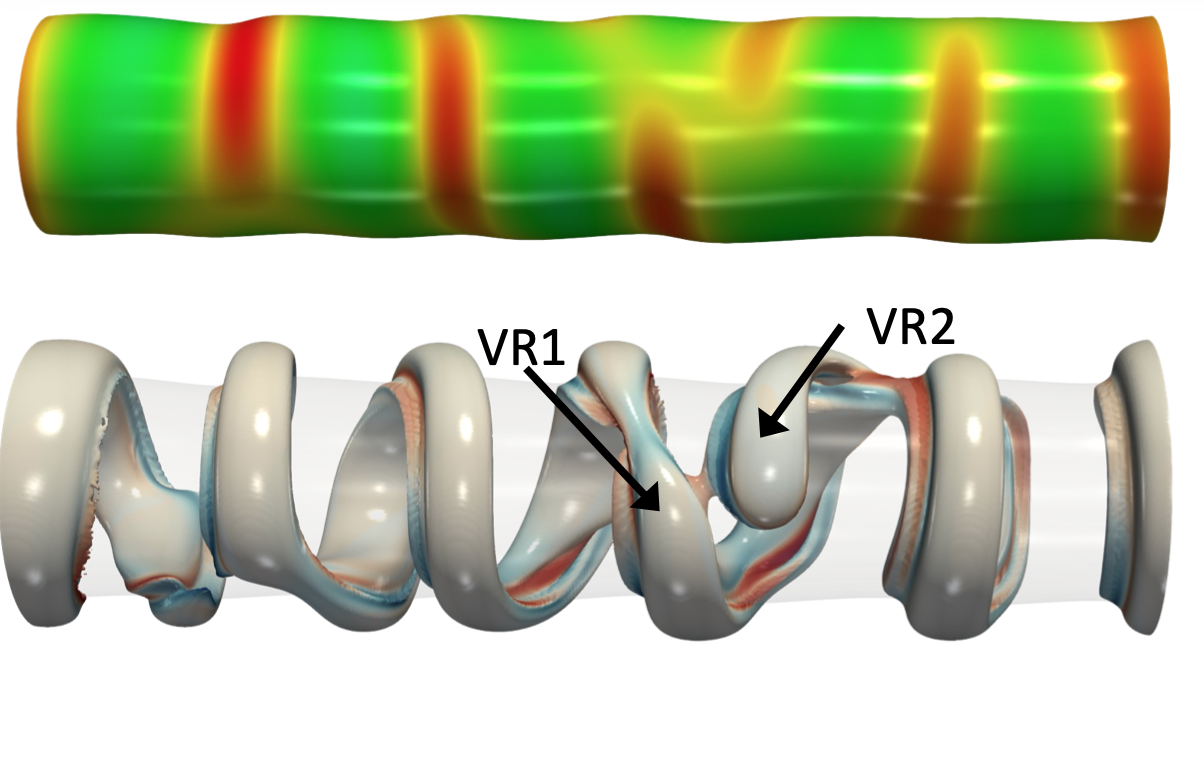} 
\includegraphics[ width=0.46\linewidth]{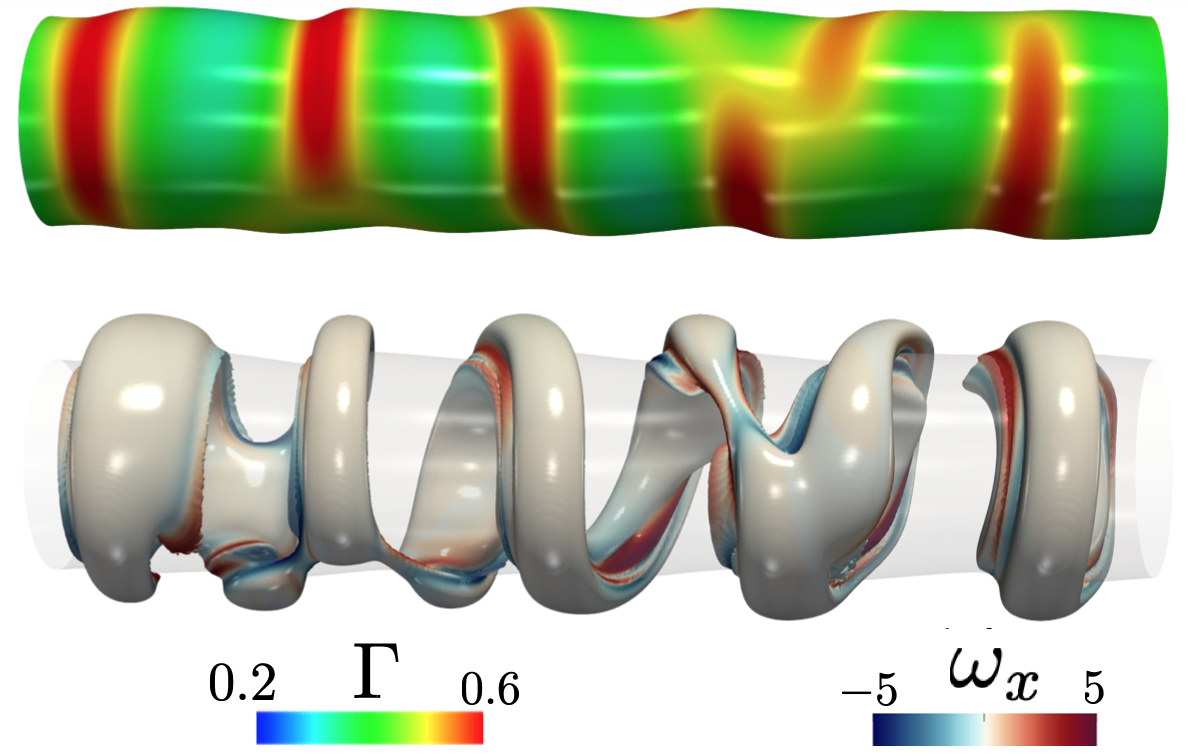}
\\ (a) \hspace{2in} (b) \\
\includegraphics[trim =  1 1 1 1, clip, 
width=0.42\linewidth]{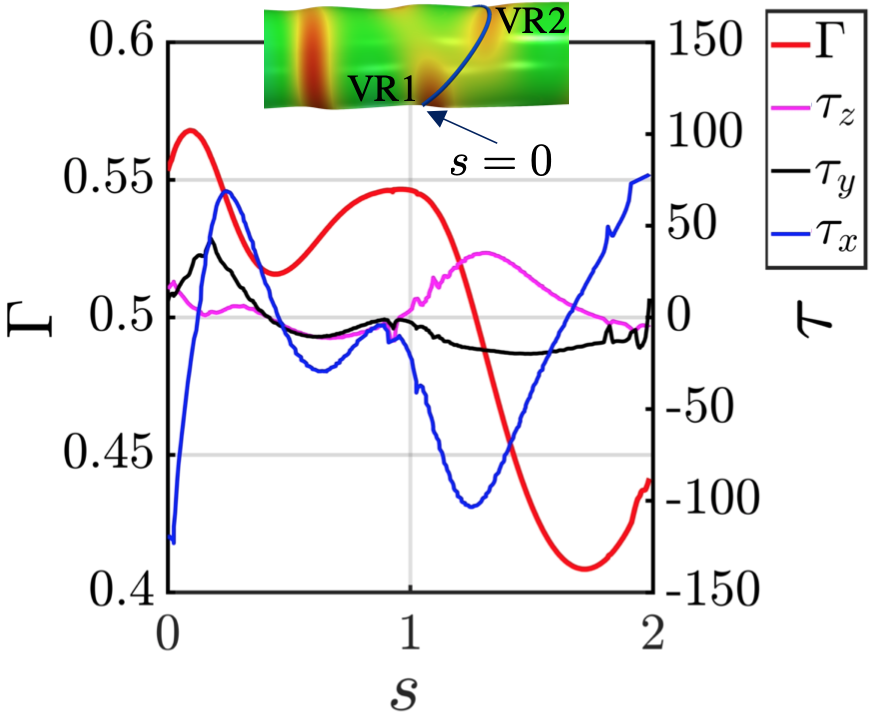} \\
(c) \\
\includegraphics[width=0.32\linewidth]{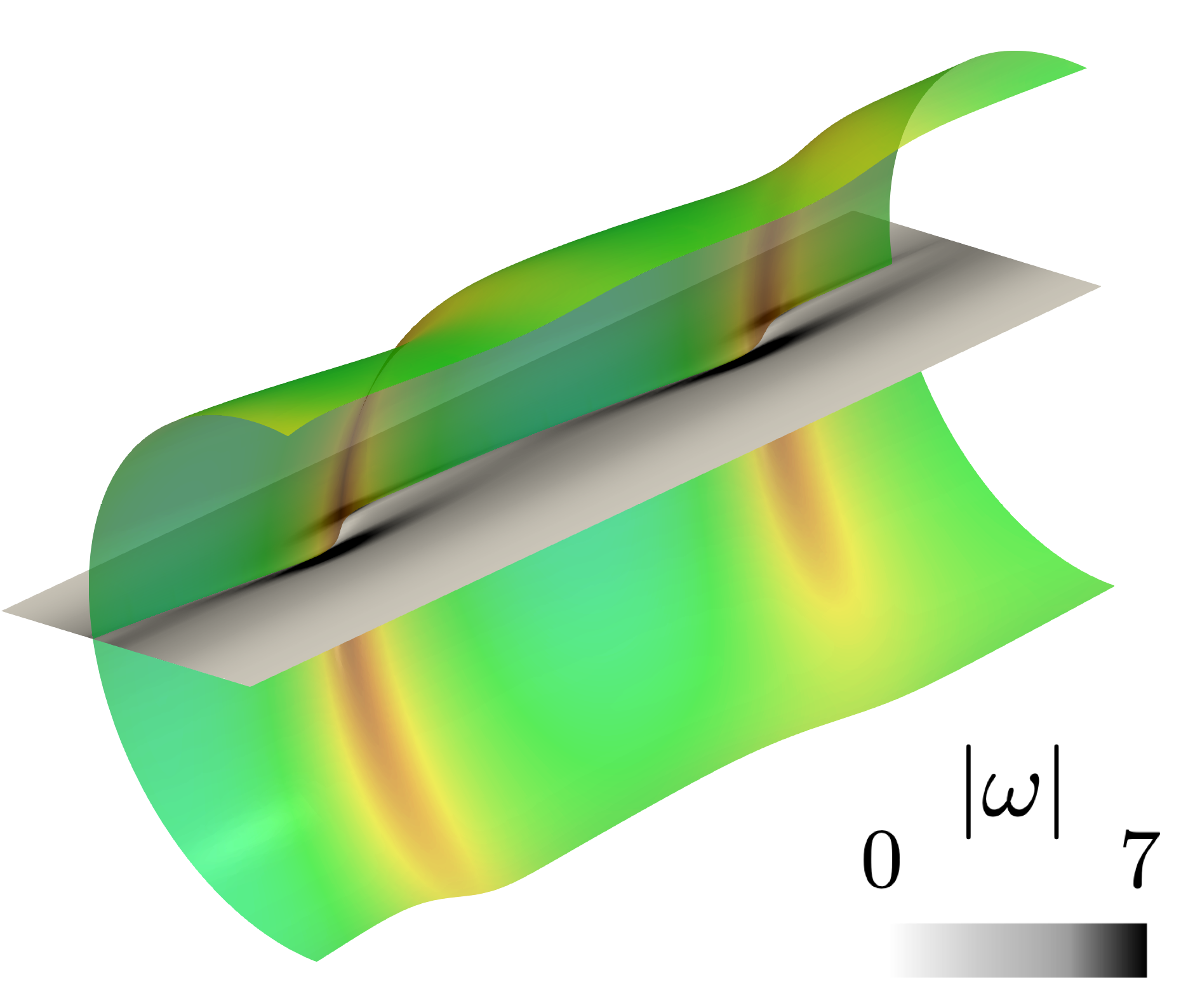} \\ (d) \\ 
\includegraphics[width=0.42\linewidth]{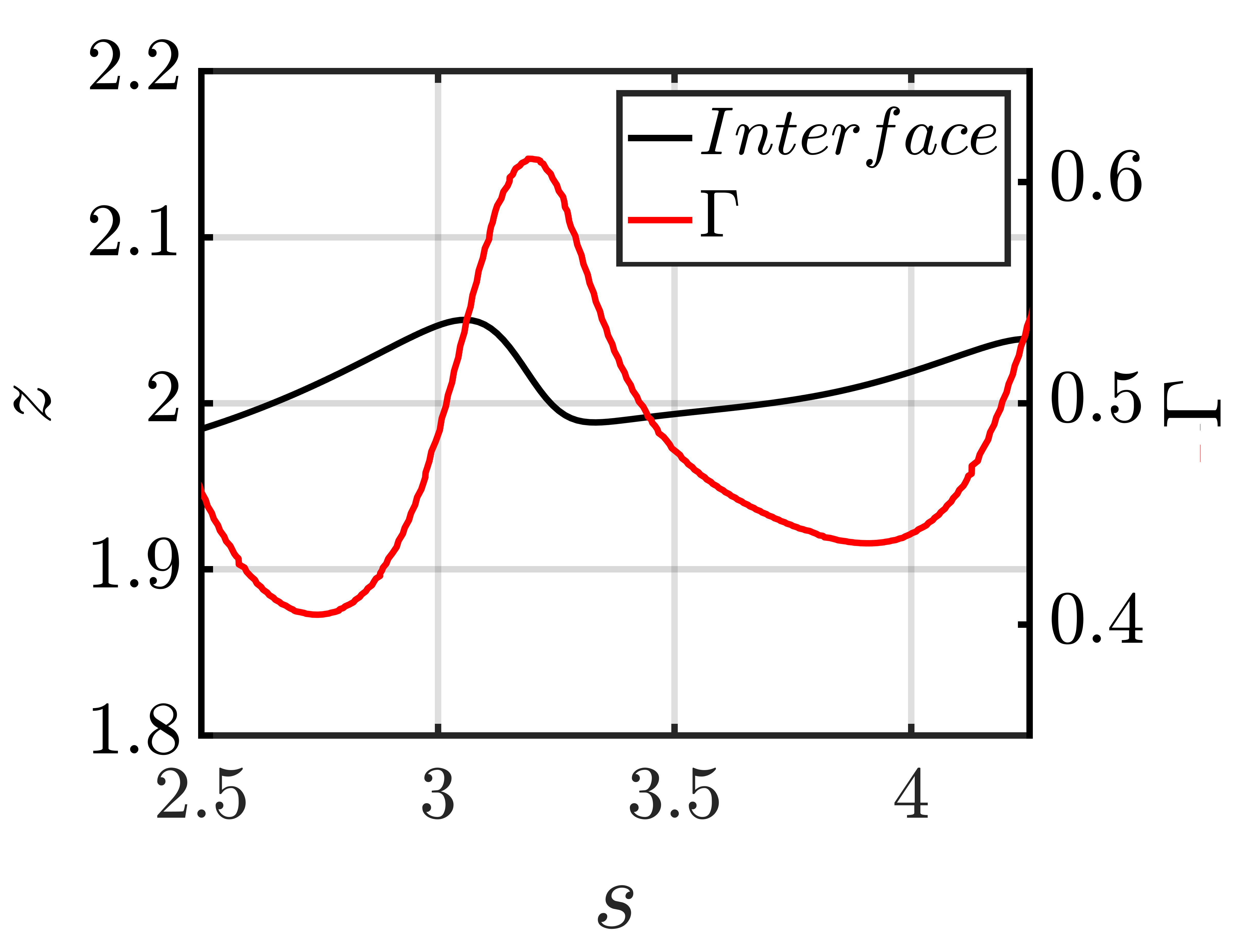} 
\includegraphics[width=0.4\linewidth]{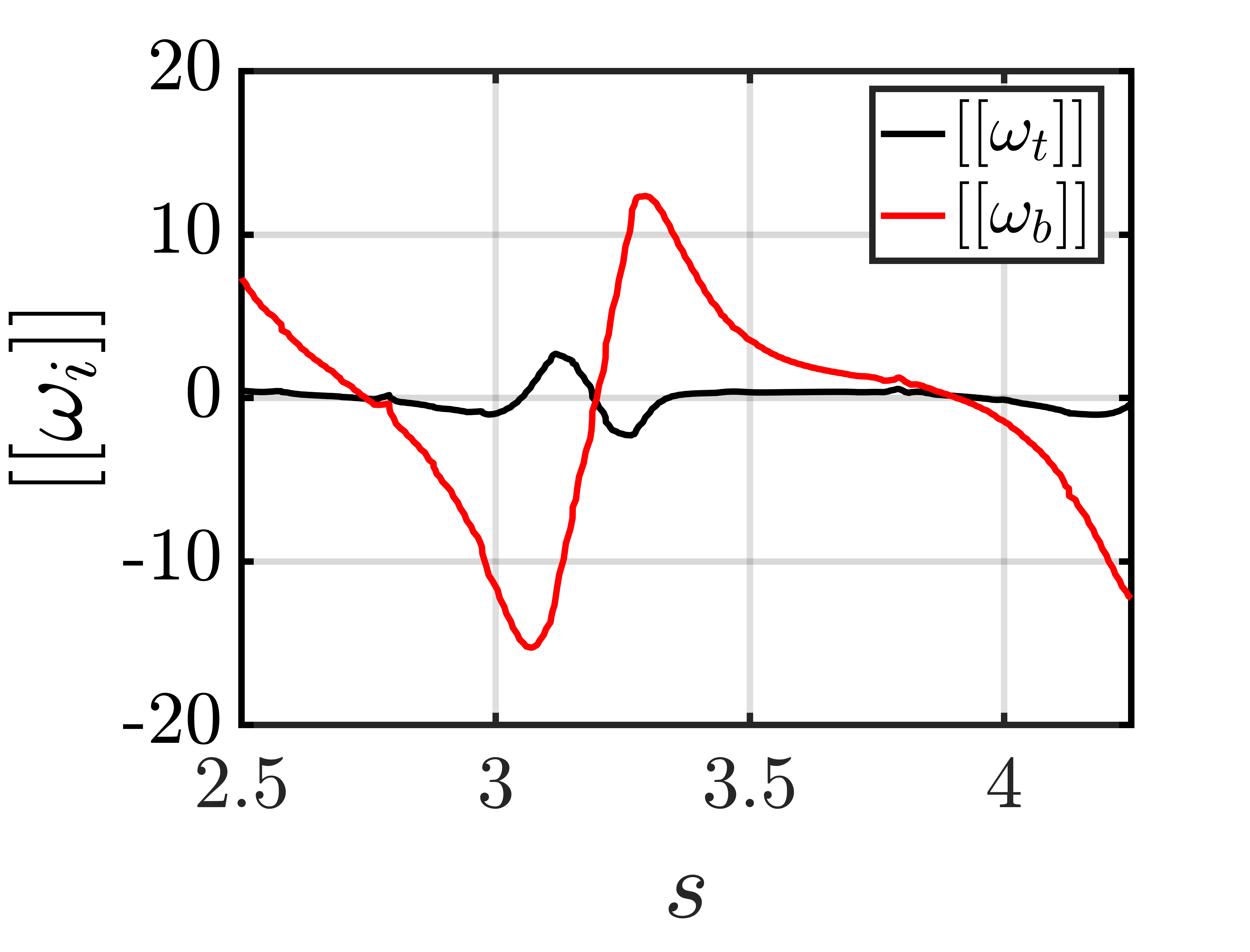} \\
(e) \hspace{2in} (f)
\end{tabular}
\end{center} 
\caption{
Effect of surfactants on the early  
interfacial dynamics  for  $Re=5000$, $We=500$, $\beta_s=0.5$, $Pe_s=100$ and $\Gamma_o=\Gamma_{\infty}/2$ at 
$t=32.03$, (a), and 32.81, (b). The top and bottom panels represent the interface coloured by $\Gamma$ and the coherent vortical structures visualised via $Q$-criterion with $Q=10$. 
Panel (c) shows a 2D representation of  $\Gamma$, and $\tau$, with respect to the arc length $s$ (see inset) at $t=32.03$.
Panel (d) shows a 2D representation of the magnitude of vorticity $|\boldsymbol{\omega}|$ in the $x$–$z$ plane ($ \textcolor{black}{y} = 2.875$) at $t=32.81$; interface location and $\Gamma$, and $[[\omega_b]]$ and $[[\omega_t]]$ vs the arc length $s$ (\textcolor{black}{e.g., $s$ corresponds to the $x$–$z$ plane ($y = 2.875$) intersecting the interface)} 
shown in (e) and (f), respectively. \textcolor{black}{The center of the jet core corresponds to z = 2.5}
\label{surf_temporal}} 
\end{figure}

\begin{figure}
\begin{center} 
\begin{tabular}{c}
\includegraphics[width=0.9\linewidth]{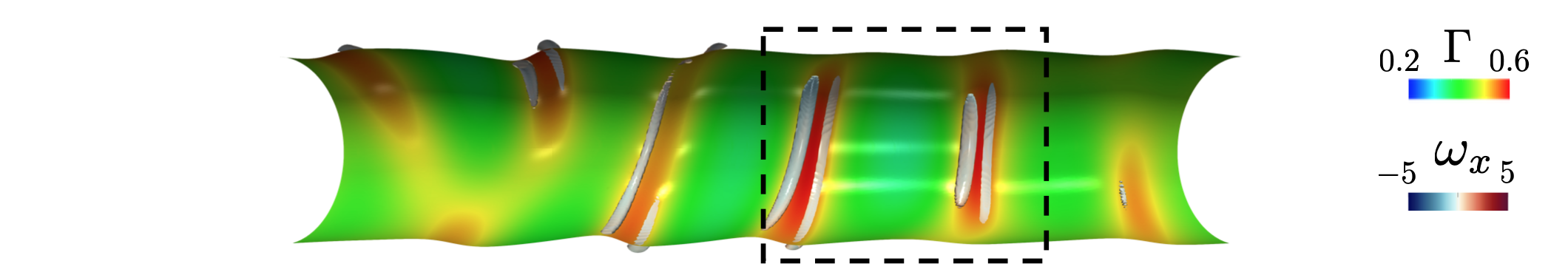}\\
(a)
\end{tabular}
\begin{tabular}{cccc}
\includegraphics[width=0.23\linewidth]{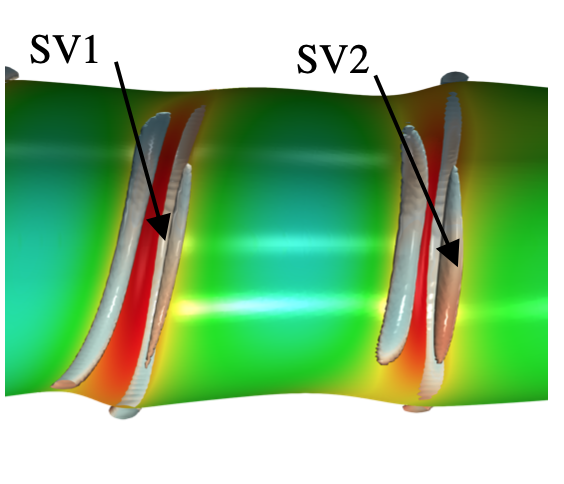} &
\includegraphics[width=0.23\linewidth]{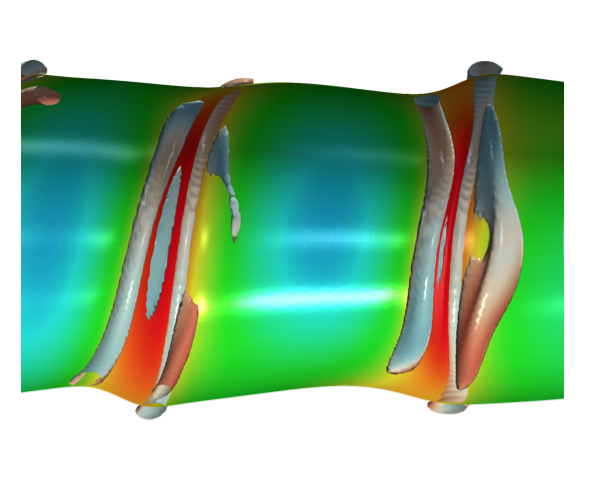} &
\includegraphics[width=0.24\linewidth]{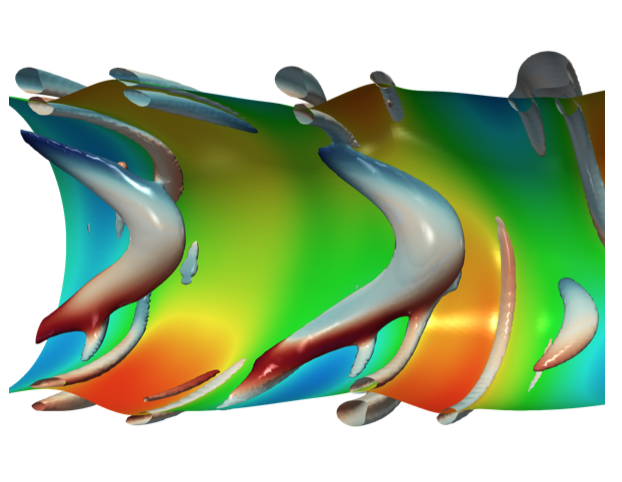} &
\includegraphics[width=0.26\linewidth]{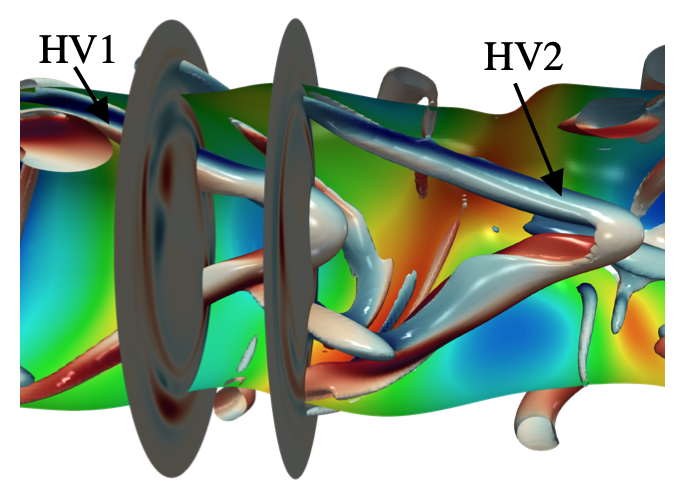} \\
(b) & (c) & (d) & (e)\\
\end{tabular}
\begin{tabular}{cc}
\includegraphics[width=0.33\linewidth]{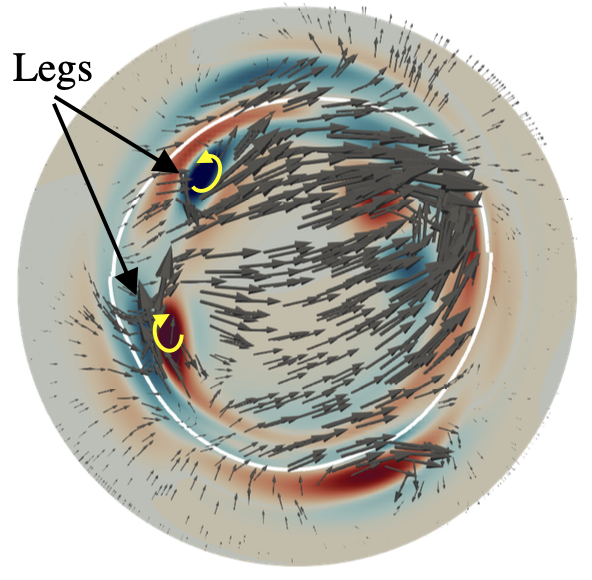} &
\includegraphics[width=0.33\linewidth]{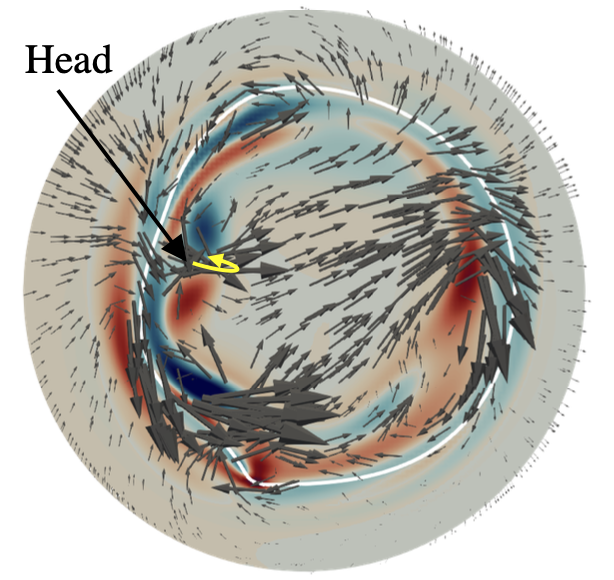} \\
(f) & (g) \\ 
\end{tabular}
\end{center} 
\caption{Surfactant-driven hairpin-vortical structures: temporal development  of the {\it HV1}  and {\it HV2} hairpin-like vortical structures via $Q$-criterion with $Q=200$ together with the interfacial location coloured by $\Gamma$
at time $t=(32.81,~33.59,~34.37,~36.71,~39.06)$, corresponding to panels (a)-(e), respectively.
In panel (e), we show the two transversal slices displayed in panels (f) and (g) which depict the 
streamwise vorticity \textcolor{black}{$\omega_x$} through the legs and head of $HV1$, respectively; arrows of in-plane velocity vectors have been added; \textcolor{black}{the white lines represent the interface location}.
The parameter values are the same as in figure \ref{surf_temporal}.
\label{HV_formation}} 
\end{figure}

Next, we turn our attention to the effect of surfactants on the flow dynamics. 
Figure \ref{surf_temporal} shows the early interfacial surfactant concentration together with the three-dimensional coherent vortical structures via the $Q$-criterion.
Similarly to the surfactant-free case, we observe the formation of a periodic array of  
quasi-axisymmetric KH-vortex rings. These rings induce  the formation of interfacial waves that are characterised by regions of radially converging and diverging motion that lead to  higher  and lower interfacial areas, and subsequently to lower and higher surfactant concentration regions, respectively; accumulation of $\Gamma$ is observed in the vicinity of the KH rings (see figure \ref{surf_temporal}a). 
Figure \ref{surf_temporal}c  presents the interfacial concentration $\Gamma$, and Marangoni stresses $\tau$ along an arc length, $s$,  
corresponding to  $t=32.03$.
We observe that the non-uniform distribution of $\Gamma$  gives rise to Marangoni-induced flow, which drives fluid motion from ring-1, `VR1', ($\tau >0$) to ring-2, `VR2', and vice versa (i.e, flow from VR2 to VR1, $\tau <0$). 
This flow is therefore accompanied by the retardation of the development of the interfacial waves and a subsequent delay of the onset of the three-dimensional instability of the jet observed in the surfactant-free case in figure \ref{clean_temporal}. 

Additionally, these Marangoni stresses promote jumps in the vorticity across the interface which we can calculate using equations \textcolor{black}{ \ref{eq:omega_b_dim}} and \textcolor{black}{\ref{eq:omega_t_dim}} in the location which coincides with the formation of vortex $SV1$ and $SV2$ from figure \ref{HV_formation} at $t=32.81$.
Figure \ref{surf_temporal}d
shows a three-dimensional representation of the interface together with an $x$-$z$ plane
at $y = 2.875$ colored by the the magnitude of vorticity, $|\boldsymbol{\omega}|$.
Figure \ref{surf_temporal}e,f show respectively the variation of the interface location and the $\Gamma$ profiles, and of the distribution of $[[\omega_b]]$ and $[[\omega_t]]$, along the arc length, $s$ \textcolor{black}{(not to be confused with $\hat{s}$ the unit vector in figure 2)}, in the plane cutting the interface shown in figure \ref{surf_temporal}d.
From figure \ref{surf_temporal}e, it is seen that the surfactant accumulates in the down-sloping region immediately downstream of an interfacial wave peak; here, the gradients in $\Gamma$, and therefore in $\sigma$, are smallest corresponding to the weakest vorticity jumps, while the largest such jumps are in the wave peak and trough regions where the $\Gamma$ (and $\sigma$) gradients are highest, as shown in figure \ref{surf_temporal}f.
Inspection of figure \ref{surf_temporal}f also shows that $[[\omega_b]] \gg [[\omega_t]]$, that is, near-interface vorticity production in the azimuthal direction is dominant. 
This acts to 
disrupt the  dynamics of vortex-pairing relative to the surfactant-free case as the `knitting process' is promoted by streamwise rather than azimuthal vorticity production and the vortex-ring deformation is replaced by vortex-reconnection and merging in the azimuthal direction in the surfactant-laden case.

\begin{figure}
\begin{center} 
\begin{tabular}{cc}
\includegraphics[width=0.5\linewidth]{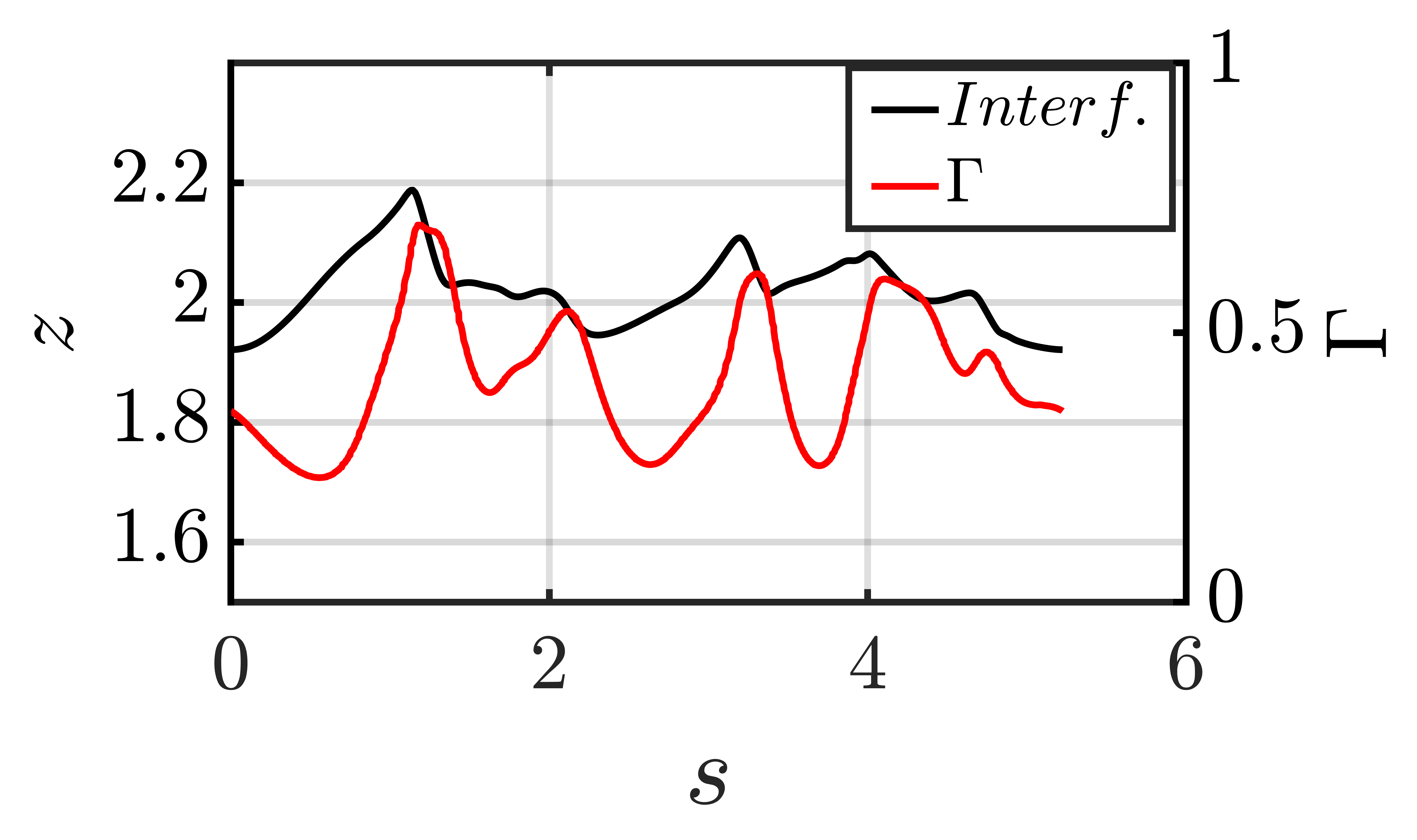} & 
\includegraphics[width=0.5\linewidth]{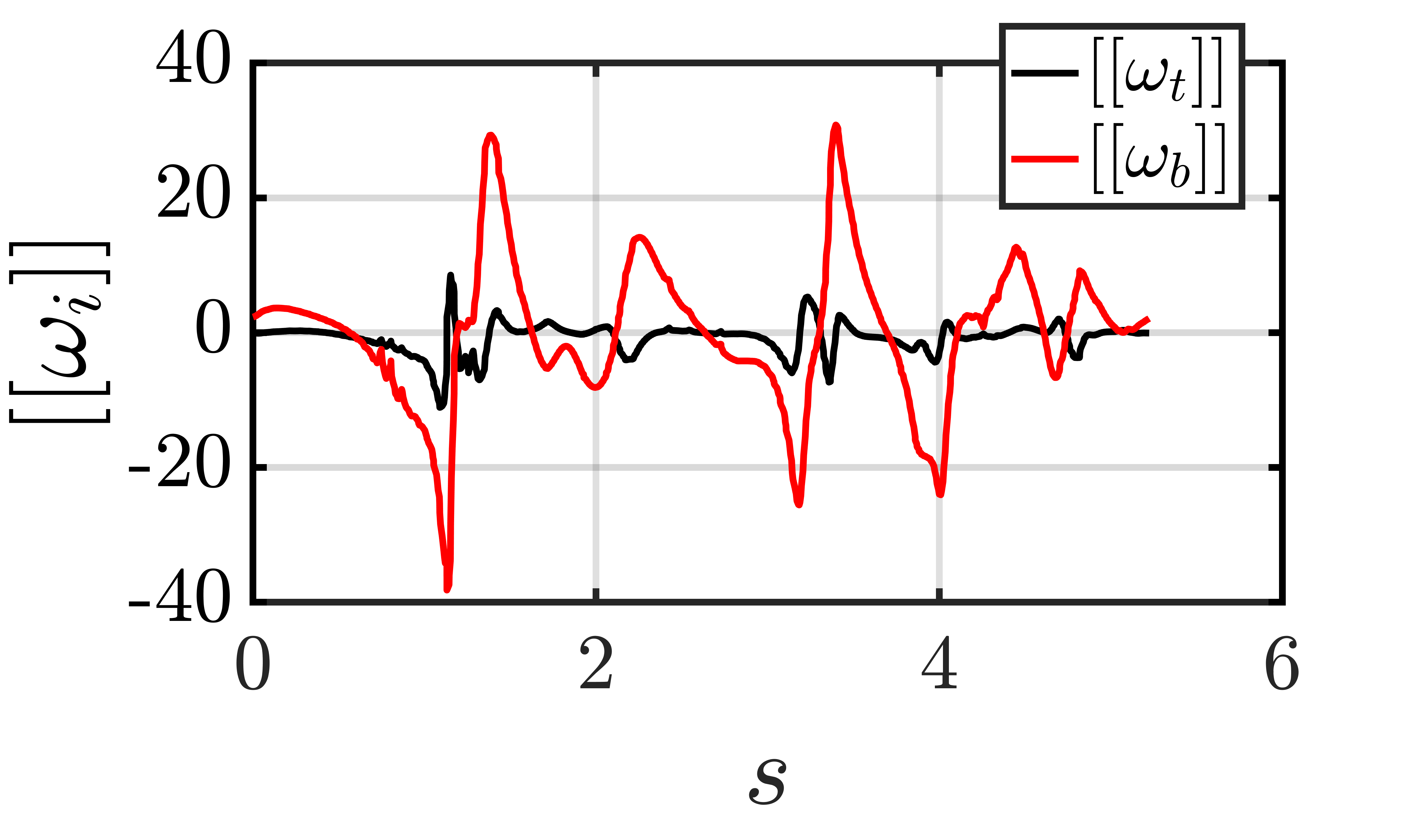} \\ 
(a) & (b) \\
\includegraphics[width=0.5\linewidth]{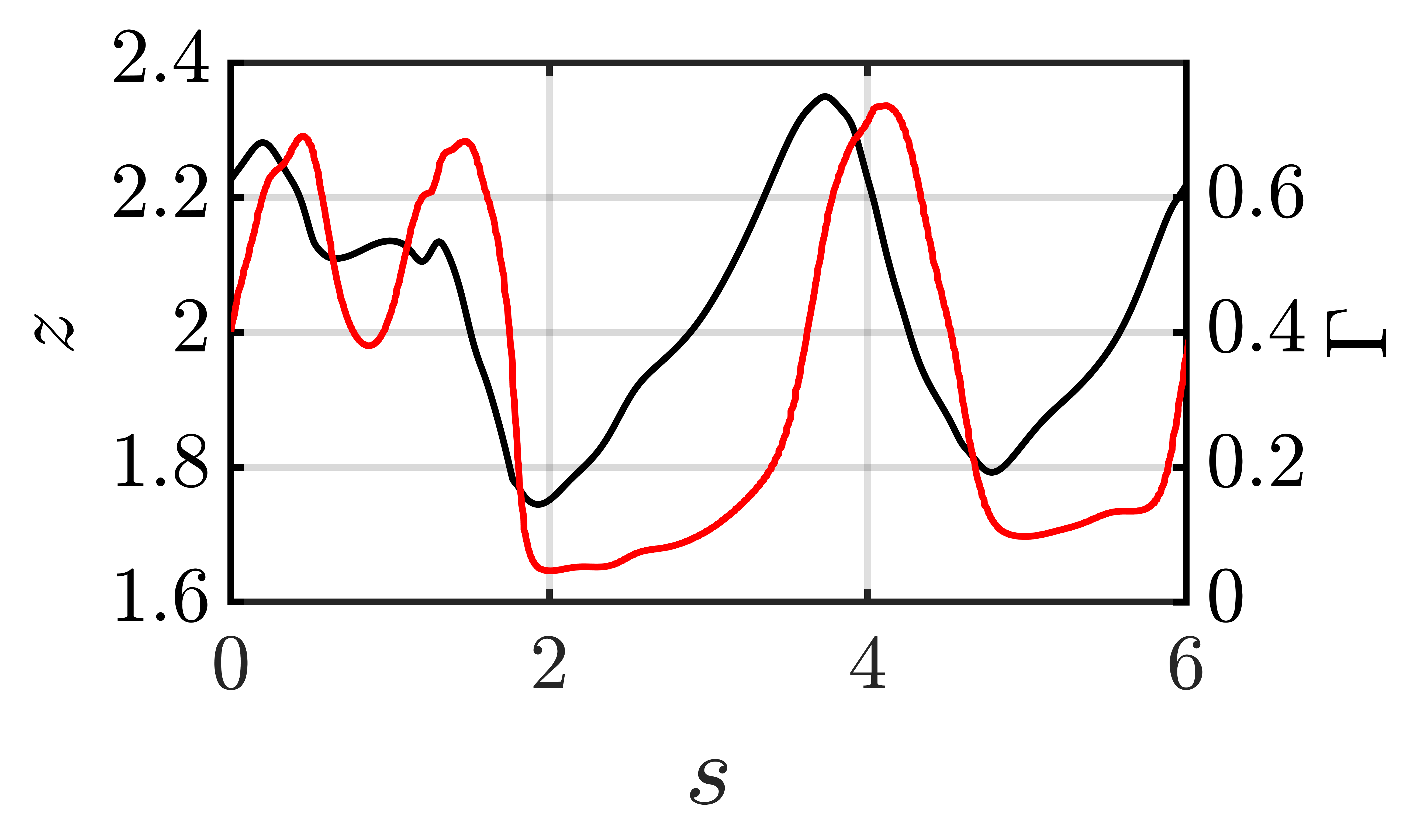} & 
\includegraphics[width=0.5\linewidth]{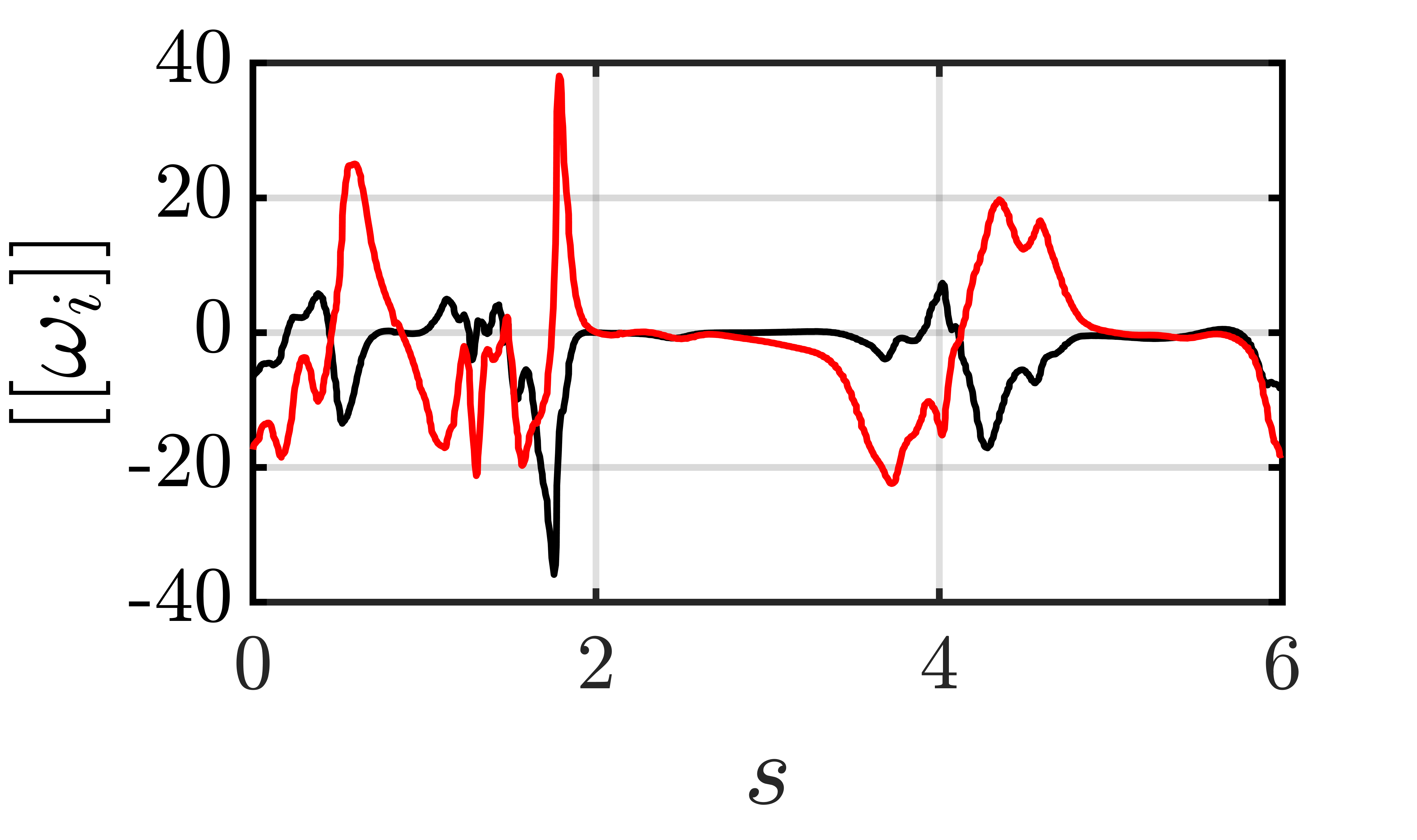} \\ 
(c) & (d) \\
\includegraphics[width=0.5\linewidth]{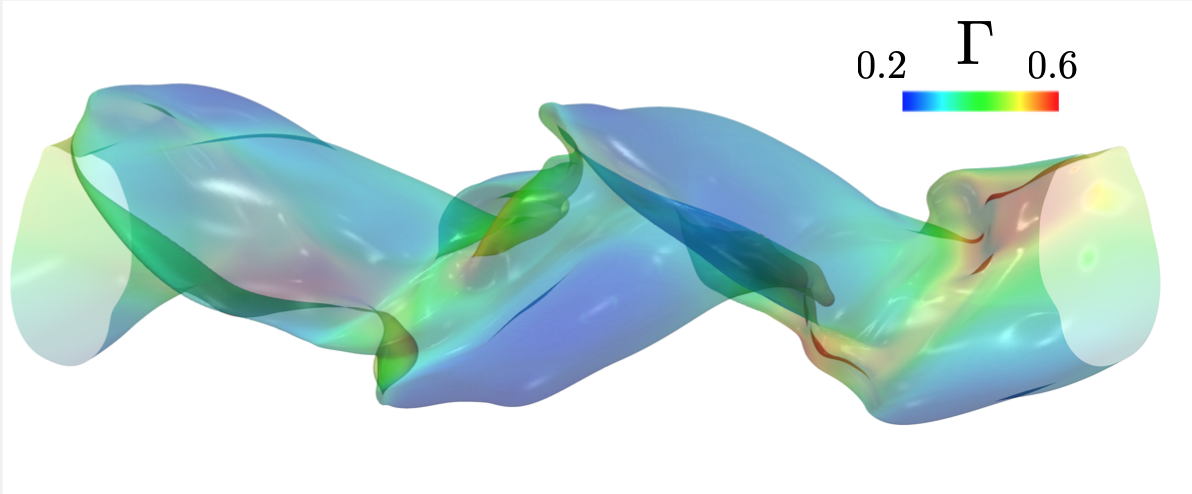} &
\includegraphics[width=0.5\linewidth]{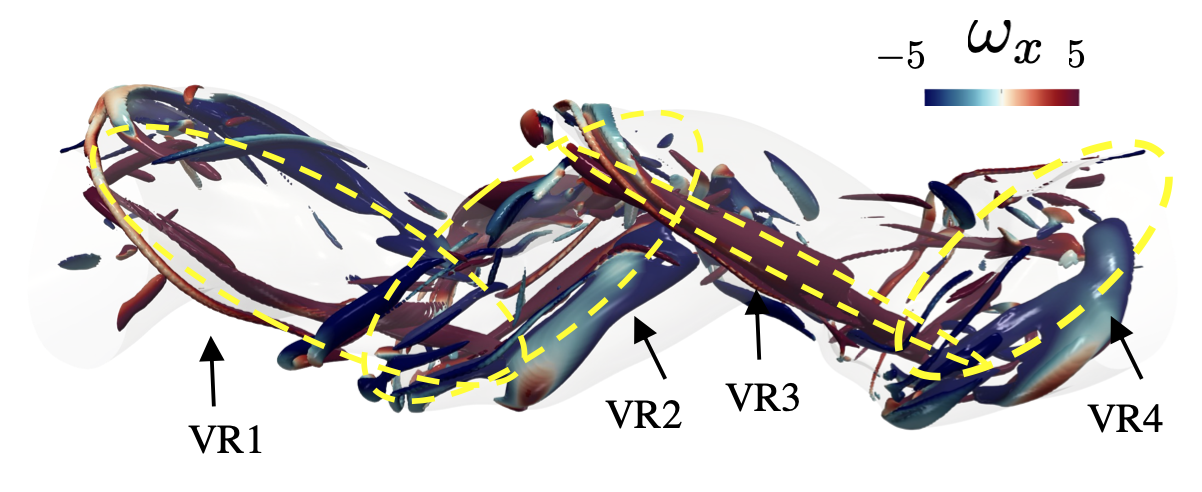}\\
(e) & (f) \\
\end{tabular}
\end{center} 
\caption{
Panels (a-b) and (c-d) show the location of the interface together with the surfactant concentration and the jumps of the vorticity across the interface for $t=36.81$ and $t=44.68$, respectively.
Panels (e-f) show a three dimensional representation of the interface location coloured by $\Gamma$ at $t=44.68$, and  vortex knitting  visualised via $Q$-criterion with value of 
$Q=1600$ where the colour represents  \textcolor{black}{$\omega_x$}, respectively.
\textcolor{black}{The arc length $s$ corresponds to the $x$–$z$ plane ($y = 2.5$) intersecting the interface.}
The center of the jet core corresponds to $z=2.5$.
The parameter values are the same as in figure \ref{surf_temporal}. 
\label{lobe_formation}} 
\end{figure}


For increasing time, figure \ref{HV_formation} shows the formation of surfactant-induced inner hairpin-like vortical structures. 
The shear stress, which is generated to balance the gradients in $\sigma$ gives rise to
counter-rotating streamwise vortices of similar magnitude to the KH rings (labelled `{\it SV1}' and `{\it SV2}' in figure \ref{HV_formation}b).
These structures grow in the $x-$direction into a  combination of streamwise vortices close to the interface, i.e. legs, and a hairpin-like head close to the center-plane of the jet (see figure \ref{HV_formation}d).
The hairpin-legs extend from the  regions of high-to-low values of $\Gamma$ on the surface,  while the hairpin-head points down in the positive $x-$direction (labelled {\it `HV1'} and {\it `HV2'} in figure \ref{HV_formation}e).
To complete the presentation of these hairpin-like vortical structures, figure  \ref{HV_formation}f,g show the direction of flow rotation of the legs and head for {\it HV1}.
For comparison, we have added  arrows to show velocity direction and  to prove that this coherent vortical structure  exhibits the same qualitative behaviour as the HV proposed by \citet{Theodorsen1955} for near-wall turbulence.
To the best of our knowledge, the formation of hairpin-like vortical structures induced by surfactant effects has not been reported yet.
We have also observed surfactant-driven outer hairpin-like vortical structures (not shown) whose heads are in the negative $x-$direction (in the frame of reference of the legs).

\begin{figure}
\begin{center} 
\begin{tabular}{c|c}
Surfactant-laden case & Surfactant-free case\\
\includegraphics[width=0.46\linewidth]{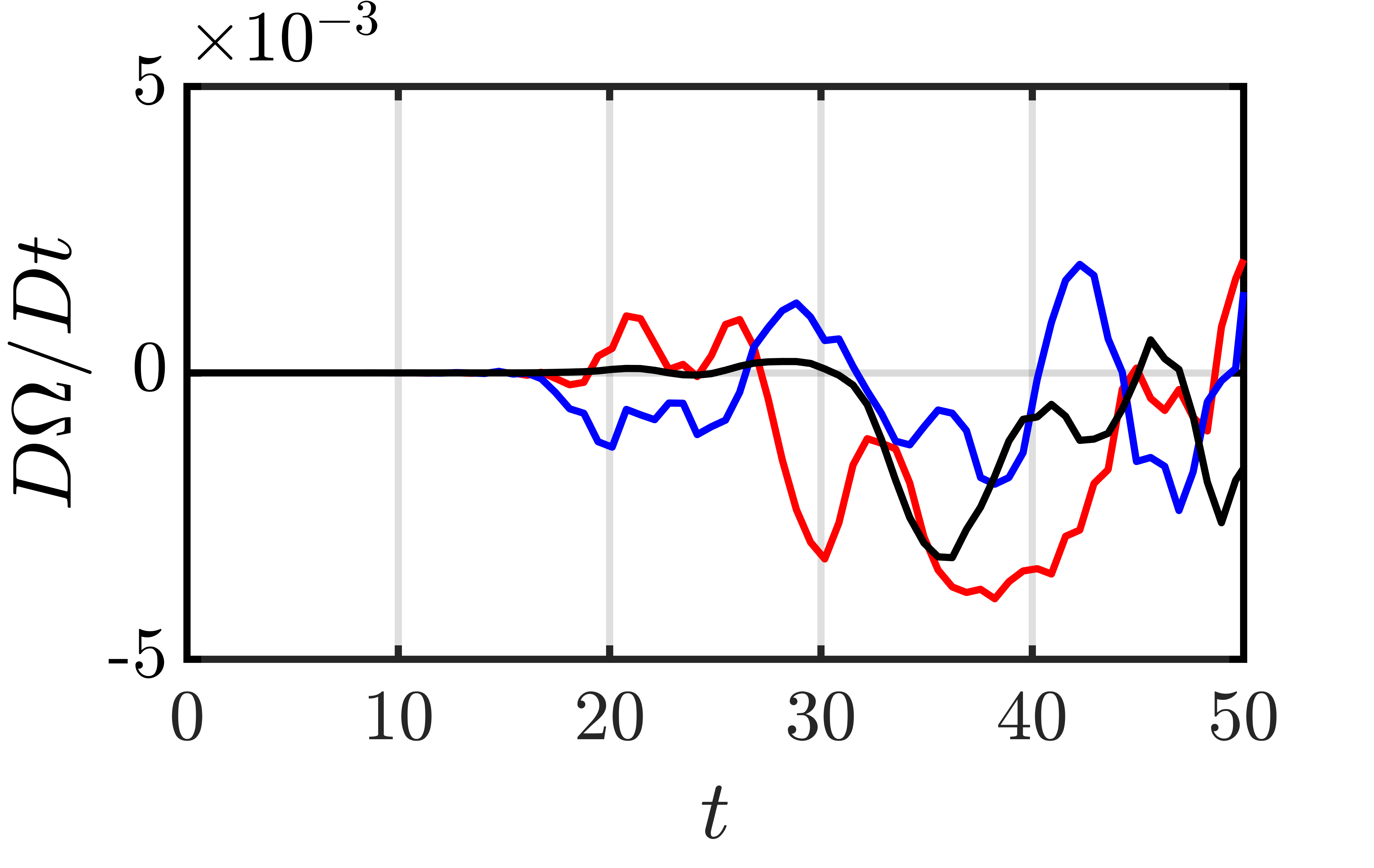} &
\includegraphics[width=0.46\linewidth]{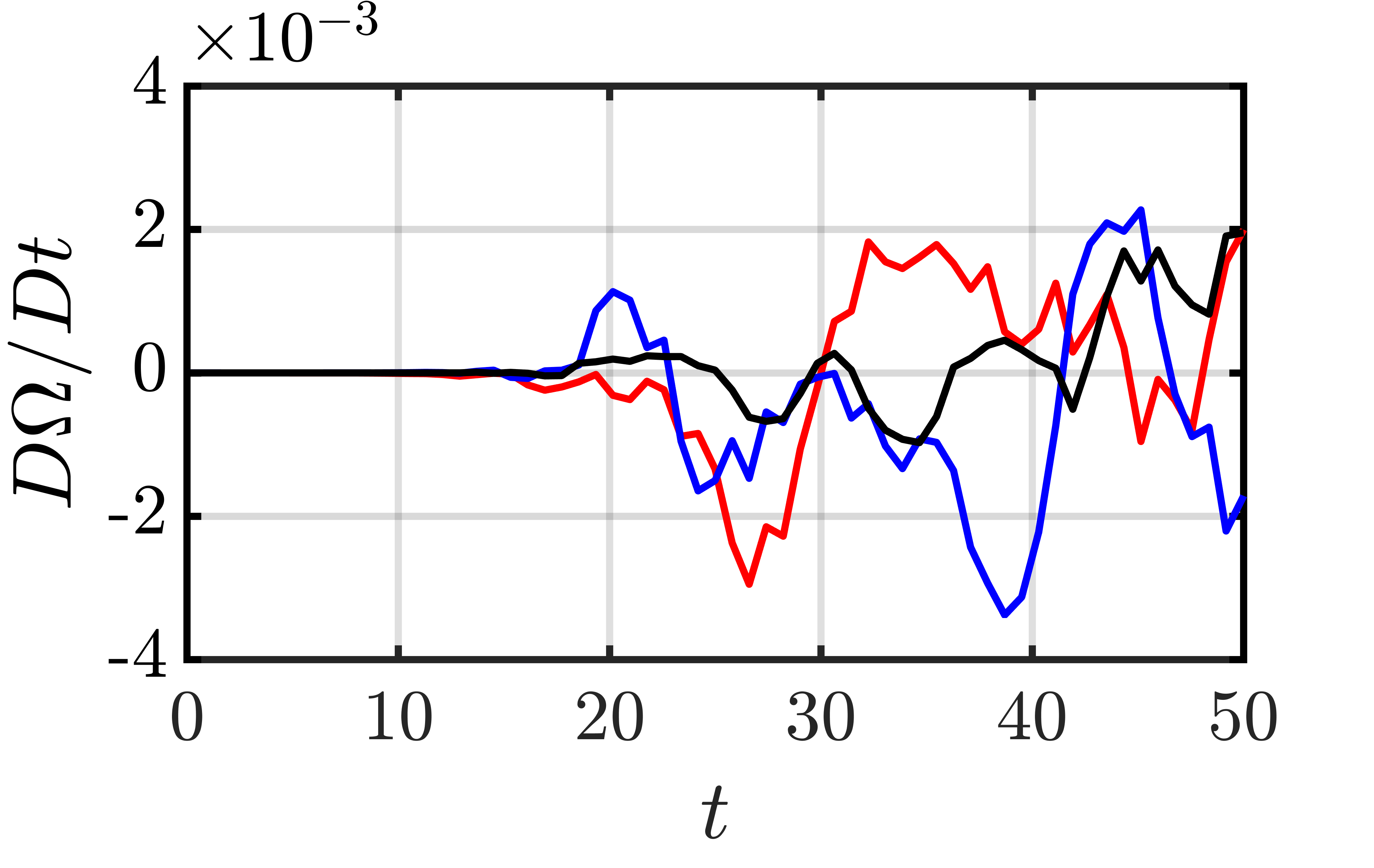} \\ 
(a) & (b) \\
\includegraphics[width=0.46\linewidth]{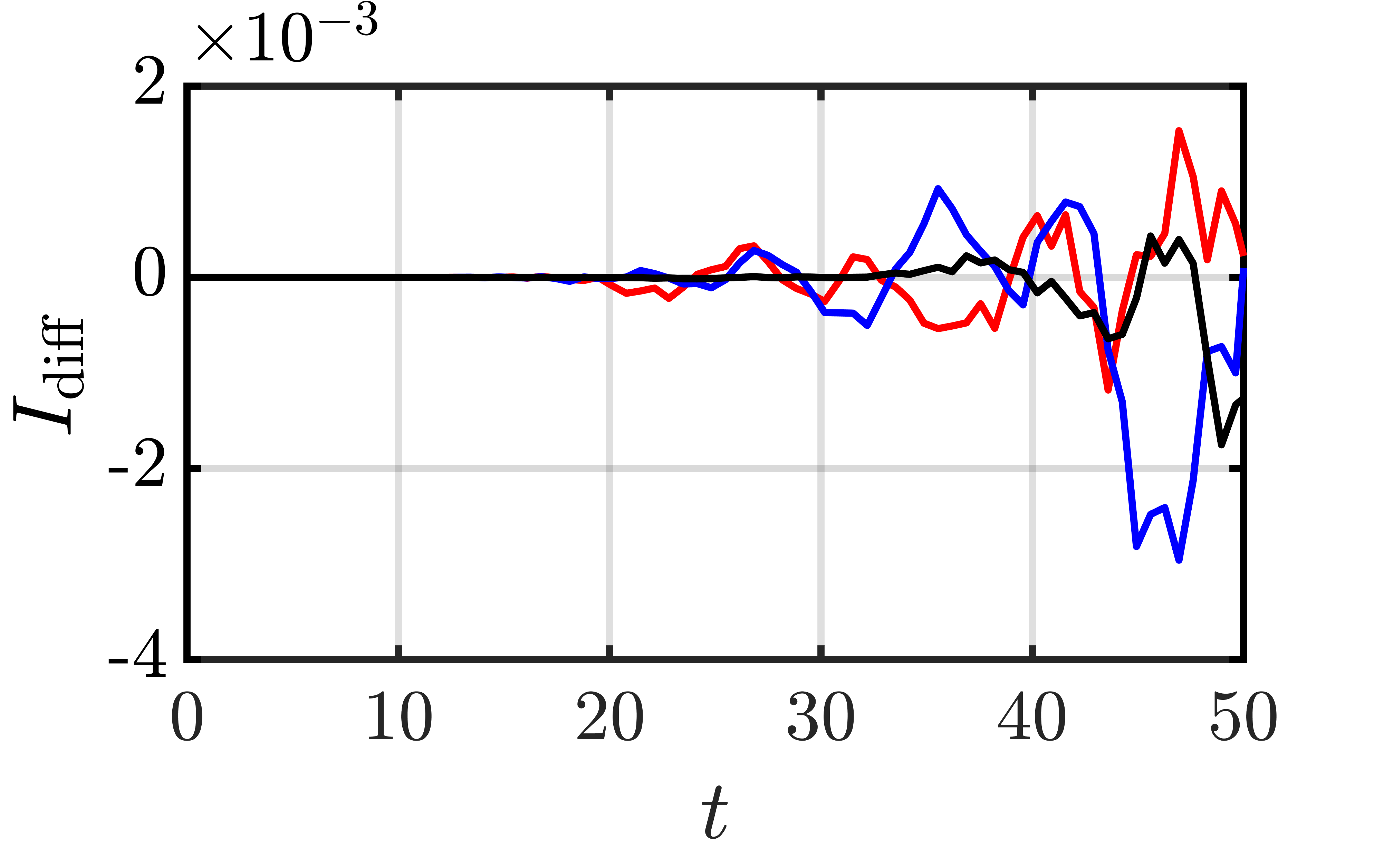} &
\includegraphics[width=0.46\linewidth]{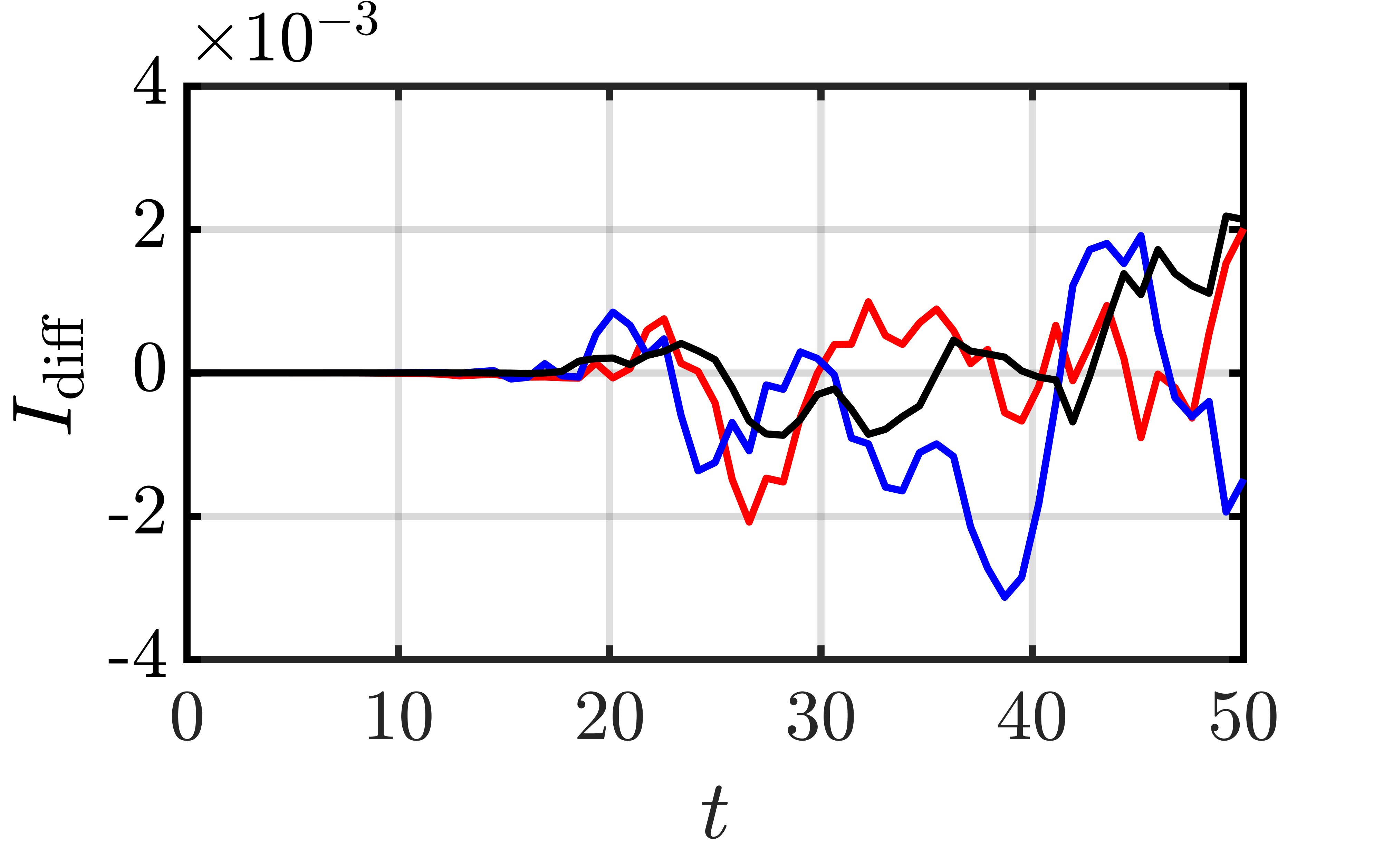} \\ 
(c) & (d) \\
\includegraphics[width=0.46\linewidth]{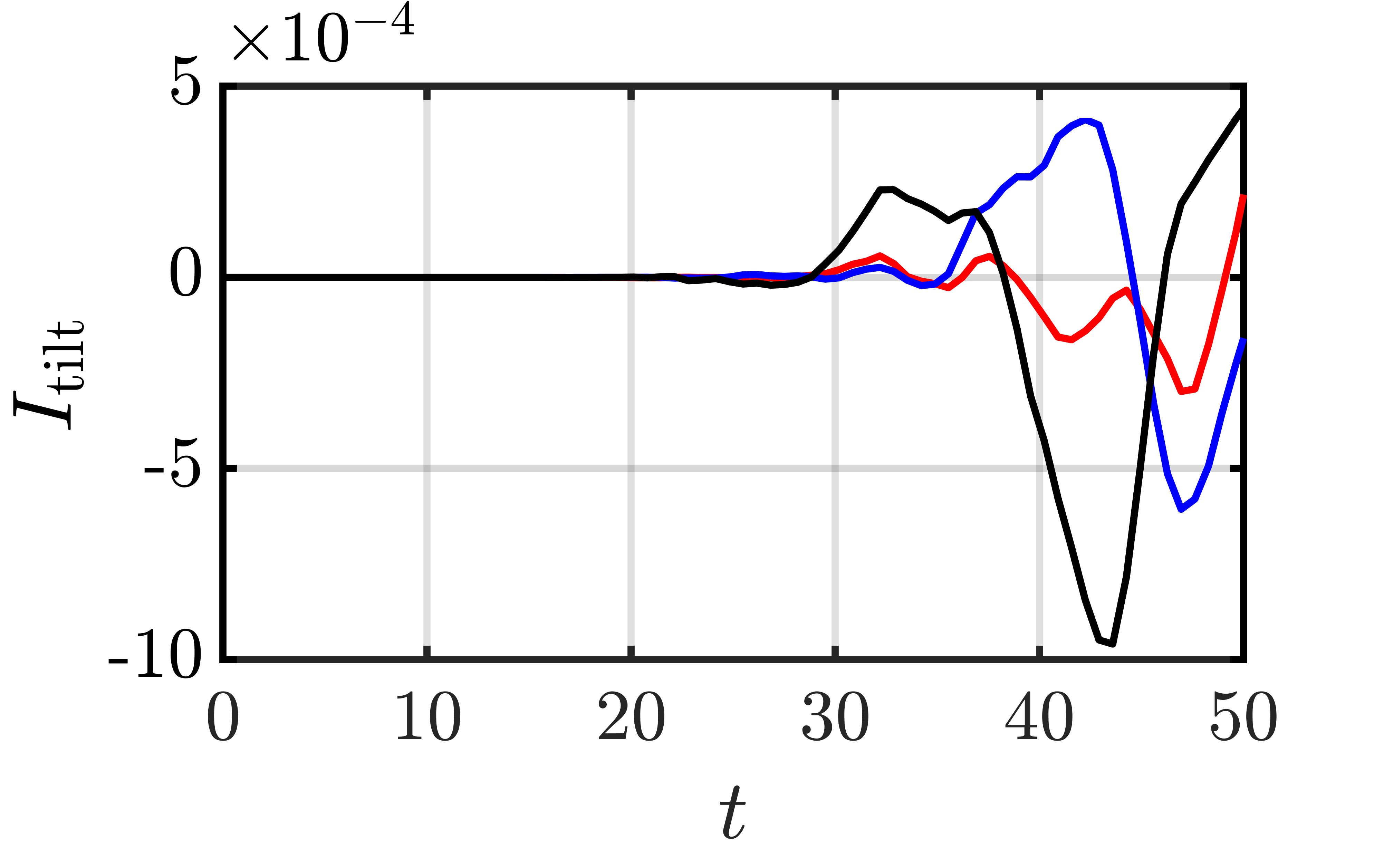} &
\includegraphics[width=0.46\linewidth]{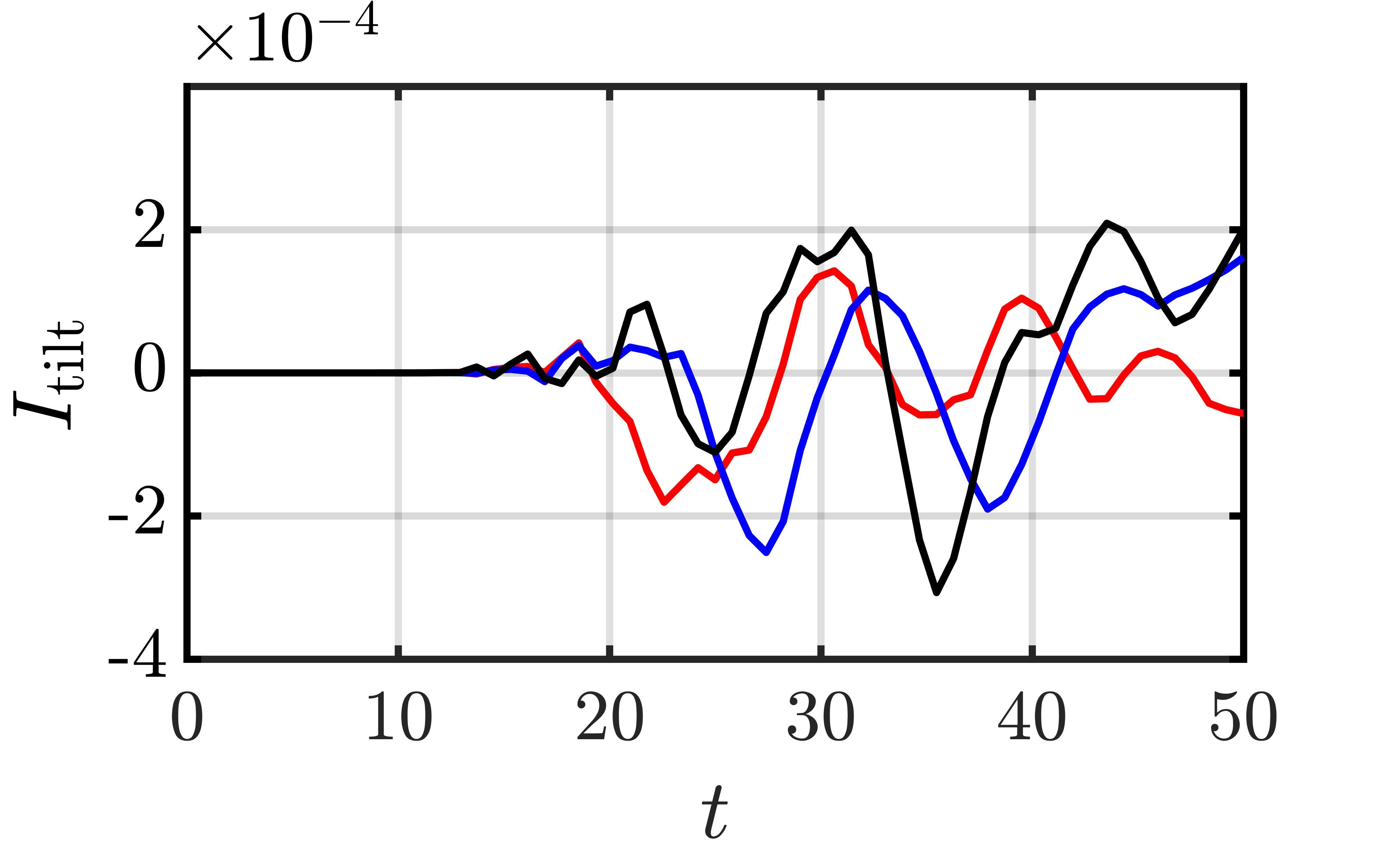} \\ 
(e) & (f) \\
\includegraphics[width=0.46\linewidth]{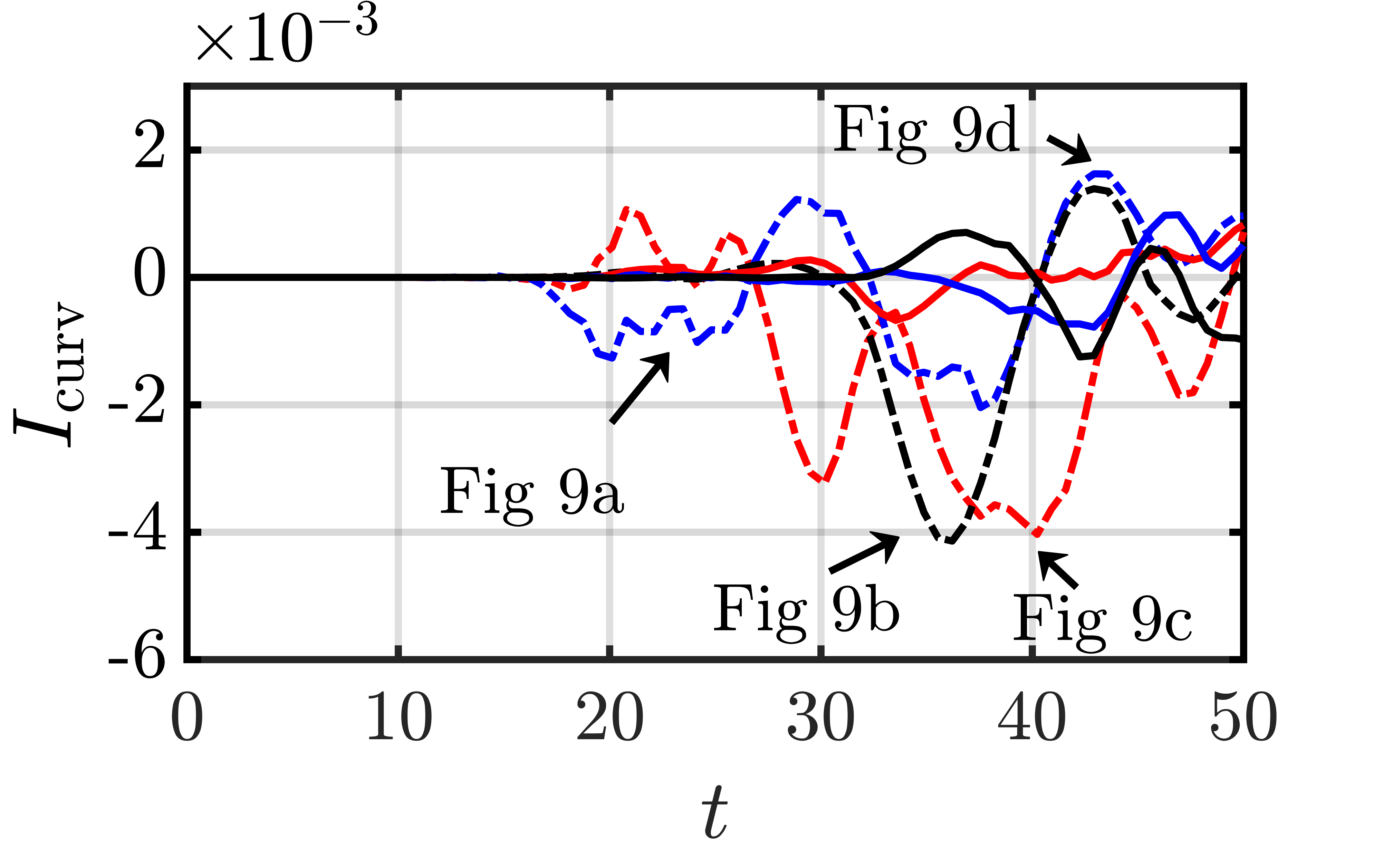} &
\includegraphics[width=0.46\linewidth]{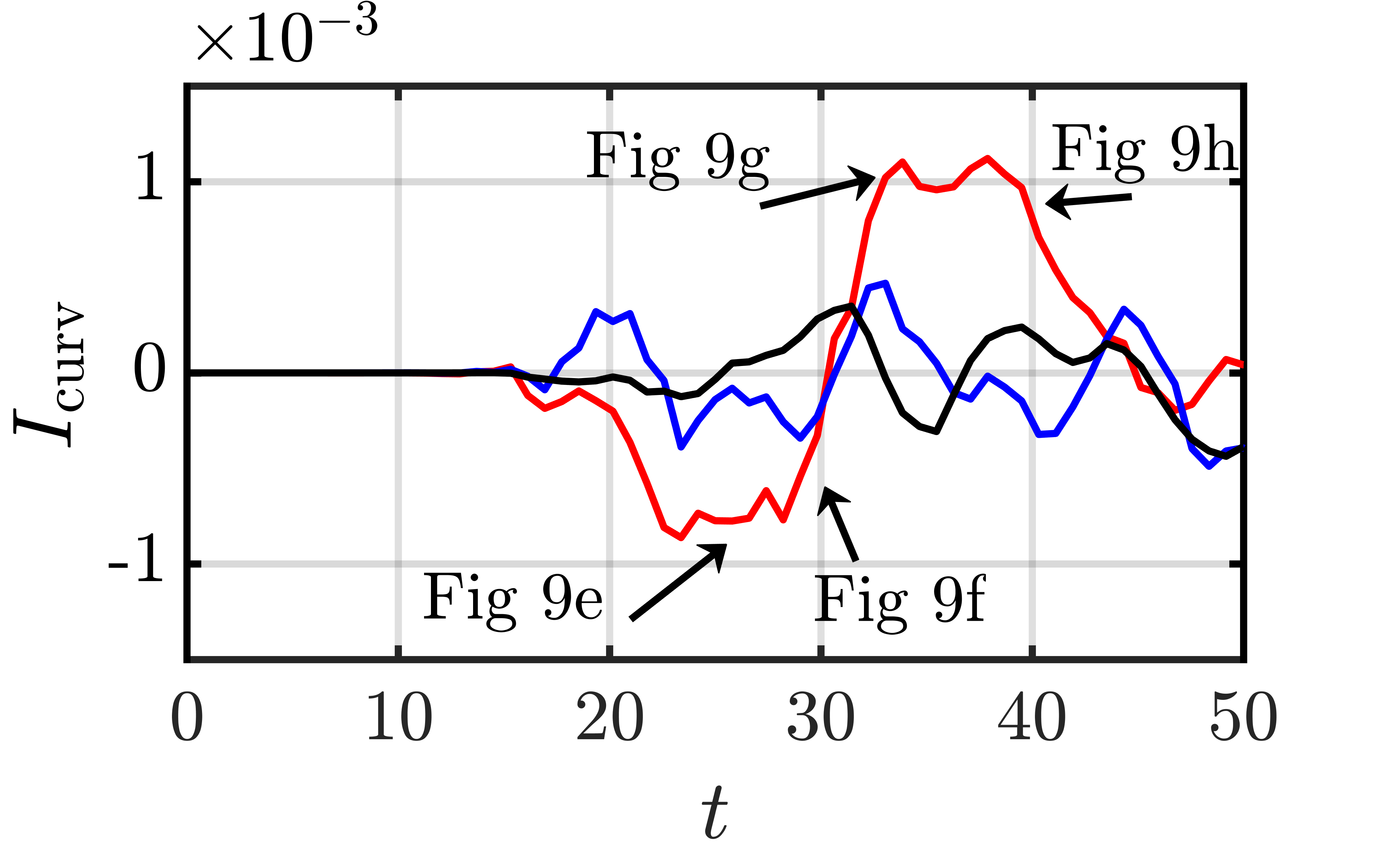} \\
(g) & (h) \\
\end{tabular}
\end{center} 
\caption{ 
Total rate of change of circulation, $\boldsymbol{\Omega}$, according to equation \ref{eq:circulation_jump_3D_simplified_dless}: $D\boldsymbol{\Omega}/Dt$,  vortex diffusion ($I_{\rm diff}$), vortex tilting ($I_{\rm tilt}$), and surface tension ($I_{\rm curv}$)
are shown in rows one to four, respectively; see equation (\ref{eq:I_define}) for the definitions of $I_{\rm tilt}$, $I_{\rm diff}$, and $I_{\rm curv}$. Surfactant-laden and surfactant-free cases correspond to left and right panels, respectively. For panel (g), we represent the  contributions that arise from the gradients of curvature (solid lines)  and  the gradients of surface tension (dashed lines) to underscore the relative importance of the Marangoni stresses.
Red, blue, and black colored lines represent component $x$, $y$ and $z$ of $D\boldsymbol{\Omega}/Dt$, $I_{\rm tilt}$, $_{\rm diff}$, and $I_{\rm curv}$. The parameters are $Re=5000$, $We=500$ (and for the surfactant-laden case) $\beta_s=0.5$, $Pe_s=100$ and $\Gamma_o=\Gamma_{\infty}/2$.
\label{fig:circulation}} 
\end{figure}

At later times, figure \ref{lobe_formation}a-d  shows the variation with arc length of the interfacial location, $\Gamma$, and  $[[\omega_t]]$ and $[[\omega_b]]$ at $t=36.51$ and $t=44.68$;  corresponding three-dimensional representations of the interface are also shown in figure \ref{lobe_formation}e,f for $t=44.68$ coloured by the magnitude of $\Gamma$ and the $Q$-criterion, respectively.   
%
The flow is accompanied by radially-converging and diverging motion due to vortex-surface-interaction; interfacial convection drives surfactant towards the inner lobes  (interfacial contraction), and away from the outer lobes (interfacial expansion).
Vorticity jumps are highest in the interfacial regions with the largest gradients in $\Gamma$. As time evolves, the ratio of these Marangoni-driven $[[\omega_t]]$ to $[[\omega_b]]$ reduces and this results in large coherent structures which merge to form
counter-rotating streamwise vortical rings that eventually `knit' with the adjacent vortex ring located in the $x-$direction (labelled `\textit{VR1-VR4}' in  figure \ref{lobe_formation}f); this pairing is similar to the surfactant-free case (in agreement with 
\citet{Urbin} and \citet{da_Silva}).
%

\begin{figure}
\begin{center} 
\begin{tabular}{cccc}
\includegraphics[trim =  600 1 600 1, clip,width=0.25\linewidth]{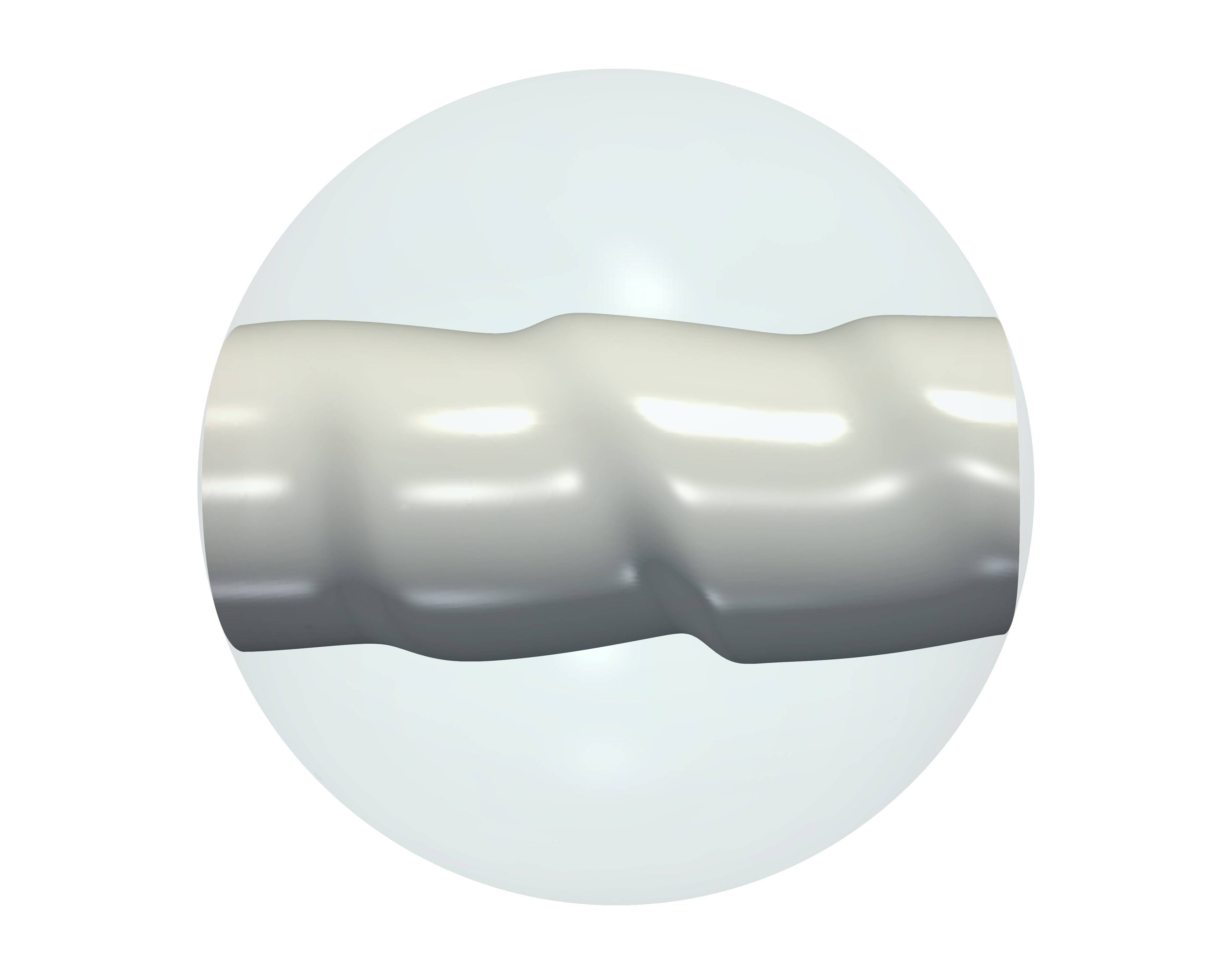}     &
\includegraphics[trim =  600 1 600 1, clip,width=0.25\linewidth]{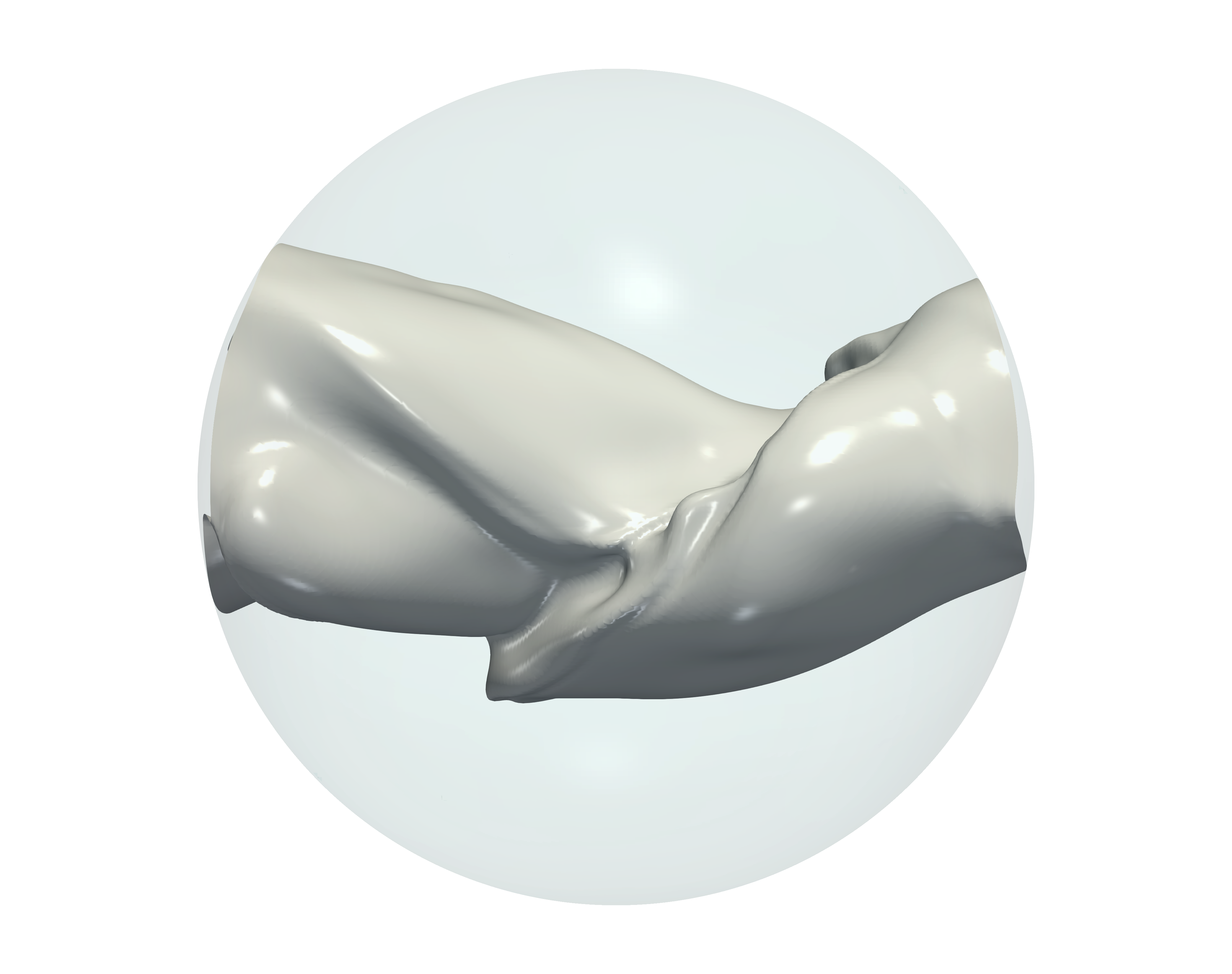}     &
\includegraphics[trim =  600 1 600 1, clip,width=0.25\linewidth]{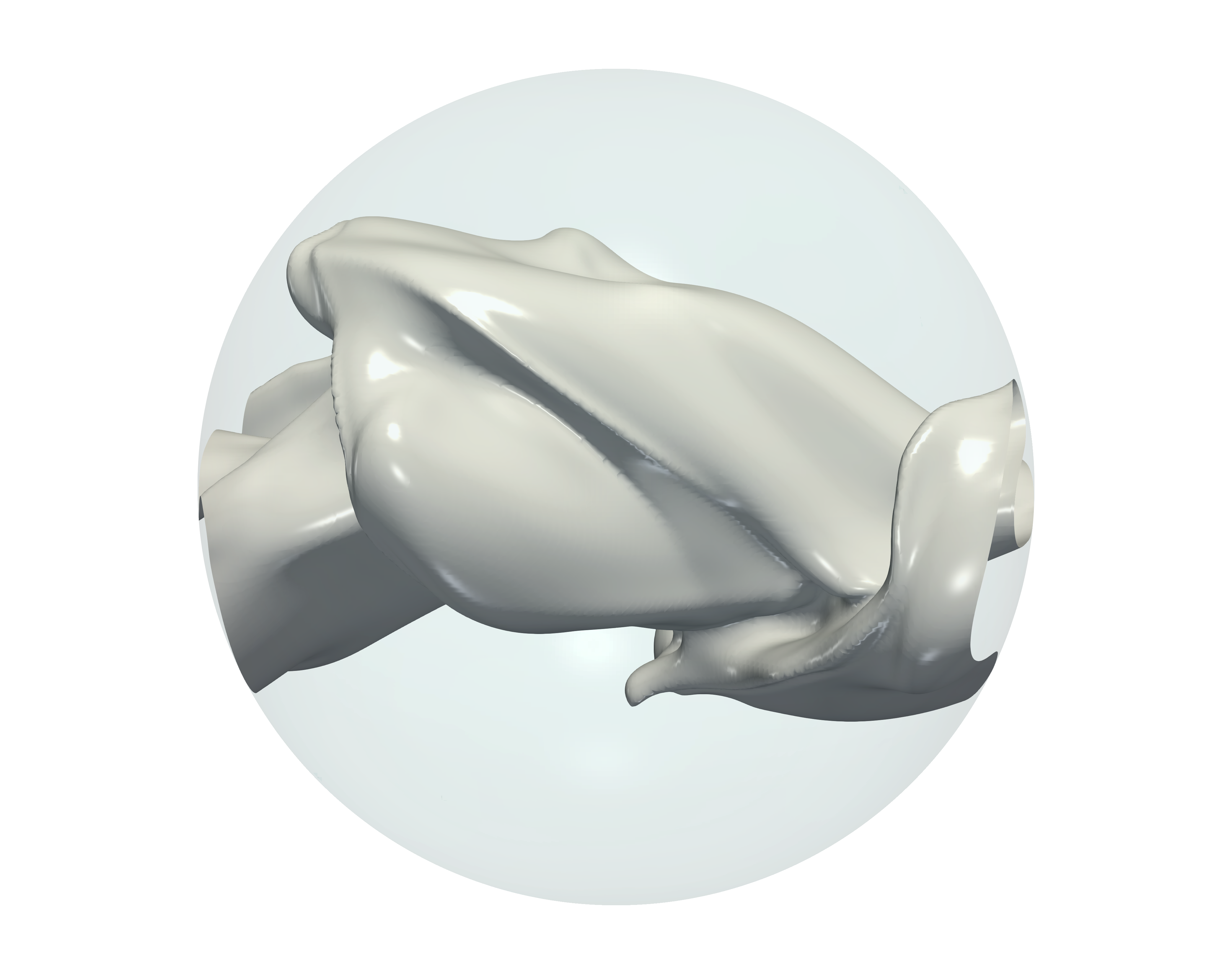}     &
\includegraphics[trim =  600 1 600 1, clip,width=0.25\linewidth]{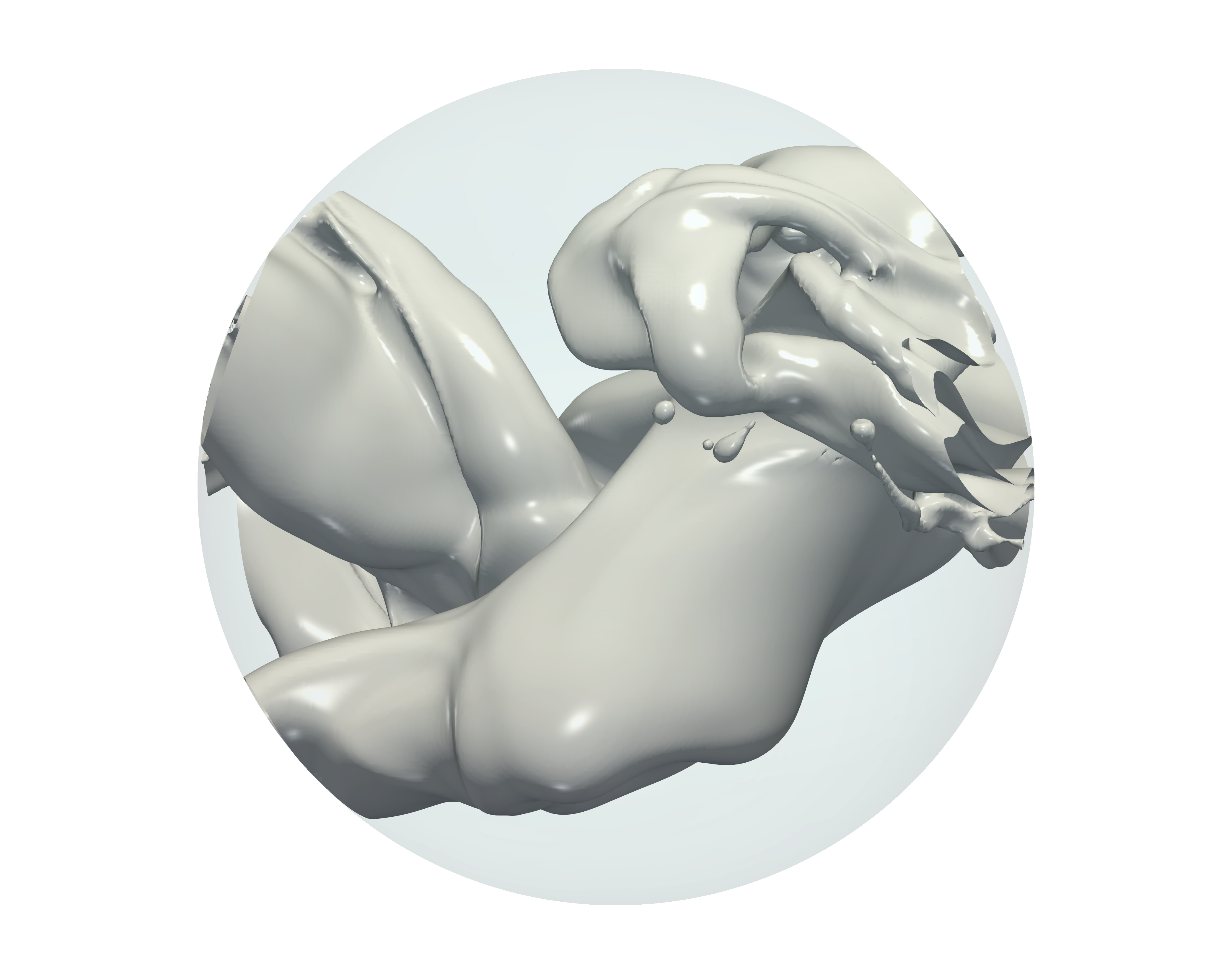}    
\\
(a) & (b) & (c) & (d)\\
\includegraphics[trim =  600 1 600 1, clip, width=0.25\linewidth]{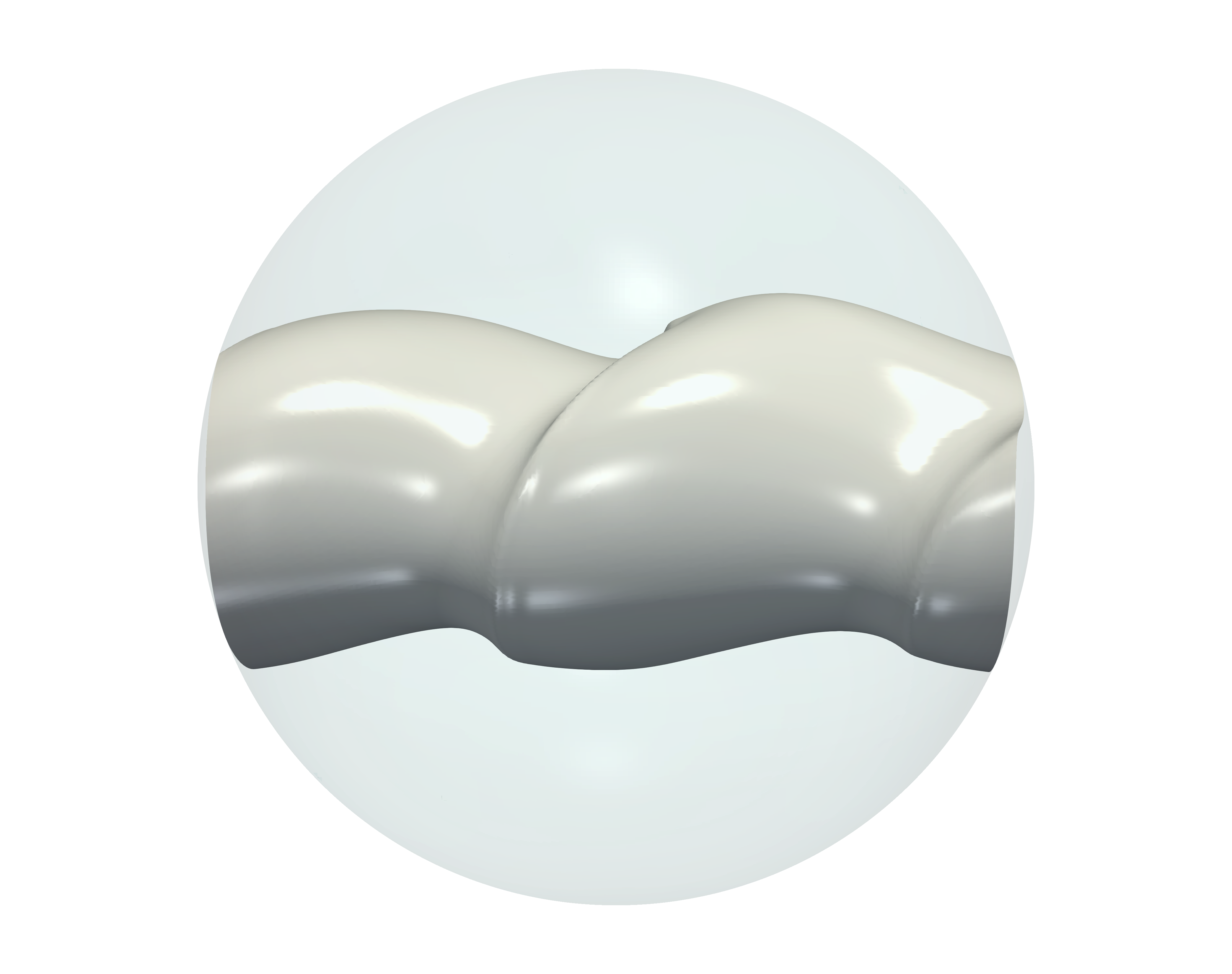}     &
\includegraphics[trim =  600 1 600 1, clip, width=0.25\linewidth]{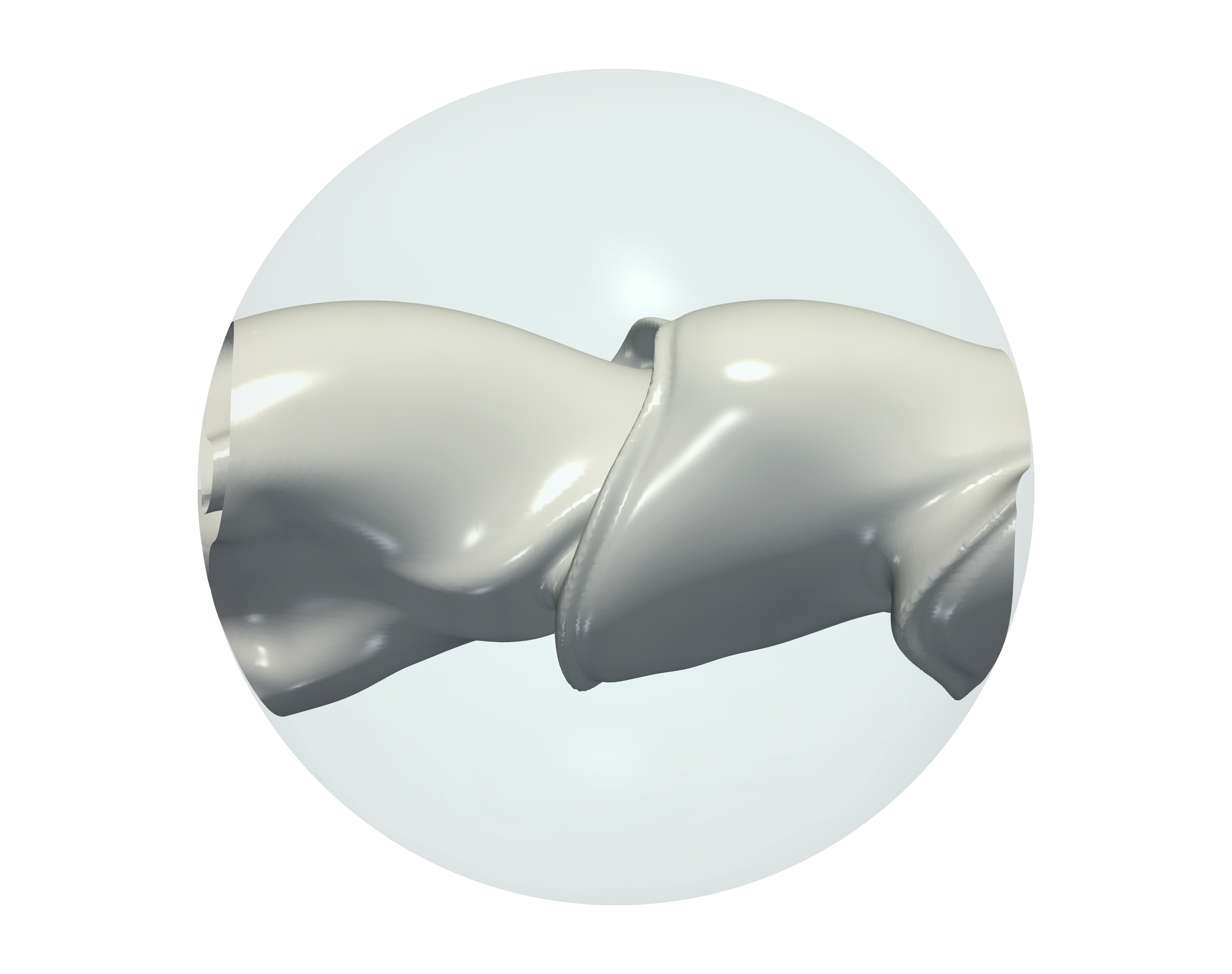}     &
\includegraphics[trim =  600 1 600 1, clip, width=0.25\linewidth]{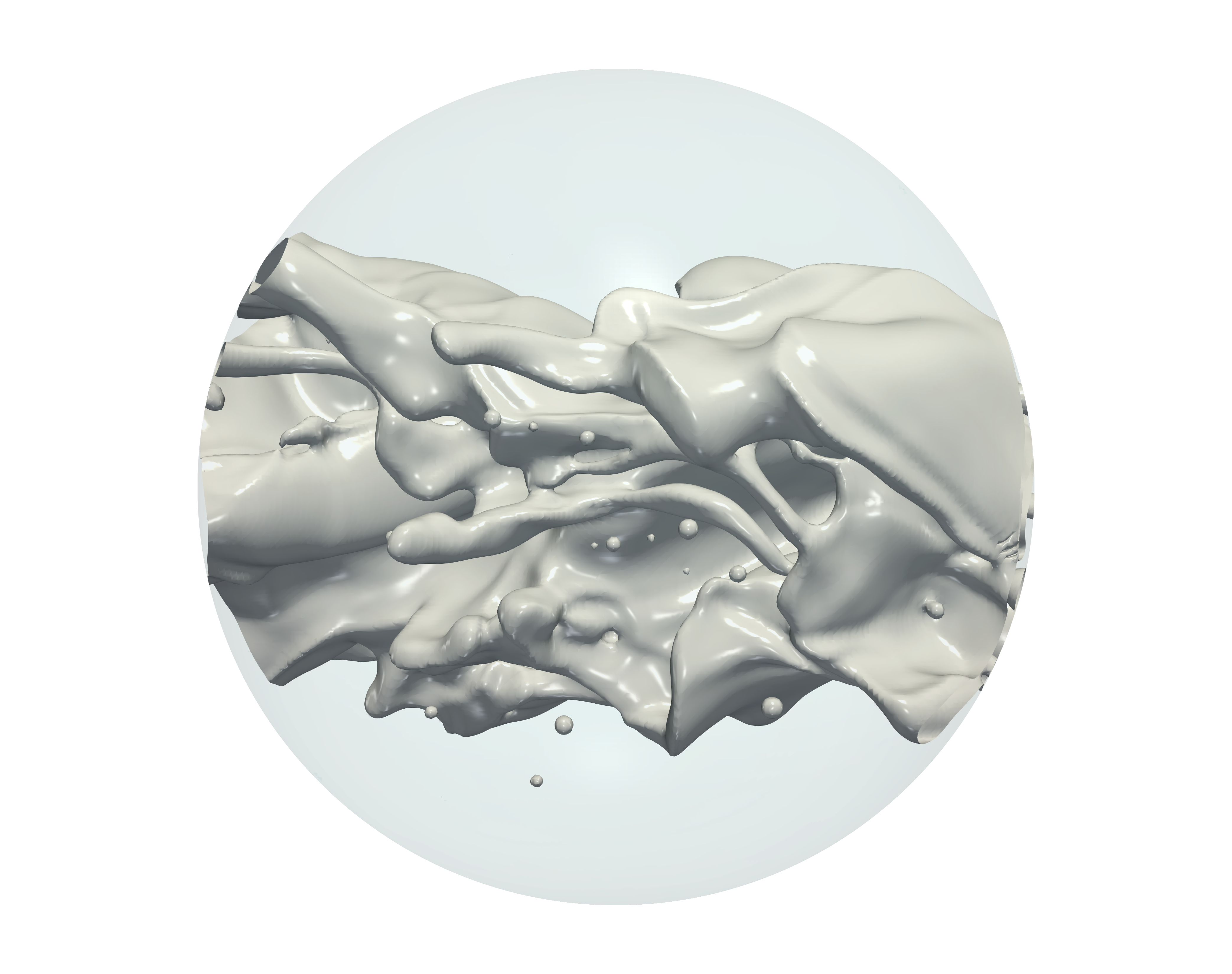}     &
\includegraphics[trim =  600 1 600 1, clip,width=0.25\linewidth]{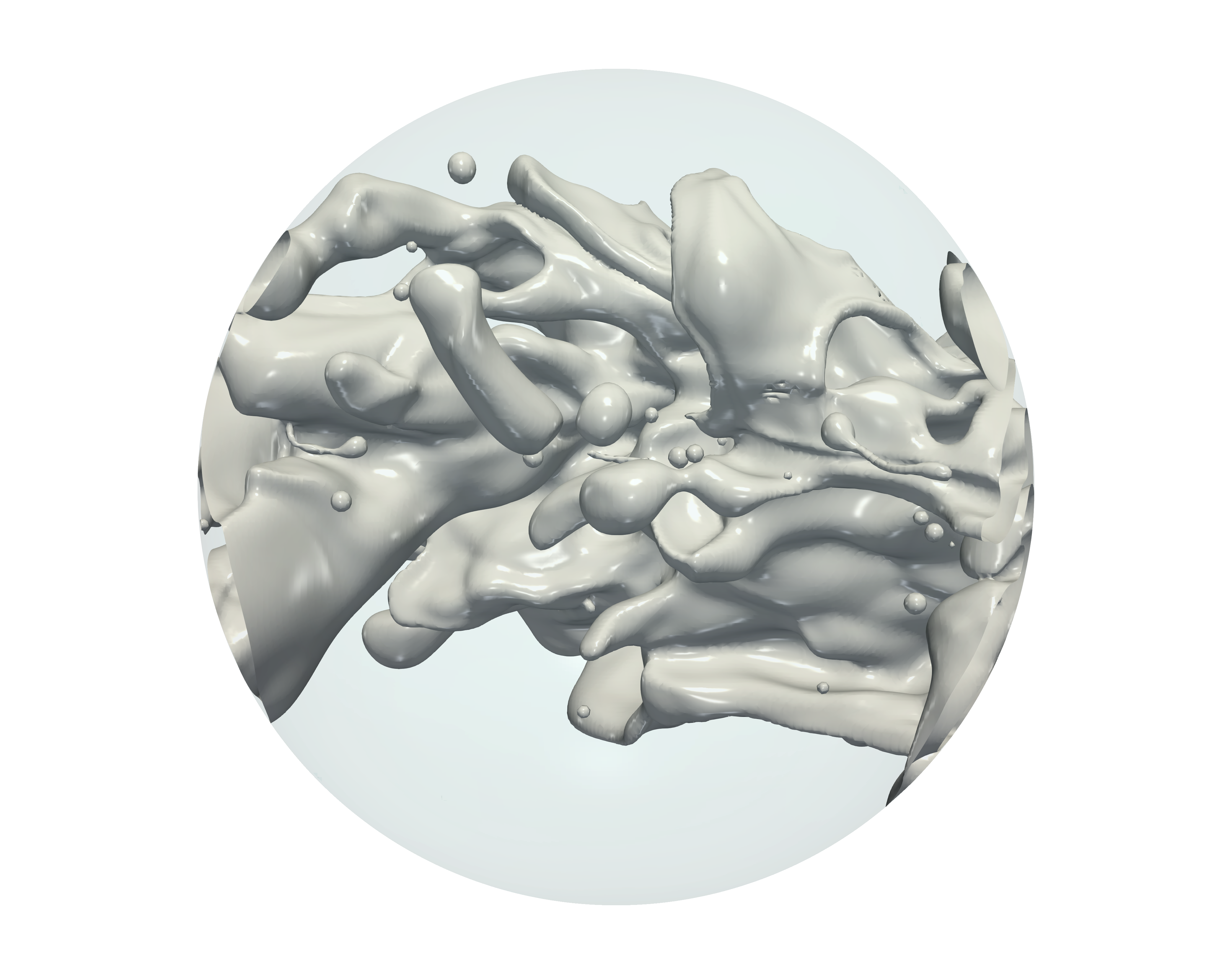}     \\
(e) & (f) & (g) & (h)\\
\end{tabular}
\end{center} 
\caption{
Spatio-temporal evolution of the control volume $V$ with a surface $\partial V$ which encloses an interface plane $I$ used to calculate the rate of change of the circulation in figure \ref{fig:circulation}.  Panels (a)-(d) and (e)-(h) correspond to the surfactant-laden and surfactant-free cases, respectively, at $t=(25.20, 35.50, 40.37, 43.75 )$ and $t=(27.12, 30.43, 34.37, 40.55 )$, and the same parameters as in figure (\ref{fig:circulation}).
} 
\label{fig:V}
\end{figure}

\begin{figure}
\begin{center} 
\begin{tabular}{cc}
\includegraphics[width=0.49\linewidth]{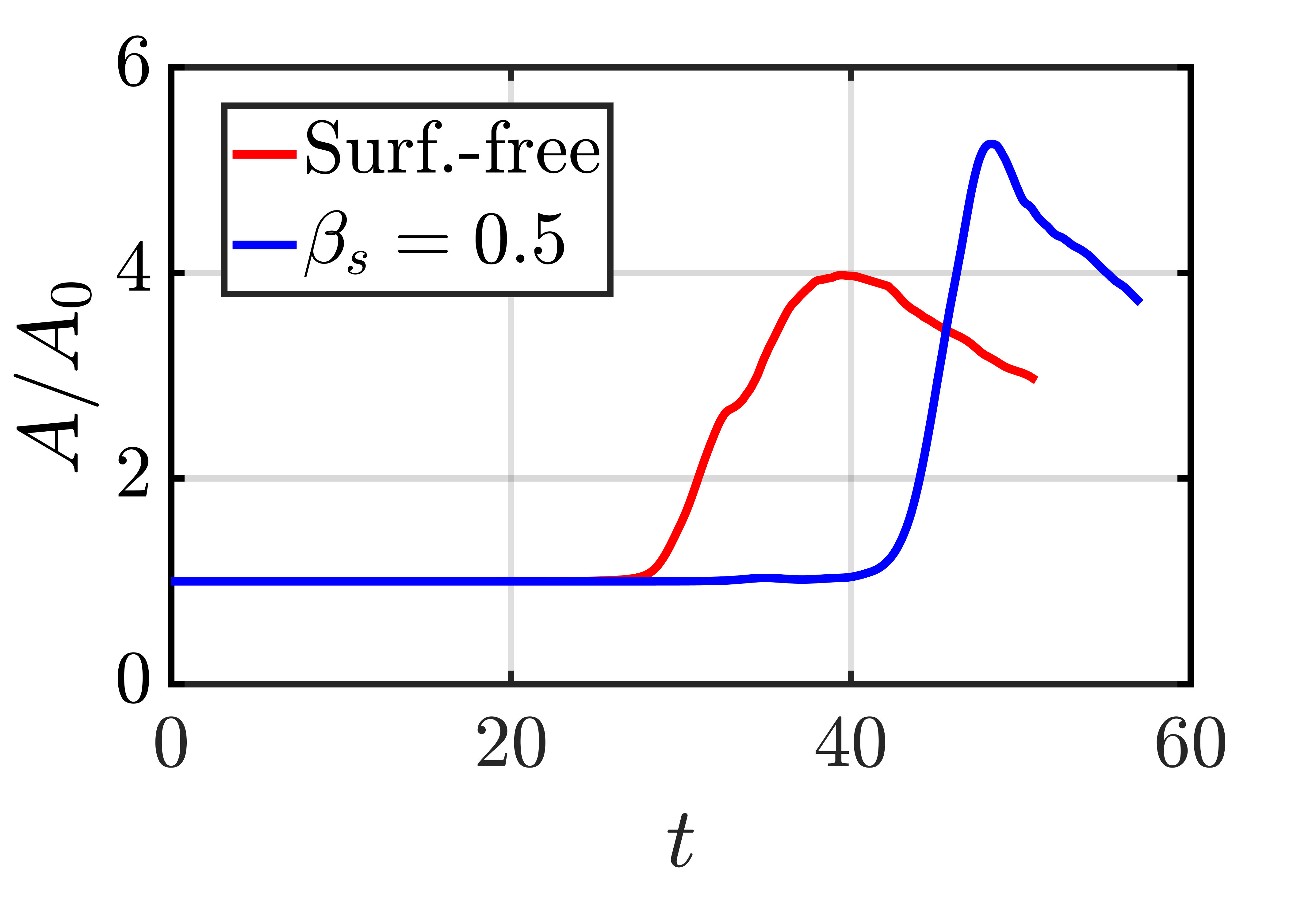}     &
\includegraphics[width=0.49\linewidth]{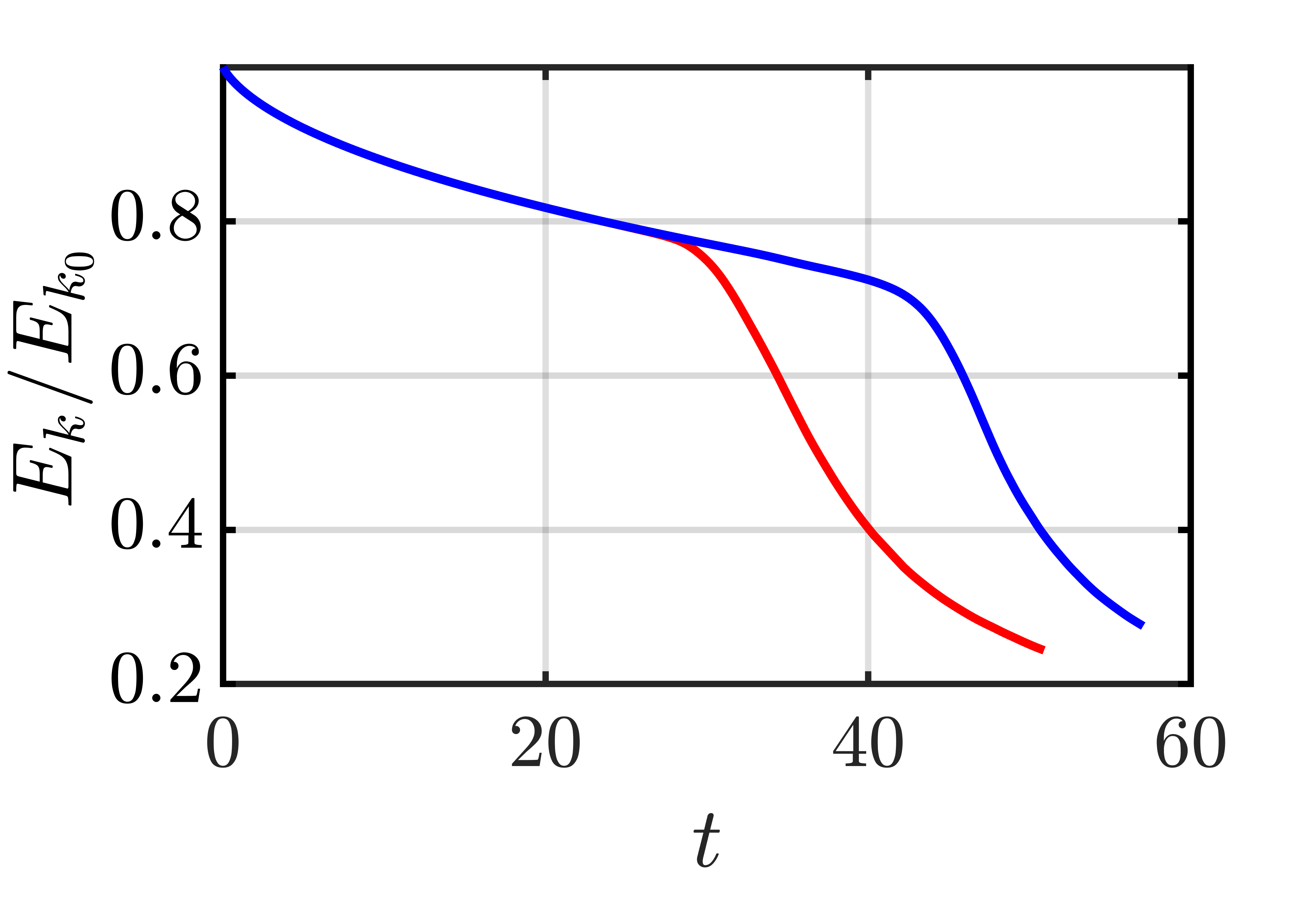}   \\
  (a)   & (b)
\end{tabular}
\begin{tabular}{c}
\includegraphics[width=0.54\linewidth]{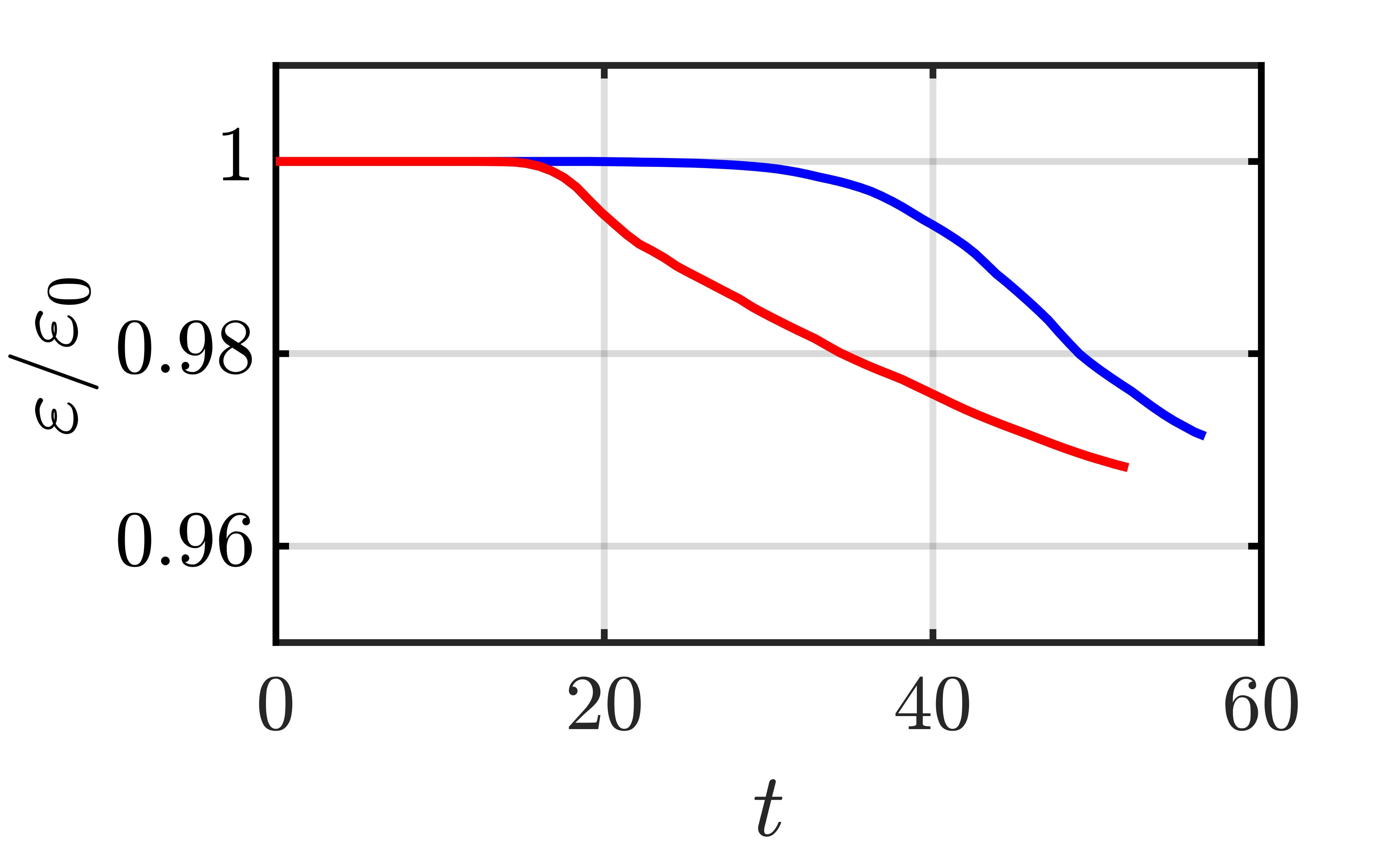}        \\
  (c)   
\end{tabular}
\end{center} 
\caption{Temporal evolution of (a) surface area, $A$,  (b) kinetic energy, $E_k=\rho\int_V \mathbf{u}^2/2 dV$, and (c) enstrophy, $\varepsilon=\int_V |\boldsymbol{\omega}^2|dV$, scaled by the initial interfacial area, $A_0$,  kinetic energy, $E_{k0}$, and enstrophy, $\varepsilon_0$, respectively. The parameter values are the same as in figure \ref{surf_temporal}. \label{metrics}} 
\end{figure}

We now examine the dynamics of the circulation $\boldsymbol{\Omega}$ by considering equation   \ref{eq:circulation_jump_3D_simplified_dless} which we express as follows:
\begin{equation}
    \frac{D\boldsymbol{\Omega}}{D\tilde{t}}=
    I_{\rm tilt} + I_{\rm diff} + I_{\rm curv},
\end{equation}
where $I_{\rm tilt}$, $I_{\rm diff}$, and $I_{\rm curv}$ are defined as
\begin{equation}
    I_{\rm tilt}\equiv \oint_{\partial \tilde{V}}\hatn\cdot\tilde{\om}\tilde{\uvec}d\tilde{S},~~
    I_{\rm diff}\equiv \frac{1}{Re}\oint_{\partial \tilde{V}}\hatn\cdot\nabla\tilde{\om}d\tilde{S},~~
    I_{\rm curv}\equiv
    \frac{1}{We}\oint_I\hatb\frac{\partial}{\partial \tilde{s}}\left(\tilde{\sigma}\left[\tilde{\kappa}_1+\tilde{\kappa}_2\right]\right)d\tilde{S},
    \label{eq:I_define}
\end{equation}
which correspond to vortex tilting/stretching, diffusion of vorticity, and circulation variation due to gradients in curvature and interfacial tension (in the case of surfactant-laden systems). Figure \ref{fig:circulation} shows the temporal evolution of $D\boldsymbol{\Omega}/Dt$, $I_{\rm tilt}$, $I_{\rm diff}$, and $I_{\rm curv}$ which 
allows us to identify the dominant physical mechanisms that contribute to the creation and dissipation of circulation. 
In figure \ref{fig:V} we also show snapshots of the three-dimensional representation of the interface  corresponding to the volume used to carry out the computations necessary to calculate $D\boldsymbol{\Omega}/Dt$ and its constituent terms for the surfactant-laden and surfactant-free cases; this allows one to pinpoint the mechanisms primarily responsible for the interfacial structures observed. 
It is clearly seen from figure \ref{fig:circulation} that during the early stages of the flow, $\boldsymbol{\Omega}$ remains approximately constant. 
Inspection of panels (c)-(h) of figure \ref{fig:circulation} shows clearly that the rate of change of circulation is dominated by the mechanisms related to vortex diffusion $I_{\rm diff}$ and curvature $I_{\rm curv}$, with vortex tilting/shielding playing a relatively minor role. It is also clear that in the surfactant-laden jet case, the Marangoni contribution to $I_{\rm curv}$ dominates that associated with curvature derivatives. This observation further bolsters the claim that Marangoni stresses drive vorticity generation in the jet dynamics. 

The snapshots depicted in figure \ref{fig:V} for the surfactant-laden (panels (a)-(d)) and surfactant-free (panels (e)-(h)) cases have been chosen carefully so as to link the various stages of jet destabilisation to the prominent changes in the temporal variation of $I_{\rm diff}$, $I_{\rm tilt}$, $I_{\rm curv}$, and $D\boldsymbol{\Omega}/Dt$. Given the dominance of $I_{\rm curv}$ over the time range considered ($0\leq t \leq 40$), we focus on the variations in this quantity and its signature effects on the interfacial shape. Inspection of figures \ref{fig:circulation}(g) and \ref{fig:V}(a) reveals that the relatively gentle interfacial undulations are linked to variations of the Marangoni contribution to $I_{\rm curv}$ in the $x-y$ plane. The development of the more complex interfacial shapes, on the other hand, is accompanied by a concomitant rise in three-dimensionality of $I_{\rm curv}$ (in addition to significant contributions from the $x-$component of $I_{\rm curv}$). In the surfactant-free case, inspection of figures (\ref{fig:circulation})(d) and (h), and (\ref{fig:V})(e)-(h) shows that the interfacial jet evolution is accompanied by large variations in the $x-$component of $I_{\rm curv}$ and vorticity diffusion characterised by $I_{\rm diff}$.

Lastly, we plot in figure \ref{metrics} the effect of surfactants on the interfacial area, kinetic energy, defined as $E_k=\rho\int_V {\bf u}^2/2 dV$, and the enstrophy, $\varepsilon=\int_V |\boldsymbol{\omega}^2|dV$, normalised by their initial values, $A_0$, $E_{k0}$, $\varepsilon_0$, respectively.
After the onset of destabilization (defined when the interfacial surface has reached $A=1.025$), we observe that
the surfactant-induced effects discussed above, which include the interfacial vorticity jumps brought about by Marangoni stresses, and their effect on the production of circulation, and jet destabilisation mechanisms associated with vortex formation and spanwise reconnection, promote the delay in increase and subsequent reduction in interfacial area; these effects also lead to a delay in the decay of the jet kinetic energy as well as its enstrophy.   
%

\section{Concluding remarks}\label{sec:Con}
 
Three-dimensional numerical simulations of jet destabilisation and atomisation in the presence of a monolayer of insoluble  surfactants have been carried out for the first time. 
A phase diagram in the space of dimensionless surfactant elasticity and Weber number 
in the inertia-dominated region  
is presented in the limiting case where there is no vorticity production associated with jumps in material properties such as fluid density and viscosity; in the present work, surface tension forces and Marangoni stress give rise to variations in vorticity and circulation in addition to the vortex tilting/shielding and diffusion mechanisms. 
We have also derived formulae for the vorticity jumps across the interface due to Marangoni stresses, and equations that provide a breakdown of the rate of production of circulation within the jet into  constituent terms which we associate 
with vortex tilting/shielding, diffusion, and gradients in interfacial curvature and surface tension. \textcolor{black}{ The present theoretical formulation is expressed as a conservation law for circulation.
 We have focused on the limiting case where there is no vorticity production associated with jumps in material properties. Future studies should examine  situations characterised by  fluids with  different material properties.
} 

Then, we have analysed in details the  vortex-interface-surfactant interactions in the flow dynamics.
At early times, the presence of surfactants induces spanwise vortex reconnections brought about Marangoni-induced flow  resulting in the delay of the onset of destabilisation to the three-dimensional interfacial instabilities.
We also show that surfactant-induced Marangoni-stresses trigger the formation of 
hairpin-like structures whose head and legs extend in the streamwise direction.
Lastly, we have attempted to link the changes in interfacial topology to the mechanisms that influence the production of vorticity and circulation demonstrating a balance between  curvature gradients and diffusion for surfactant-free jets, and the dominance of Marangoni stresses in the surfactant-laden cases. 

The present results have been obtained for insoluble surfactants, and we acknowledge that experimental and numerical studies feature soluble surfactants which are dissolved in the liquid that issues from a nozzle to form the jet \citep{Sijs_2021,phd_thesis_CRCA}.
\textcolor{black}{
It is well known that the addition of surfactant-solubility will lead to additional richness and complexity. Although they do not affect the governing equations that describe the bulk fluid, they will change the boundary conditions that constrain them, resulting in a change in the flow dynamics. We can anticipate that a change of flow in the vicinity of the interface will have a detrimental effect on the coherent structures that emerge, subsequently affecting the close interplay between interface-vorticity-surfactant. These challenges will be the subject of future work. 
}  
 \\

Declaration of Interests. The authors report no conflict of interest.\\

This work is supported by the Engineering and Physical Sciences Research Council, United Kingdom, through the EPSRC MEMPHIS (EP/K003976/1) and PREMIERE (EP/T000414/1) Programme Grants. O.K.M. acknowledges funding from PETRONAS and the Royal Academy of Engineering for a Research Chair in Multiphase Fluid Dynamics. We acknowledge HPC facilities provided by the Research Computing Service (RCS) of Imperial College London for the computing time. AAC-P acknowledge the support from the Royal Society through a University Research Fellowship (URF/R/180016), an Enhancement Grant (RGF/EA/181002) and two NSF/CBET-EPSRC grants (Grant Nos. EP/S029966/1 and EP/W016036/1). D.J. and J.C. acknowledge support through HPC/AI computing time at the Institut du Developpement et des Ressources en Informatique Scientifique (IDRIS) of the Centre National de la Recherche Scientifique (CNRS), coordinated by GENCI (Grand Equipement National de Calcul Intensif) Grant 2022 A0122B06721. 

\appendix

\section{Kinematics}
\label{sec:app_kin}
We first develop an expression for $D\hatt/Dt$. We consider the motion of an infinitesimal fluid parcel in the plane of the interface, which is treated as a material surface. The position vector is $\mathbf{x}=\mathbf{x}(s,b,t)$ where $s$ and $b$ represent arc length distances along the $\hatt$ and $\hatb$ directions, respectively. At time, $t+\delta t$, to leading order in $\delta t$, we can write the following expression for the tangent to the interface at the fluid parcel which at time $t$ was located at $\mathbf{x}(0,0,t)$
\begin{equation}
    \tilde{\mathbf{t}}(t+\delta t)=\frac{\partial\mathbf{x}}{\partial s}(0,0,t)+\nabla \uvec\cdot\left(\frac{\partial\mathbf{x}}{\partial s}+\frac{\partial\mathbf{x}}{\partial b}\right)\delta t.
\end{equation}
Noting that $\hatt=\partial\mathbf{x}/\partial s$,  $\mathbf{b}=\partial\mathbf{x}/\partial b$, $\hatt\cdot\nabla\uvec=\partial \uvec/\partial s$, and $\hatb\cdot\nabla\uvec=\partial \uvec/\partial b$, this equation can be re-expressed as
\begin{equation}
    \tilde{\mathbf{t}}(t+\delta t)=\hatt+\left(\frac{\partial\uvec}{\partial s}+\frac{\partial\uvec}{\partial b}\right)\delta t.
\end{equation}
The magnitude of $\tilde{\mathbf{t}}(t+\delta t)$ is given by
\begin{equation}
    |\tilde{\mathbf{t}}|=1+\hatt\cdot\left(\frac{\partial \uvec}{\partial s}+\frac{\partial \uvec}{\partial b}\right)\delta t + O(\delta t)^2,
\end{equation}
and normalisation of $\tilde{\mathbf{t}}(t+\delta t)$ by this magnitude gives the following tangent unit vector $\hatt(t+\delta t)$
\begin{eqnarray}
\hatt(t+\delta t)&=&\frac{\hatt+\left(\frac{\partial\uvec}{\partial s}+\frac{\partial\uvec}{\partial b}\right)\delta t}{|\tilde{\hatt}(t+\delta t)|}=\frac{\hatt+\left(\frac{\partial\uvec}{\partial s}+\frac{\partial\uvec}{\partial b}\right)\delta t}{1+\hatt\cdot\left(\frac{\partial \uvec}{\partial s}+\frac{\partial \uvec}{\partial b}\right)\delta t}\nonumber\\
&=&\left(\hatt+\left(\frac{\partial\uvec}{\partial s}+\frac{\partial\uvec}{\partial b}\right)\delta t\right)\left(1-\hatt\cdot\left(\frac{\partial\uvec}{\partial s}+\frac{\partial\uvec}{\partial b}\right)\delta t+O(\delta t)^2\right)\nonumber\\
&=&\hatt+\left(\frac{\partial\uvec}{\partial s}+\frac{\partial\uvec}{\partial b}-\hatt\left(\hatt\cdot\left(\frac{\partial\uvec}{\partial s}+\frac{\partial\uvec}{\partial b}\right)\right)\right)\delta t + O(\delta t)^2.
\end{eqnarray}
From this expression, we can arrive at an approximate formula for $D\hatt/Dt$:
\begin{equation}
    \frac{D\hatt}{Dt}\sim \frac{\hatt(t+\delta t)-\hatt(t)}{\delta t}=\frac{\partial\uvec}{\partial s}+\frac{\partial\uvec}{\partial b}-\hatt\left(\hatt\cdot\left(\frac{\partial\uvec}{\partial s}+\frac{\partial\uvec}{\partial b}\right)\right)+O(\delta t).
    \label{eq:DtDt_3D}
\end{equation}

We now insert the following expression for $\uvec$ into $\uvec\cdot D\hatt/Dt$
\begin{equation}
    \uvec=(\uvec\cdot\hats)\hats+(\uvec\cdot\hatt)\hatt+(\uvec\cdot\hatb)\hatb,
    \label{eq:uvec}
\end{equation}
which yields
\begin{equation}
    \uvec\cdot\frac{D\hatt}{Dt}=(\uvec\cdot\hats)\hats\cdot\frac{D\hatt}{Dt}+(\uvec\cdot\hatb)\hatb\cdot\frac{D\hatt}{Dt}.
    \label{eq:udotDtDt}
\end{equation}
Substitution of Eq. (\ref{eq:DtDt_3D}) into $\hats\cdot D\hatt/Dt$ and $\hatb\cdot D\hatt/Dt$ gives
\begin{eqnarray}
    \hats\cdot\frac{D\hatt}{Dt}&=&\hats\cdot\left(\frac{\partial\uvec}{\partial s}+\frac{\partial\uvec}{\partial b}\right),\label{eq:sdotDtDt}\\
    \hatb\cdot\frac{D\hatt}{Dt}&=&\hatb\cdot\left(\frac{\partial\uvec}{\partial s}+\frac{\partial\uvec}{\partial b}\right),\label{eq:bdotDtDt}
\end{eqnarray}
where we have made use of $\hats\cdot\hatt=0$ and $\hatb\cdot\hatt=0$. We can re-express the RHS of Eqs. (\ref{eq:sdotDtDt}) and (\ref{eq:bdotDtDt}) as follows
\begin{eqnarray}
    \hats\cdot\frac{\partial\uvec}{\partial s}&=&\frac{\partial}{\partial s}(\hats\cdot\uvec)-\uvec\cdot\frac{\partial\hats}{\partial s},\label{eq:sduds}\\
    \hats\cdot\frac{\partial\uvec}{\partial b}&=&\frac{\partial}{\partial b}(\hats\cdot\uvec)-\uvec\cdot\frac{\partial\hats}{\partial b},\label{eq:sdudb}\\
    \hatb\cdot\frac{\partial\uvec}{\partial s}&=&\frac{\partial}{\partial s}(\hatb\cdot\uvec)-\uvec\cdot\frac{\partial\hatb}{\partial s},\label{eq:bduds}\\
    \hatb\cdot\frac{\partial\uvec}{\partial b}&=&\frac{\partial}{\partial b}(\hatb\cdot\uvec)-\uvec\cdot\frac{\partial\hatb}{\partial b}.\label{eq:bdudb_1}
\end{eqnarray}
Inserting Eq. (\ref{eq:uvec}) into the second term on the RHS of Eqs. (\ref{eq:sduds})-(\ref{eq:bdudb_1}), we obtain
\begin{eqnarray}
    \uvec\cdot\frac{\partial\hats}{\partial s}&=&\kappa_1(\uvec\cdot\hatt),\label{eq:udsds}\\
    \uvec\cdot\frac{\partial\hats}{\partial b}&=&\kappa_2(\uvec\cdot\hatb),\label{eq:udsdb}\\
    \uvec\cdot\frac{\partial\hatb}{\partial s}&=&0,\label{eq:udbds}\\
    \uvec\cdot\frac{\partial\hatb}{\partial b}&=&-\kappa_2(\uvec\cdot\hats), \label{eq:udbdb_1}
\end{eqnarray}
where the curvatures $\kappa_1$ and $\kappa_2$ are defined as follows
\begin{equation}
    \kappa_1=\hatt\cdot\frac{\partial\hats}{\partial s}, ~~~~~
    \kappa_2=\hatb\cdot\frac{\partial\hats}{\partial b}.
    \label{eq:kappa1_kappa2}
\end{equation}
In deriving Eqs. (\ref{eq:udsds})-(\ref{eq:udbdb_1}), we have noted that $\hatt\neq\hatt(b)$ and $\hatb\neq\hatb(s)$. Substitution of Eqs. (\ref{eq:udsds})-(\ref{eq:udbdb_1}) into Eqs. (\ref{eq:sduds})-(\ref{eq:bdudb_1}) and the resultant relations into Eqs. (\ref{eq:sdotDtDt}) and (\ref{eq:bdotDtDt}) respectively yields the following expressions for $\hats\cdot(D\hatt/Dt)$ and $\hatb\cdot(D\hatt/Dt)$
\begin{eqnarray}
    \hats\cdot\frac{D\hatt}{Dt}&=&\frac{\partial}{\partial s}(\uvec\cdot\hats)+\frac{\partial}{\partial b}(\uvec\cdot\hats)-\kappa_1 (\uvec\cdot\hatt)-\kappa_2(\uvec\cdot\hatb)\label{eq:sdotDtDt_2},\\
    \hatb\cdot\frac{D\hatt}{Dt}&=&\frac{\partial}{\partial s}(\uvec\cdot\hatb)+\frac{\partial}{\partial b}(\uvec\cdot\hatb)+\kappa_2 (\uvec\cdot\hats).
    \label{eq:bdotDtDt_2}
\end{eqnarray}
Substitution of Eqs. (\ref{eq:sdotDtDt_2}) and (\ref{eq:bdotDtDt_2}) into Eq. (\ref{eq:udotDtDt}) and re-arranging yields
\begin{equation}
    \uvec\cdot\frac{D\hatt}{Dt}=\frac{1}{2}\frac{\partial}{\partial s}\left[(\uvec\cdot\hats)^2+(\uvec\cdot\hatb)^2\right] + \frac{1}{2}\frac{\partial}{\partial b}\left[(\uvec\cdot\hats)^2+(\uvec\cdot\hatb)^2\right]  -\kappa_1(\uvec\cdot\hatt)(\uvec\cdot\hats). 
    \label{eq:uDtDt_appendix}
\end{equation}


\section{Near-interface normal stress jump}
\label{sec:app_pressure}
In order to generate a 3D version of the pressure gradient term in Eq. (\ref{eq:circulation_jump_3D_2}), we first consider the jump in the normal stress across the plane of the interface:
\begin{equation}
    p_2-p_1=-\sigma(\kappa_1+\kappa_2)+[[\mu \hats\cdot \mathbf{D}\cdot \hats],
    \label{eq:p2p1}
\end{equation}
where $\kappa_1$ and $\kappa_2$ are given by Eqs. (\ref{eq:kappa1_kappa2}). 
Substitution of Eq. (\ref{eq:uvec}) into $\nabla\cdot \uvec=0$ yields
\begin{eqnarray}
    \hats\cdot\nabla\uvec\cdot\hats&=&-\hatt\cdot\nabla\uvec\cdot\hatt-\hatb\cdot\nabla\uvec\cdot\hatb\nonumber\\
    &=&-\hatt\cdot\frac{\partial \uvec}{\partial s}-\hatb\cdot\frac{\partial\uvec}{\partial b},
    \label{eq:continuity_3D}
\end{eqnarray}
where we have set $\hatt\cdot \nabla\uvec=\partial \uvec/\partial s$ and $\hatb\cdot\nabla\uvec=\partial\uvec/\partial b$. We can re-express $\hatt\cdot (\partial \uvec/\partial s)$ and $\hatb\cdot (\partial \uvec/\partial b)$ as follows
\begin{eqnarray}
    \hatt\cdot\frac{\partial \uvec}{\partial s}&=&\frac{\partial}{\partial s}(\uvec\cdot\hatt)-\uvec\cdot\frac{\partial\hatt}{\partial s},\label{eq:tduds}\\
    \hatb\cdot\frac{\partial \uvec}{\partial b}&=&\frac{\partial}{\partial b}(\uvec\cdot\hatb)-\uvec\cdot\frac{\partial\hatb}{\partial b}.\label{eq:bdudb}
\end{eqnarray}
Substitution of Eq. (\ref{eq:uvec}) into $\uvec\cdot (\partial \hatt/\partial s)$ and $\uvec\cdot (\partial\hatb/\partial s)$ leads to
\begin{eqnarray}
    \uvec\cdot\frac{\partial\hatt}{\partial s}&=&-\kappa_1(\uvec\cdot\hats),\label{eq:udtds}\\
    \uvec\cdot\frac{\partial\hatb}{\partial b}&=&-\kappa_2(\uvec\cdot\hats),\label{eq:udbdb}
\end{eqnarray}
where, again, we have made use of the fact that $\hatt\neq\hatt(b)$ and $\hatb\neq\hatb(s)$.
Substitution of Eqs. (\ref{eq:udtds}) and (\ref{eq:udbdb}) into Eqs. (\ref{eq:tduds}) and (\ref{eq:bdudb}) and the resultant relations into Eq. (\ref{eq:continuity_3D}) gives
\begin{equation}
    \hats\cdot\nabla\uvec\cdot\hats=-\frac{\partial}{\partial s}\left[(\uvec\cdot\hatt)+(\uvec\cdot\hatb)\right]-(\kappa_1+\kappa_2)(\uvec\cdot\hats).
\end{equation}
Since $\hats\cdot\mathbf{D}\cdot\hats=2\hats\cdot\nabla\uvec\cdot\hats$, it follows that 
\begin{equation}
    \hats\cdot\mathbf{D}\cdot\hats=-2\frac{\partial}{\partial s}\left[(\uvec\cdot\hatt)+(\uvec\cdot\hatb)\right]-2(\kappa_1+\kappa_2)(\uvec\cdot\hats).
    \label{eq:sDs}
\end{equation}
Substitution of this equation into Eq. (\ref{eq:p2p1}) yields the following expression for the pressure jump
\begin{equation}
    [[p]]=-\sigma(\kappa_1+\kappa_2)-2[[\mu\left(\frac{\partial}{\partial s}\left[(\uvec\cdot\hatt)+(\uvec\cdot\hatb)\right]+(\kappa_1+\kappa_2)(\uvec\cdot\hats)\right)]].
    \label{eq:pjump_3D_appendix}
\end{equation}


\bibliographystyle{jfm}
\bibliography{jfm-instructions.bib}

\end{document}